\documentclass[envcountsame]{llncs}

\usepackage{graphicx}
\usepackage{amsmath}
\usepackage{color} % used ony for \todo{} command

\newcommand{\tmbraw}[3]{(\ensuremath{#1}, \ensuremath{#2}, \texttt{#3})}
\newcommand{\tmb}[3]{\tmbraw{#1}{#2}{#3}}

\newcommand{\token}[2]{\texttt{#1}$^{#2}$}

\begin{document}

\title{Anonymous On-line Communication Between Program Analyses}
\subtitle{(Technical report / Specification)}

\author{Marek Trt\'{i}k}
\institute{
    VERIMAG, Grenoble\\
    \email{trtikm@gmail.com}\\
    \today
    }

\maketitle

\begin{abstract}

We propose a light-weight client-server model of communication between existing
implementations of different program analyses. The communication is on-line and
anonymous which means that all analyses simultaneously analyse the same program
and an analysis does not know what other analyses participate in the
communication. The anonymity and model's strong emphasis on independence of
analyses allow to preserve almost everything in existing implementations. An
analysis only has to add an implementation of a proposed communication protocol,
determine places in its code where information from others would help, and then
check whether there is no communication scenario, which would corrupt its
result. We demonstrate functionality and effectiveness of the proposed
communication model in a detailed case study with three analyses: two abstract
interpreters and the classic symbolic execution. Results of the evaluation on
SV-COMP benchmarks show impressive improvements in computed invariants and
increased counts of successfully analysed benchmarks.

\end{abstract}

\section{Introduction}
\label{sec:Introduction}

%Computer scientists have implemented a lot of program analysis tools so far.
%Unfortunately, each of them has limited capabilities. That is a good reason for
%development of new tools with better analyses. Nevertheless, it is also good to
%reuse current tools, if there is a way how to push their limits further.

Our experience with several program analysis tools suggests that a piece of
information provided to an analysis at a right place may (substantially) improve
its result. Although it is usually not difficult to identify such places
(e.g.~evaluation of highly over-approximated program operators, like
bit-operators, pointer arithmetic, etc.) a problem is where to get the
information. The additional information may be known to another program
analysis.

We propose a light-weight communication model allowing a program analysis to
issue a query to other analyses (analysing the same program in the same time)
for an additional information which might hopefully push it towards to a better
result. The communication is in the style client-server. The server mediates the
communication between individual analyses (clients). Each client performs his
work on his own private data (like internal program representation, memory
model, etc.). Nevertheless, the communication itself has to be performed in term
all clients understand. Client's query thus typically begins with a conversion
of his internal data to common terms, and it ends by the opposite conversion.

The proposed communication model is based on the following key features which
together distinguishes our work from previous approaches in the field: %
\begin{itemize}
\item \emph{Independence of clients}: Each client implements his communication
protocol without consideration of other clients. This implies that in order to
add a new client into the communication, implementations of all former clients
remain exactly the same, i.e.~no line of code has to be changed, added, or
removed. Also, if we have $ n > 0 $ clients with their communication code
already established, then code of no client has to be changed in order to run
any of $ 2^n - 1 $ their possible combinations. %
\item \emph{Asynchronous execution of clients}: The communication model casts no
requirements for synchronization of computational steps of clients. It means
that some clients may perform forward analysis, other backward, and even speed
in which different clients explore their internal program representations may
differ. %
\item \emph{Reuse of current implementations}: The model attempts to preserve
both algorithms and related data structures of analyses which clients perform.
So, there is no need for their re-implementations. A client only has to add an
implementation of the communication protocol, determine places in its code where
to use information from others, and check that there is no communication
scenario which would corrupt his results.
\end{itemize}

An important part of the paper is also the case study, which represents the
first instance of the proposed communication model. There are three clients in
the case study. Two of them perform ``intervals'' and ``polyhedrons'' abstract
interpretations~\cite{AI_CousotCousot77} and the third is the classic symbolic
execution~\cite{SE_King76}. The description of the case study in this paper
serves as a detailed example. It is meant to be used as guide for building other
instances of the model. The evaluation of the case study was performed on
SV-COMP~\cite{SVCOMPURL} benchmarks for five combinations of clients. Results of
individual combination were compared per client in terms of strengthened
invariants (for abstract interpreters) and impact of the communication on
success/fail termination states (for all clients). Although results of
individual configurations are quite impressive for themselves, we further
observed that \emph{each} configuration can actually bring as \emph{not}
negligible improvements over others. This opens a new (originally unexpected)
application of the approach: Given $ n > 0 $ clients and a program to be
analysed, we build as many combinations from all $ 2^n - 1 $ as possible (we are
only limited by resources, like number of computers, threads, time, etc.) and we
analyse the program for each of them. Then we merge all results.

We explain the communication model in Sec.~\ref{sec:Communication}. We first
describe common terms in which the communication is performed and then we
present communication protocols for clients and the server. The case study is
presented in Sec.~\ref{sec:CaseStudy}. It starts with a description of all three
clients and the rest is dedicated to a detailed evaluation.

\section{Communication model}
\label{sec:Communication}

A \emph{client} is a program analysis tool or a program analysis inside a tool
which is able to  communicate with other clients during analysis of a given
program. A \emph{server} is a program utility mediating the communication
between clients. A client can only communicate with the server and has no
information about other clients, except their count. Data exchanged between the
server and a client are received in exactly the same order as they are sent.
There is a \emph{time-out} for whole the communication common to all clients and
the server. There is a single program analysed by all clients.

Given a program written in a certain programming language, a \emph{concrete
program state} is  an element of the concrete semantics of the language, an
\emph{abstract state space} is any subset of a client's interpretation
(e.g.~abstraction or generalisation) of the semantics of the language, and 
an \emph{abstract program state} is an element of an abstract state space.

Let us suppose a client receives an information from other clients (via the
server) that ``at address 1234 there is stored an integer greater than zero''. This
information is useless for the client until it is coupled with the following:
\begin{enumerate}
\item An actual value of the instruction counter. %
\item A set of program paths (from the program's entry to the actual instruction
counter) for which the information was inferred. %
\item A meaning of the address 1234 in order to map it into a particular address
in client's representation of program's memory.
\end{enumerate}

Points 1. and 2. are discussed in Sec.~\ref{sec:CanonicalProgram}, the point 3.
in Sec.~\ref{sec:CanonicalMemory}, and a concrete structure of the received
information in Sec.~\ref{sec:ClientProtocol}. The communication is described in
Sec.~\ref{sec:ClientProtocol} and \ref{sec:ServerProtocol} as client's and
server's communication protocols.

\subsection{Canonical program}
\label{sec:CanonicalProgram}

Given a program written in a certain programming language, a \emph{canonical
program} is a model of program's instruction counter designed and used by
clients for purposes of their communication. Before start of the communication
each client must have his copy of the canonical program.

There is a program representation which is very popular among clients: a
control-flow graph. We present a ``default'' recipe how clients may build a
canonical program from a control-flow graph of an elected client. Clients may
later separate the resulting implementation into a stand-alone utility program.

\begin{figure}[!htb]
\begin{tabular}{c}
\begin{tabular}{c|c}
\begin{tabular}{c|| c|c|c|c|c| c|c|c|c|c| c|c|c|c|c| c|c}
Token idx. &
1 & 2 & 3 & 4 & 5 &
6 & 7 & 8 & 9 & 10 &
11 & 12 & 13 & 14 & 15 &
16 & 17
\\ \hline\hline
Sample 1 & 
\texttt{a} & \texttt{=} & \texttt{0} & \texttt{;} & \texttt{b} &
\texttt{=} & \texttt{1} & \texttt{;} & & & & & & & & &
\\ \hline
Sample 2 & 
\texttt{if} & \texttt{(} & \texttt{a} & \texttt{>} & \texttt{0} &
\texttt{)} & \texttt{b} & \texttt{=} & \texttt{1} & \texttt{;} &
\texttt{else} & \texttt{b} & \texttt{=} & \texttt{2} & \texttt{;} & &
\\ \hline
Sample 3 & 
\texttt{if} & \texttt{(} & \texttt{a} & \texttt{>} & \texttt{0} &
\texttt{||} & \texttt{a} & \texttt{<} & \texttt{10} & \texttt{)} &
\texttt{b} & \texttt{=} & \texttt{1} & \texttt{;} & & &
\\ \hline
Sample 4 & 
\texttt{for} & \texttt{(} & \texttt{i} & \texttt{=} & \texttt{0} &
\texttt{;} & \texttt{i} & \texttt{<} & \texttt{10} & \texttt{;} &
\texttt{++} & \texttt{i} & \texttt{)} & \texttt{++} & \texttt{a} &
\texttt{;} &
\\ \hline
Sample 5 & 
\texttt{if} & \texttt{(} & \texttt{a} & \texttt{>} & \texttt{0} &
\texttt{)} & \texttt{goto} & \texttt{L} & \texttt{;} & \texttt{++} &
\texttt{a} & \texttt{;} & \texttt{L} & \texttt{:} & \texttt{--} &
\texttt{b} & \texttt{;}
\\
\end{tabular}
~~ & ~~
\begin{tabular}{c}
\includegraphics[scale=0.6]{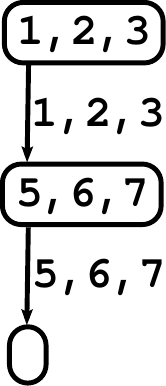}
\includegraphics[scale=0.6]{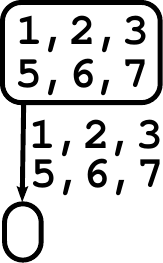}
\end{tabular}
\\
(a) & (b)
\end{tabular}
\\ \\
\begin{tabular}{c|c|c|c}
\includegraphics[scale=0.6]{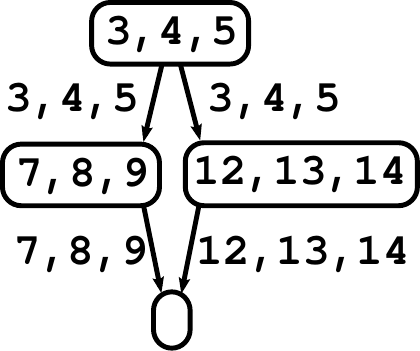} &
\includegraphics[scale=0.6]{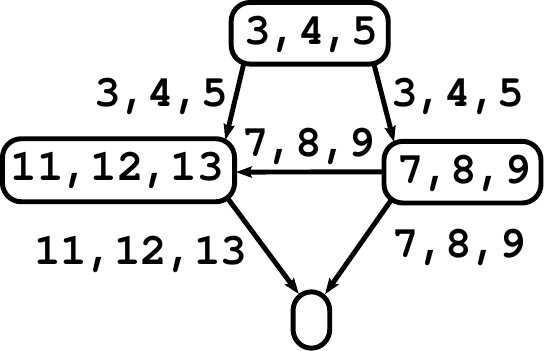} &
\includegraphics[scale=0.6]{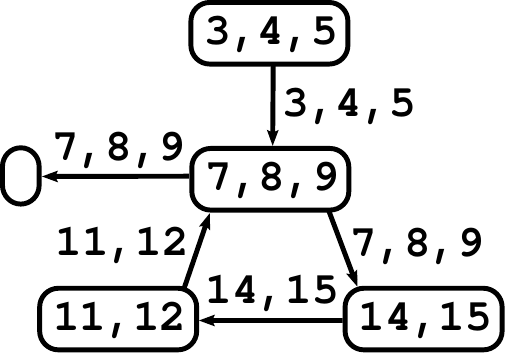} &
\includegraphics[scale=0.6]{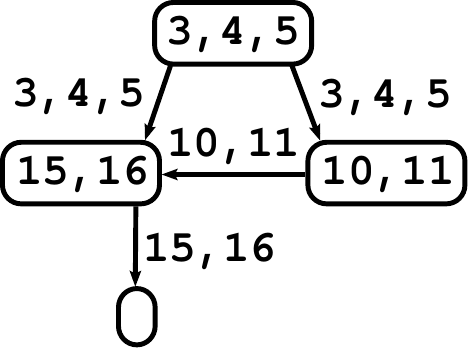}
\\
(c) & (d) & (e) & (f)
\end{tabular}
\end{tabular}
\caption{(a) Five C samples and indices of their lexical tokens. (b)-(f) A
possible canonical program for samples 1-5 respectively. Labels on edges
represent back-mapping of the input control-flow graph and labels inside nodes
represent the labeling in the final canonical program. Note that (b) shows 2
possible versions of a canonical program.}
\label{fig:TowardsCanonicalProgram}
\end{figure}

We assume in the recipe that the input control-flow graph satisfies the
following properties (which might require some pre- and/or post-processes of a
raw control-flow graph produced by the client):

\begin{itemize}
\item The instruction counter is modeled by nodes and edges represent possible
transitions of the counter during execution of the program. %
\item Each sub-program is represented by a separate component with a single
entry node and a single exit node. %
\item The control-flow graph is coupled with a set of all edges representing
calls of sub-programs. Let us denote it ``set of calls''.  %
\item The control-flow graph was constructed on a syntactically reshaped version
of the original program, where each lexical token was put into a separate line.
For C programs there are tools available for this functionality including
computation of the back-mapping of lines to the original program
\cite{BugstURL,CPAcheckerURL}. This reshaping is used to achieve one-to-one
mapping between program lines and indices of lexical tokens (since these indices
may not be directly available to a client). In
Fig.~\ref{fig:TowardsCanonicalProgram}~(a) there are five C samples with tokens
indexed. In the reshaped program these indices are equal to program lines. %
\item The control-flow graph is coupled with a ``back-mapping'' to the reshaped
source code of the analysed program, which maps each edge to a set of all those
program lines (i.e.~lexical tokens) s.t. an effect of the code at those lines is
associated with the edge. It does not imply that index of every program's token
must appear in the range of the back-mapping. Typically, indices of tokens whose
purpose is captured by the shape of the control-flow graph (e.g.~semicolons,
some brackets, keywords like \texttt{if}, \texttt{else}, \texttt{for}, etc.)
must not appear in the back-mapping. Such conventions are dependent on agreement
of clients. We require the back-mapping has the following property: If two edges
have non-empty intersection of their sets of lines, then the edges have the same
head node. Back-mappings for all 5 samples of
Fig.~\ref{fig:TowardsCanonicalProgram}~(a) are depicted in
Fig.~\ref{fig:TowardsCanonicalProgram}~(b)-(f) as labels of edges. Please,
ignore labels of nodes. A code associated with an edge may consist of one or
more statements (e.g.~a sequence of assignments, whole \texttt{if-then-else}
statement, or even several loops). This level of details is given by the input
control-flow graph. Of course, more detailed control-flow graphs are preferable.
For sample 1 we present at Fig.~\ref{fig:TowardsCanonicalProgram}~(b) a detailed
version (left) and a coarse one (right). %
\end{itemize}

\noindent
The recipe proceeds in the following three steps:

\begin{enumerate}
\item Discard all labels of all nodes and edges of the control-flow graph. %
\item Label each edge in the set of calls as \emph{call} edge and label each
edge with an out-degree greater then 1 as \emph{branching} edge. %
\item Label each node by a set of indices of program's lexical tokens. We
compute node's label as the union of values of the back-mapping function for
each out-edge from the node.
\end{enumerate}

Observe that labels of exit nodes of all sub-programs are necessarily empty. Now we can define meaning of a node of a canonical program:

\vspace{2mm} %
\emph{A statement that the instruction counter is at a certain node of a
canonical program means that the instruction counter is at the position in the
source code s.t. all lines of an instruction to be executed next belong to the
label of the node.} %
\vspace{2mm}

Canonical programs for all five samples of
Fig.~\ref{fig:TowardsCanonicalProgram}~(a) are depicted in
Fig.~\ref{fig:TowardsCanonicalProgram}~(b)-(f) (please ignore labels of edges;
labels of branching edges are omitted).

In the remainder of the paper we use the term \emph{node} as a representation of
a concrete value of the instructions counter in a canonical program. Similarly,
we use the term \emph{edge} for a possible transition of the counter between
nodes. Finally, we distinguish two kinds of edges: \emph{call} and
\emph{branching}, with obvious meanings.

\subsubsection{Mapping between canonical and internal program representation}~\\
\label{sec:MappingCanonicalInternalProgram}

\noindent Another important purpose of labels of canonical program's nodes is to
allow individual clients to build a mapping between nodes of the canonical
program and their own internal program representations (on which analyses are
actually performed). A construction of such mapping requires that all clients
build their internal program representations from the reshaped program instead
of the original one. Not all nodes of the canonical program has to appear in the
mapping. Nevertheless, a client should attempt to establish the mapping for as
much nodes as possible, since he may issue communication queries (or provide a
useful information to others) only at nodes with the mapping.

%A client may sill report results of his analysis in terms of the original program
%via the back-mapping from the reshaped to the original program. 

Since labels of nodes of a canonical program partition all tokens of the program
(possibly expect some auxiliary tokens as mentioned above), clients may store
only incomplete information about program lines in their internal program
representations. This information may differ from client to client. For example,
given a C statement \texttt{a=0;} one client may identify it only by the line of
its first token, i.e.~the line of \texttt{a}, another by the line of the last
token, i.e.~of \texttt{0} (or even of \texttt{;}), and another by the line of
the root operator, i.e.~of \texttt{=}. Moreover, client may perform several
equivalent program transformations, e.g.~code optimisations. Consider for
example a client which first translates a C program to LLVM assembly with
several optimisations enabled, and then he builds his internal program
representation from the optimised assembly. Such transformation typically reduce information about program lines.

It is very difficult (or impossible) to provide a general algorithm for
construction of such mapping, since internal program representation of a client
may be arbitrary. Here we provide a ``default'' recipe for construction of the
mapping for clients whose internal program representation has a form of a
control-flow graph in which nodes model the instruction counter:

Let $ (h,t) $ be an edge of an internal program representation s.t. there is a
source code line associated with the edge, and let $ n $ be a node of the
canonical program containing the line in its label. We extend the mapping in any
of the following three cases:
\begin{enumerate}
\item If in-degrees of both $ n $ and $ h $ are zeros, then link $ h $ with $ n
$. %
\item If $ t $ is an exit node and $ n $ has a successor with the empty label
and with the out-degree 0, then link $ t $ with the successor of $ n $. %
\item If all successor edges of $ t $ have lines associated and all the lines
belong to the label of a single successors node $ n' $ of the node $ n $ and $ n
\neq n' $, then link $ t $ with $ n' $.
\end{enumerate}

\begin{figure}[!htb]
\begin{tabular}{c}
\begin{tabular}{l}
\token{int}{1} \token{btree\_contains}{2}\token{(}{3}\token{int}{4}
\token{key}{5}\token{,}{6}\token{btree\_node}{7}\token{*}{8}
\token{node}{9}\token{)}{10}\token{\{}{11}
\\ %
\token{~~int}{12} \token{i}{13}\token{=}{14}\token{0}{15}\token{;}{16}
\\ %
\token{~~while}{17}\token{(}{18}\token{i}{19} \token{<}{20}
\token{node}{21}\token{->}{22}\token{nkeys}{23} \token{\&\&}{24}
%\\ %
%\texttt{~~~~~~~~~}
\token{node}{25}\token{->}{26}\token{key}{27}\token{[}{28}\token{i}{29}\token{]}{30}
\token{<}{31}\token{key}{32}\token{)}{33}
\\ %
\token{~~~~++}{34}\token{i}{35}\token{;}{36}
\\ %
\token{~~if}{37}\token{(}{38}\token{i}{39} \token{<}{40}
\token{node}{41}\token{->}{42}\token{nkeys}{43} \token{\&\&}{44}
%\\ %
%\texttt{~~~~~~}
\token{node}{45}\token{->}{46}\token{key}{47}\token{[}{48}\token{i}{49}\token{]}{50}
\token{==}{51}\token{key}{52}\token{)}{53}
\\ %
\token{~~~~return}{54} \token{1}{55}\token{;}{56}
\\ %
\token{~~if}{57}\token{(}{58}
\token{node}{59}\token{->}{60}\token{child}{61}\token{[}{62}\token{i}{63}
\token{+}{64}\token{1}{65}\token{]}{66}
\token{==}{67}\token{NULL}{68}\token{)}{69}
\\ %
\token{~~~~return}{70} \token{0}{71}\token{;}{72}
\\ %
\token{~~return}{73} \token{btree\_contains}{74}\token{(}{75}\token{key}{76}\token{,}{77}
\token{node}{78}\token{->}{79}\token{child}{80}\token{[}{81}\token{i}{82}
\token{+}{83}\token{1}{84}\token{]}{85}\token{)}{86}\token{;}{87}
\\ %
\token{\}}{88}
\end{tabular}

\\
(a)
\\\\
\begin{tabular}{ccc}
\includegraphics[scale=0.6]{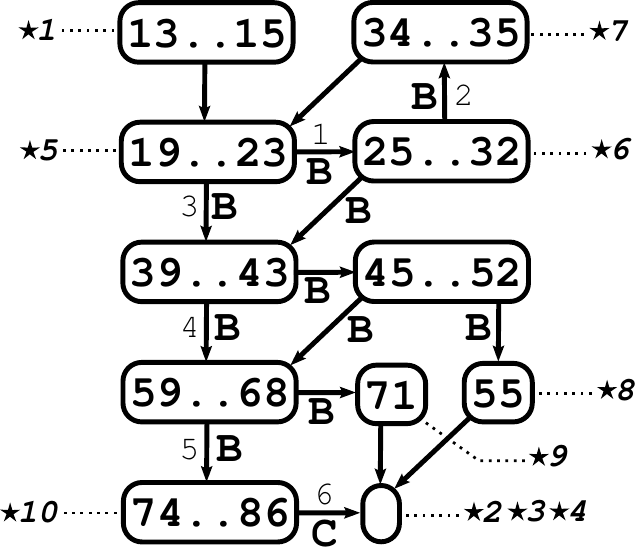}
& ~~~~~~~~~ &
\includegraphics[scale=0.6]{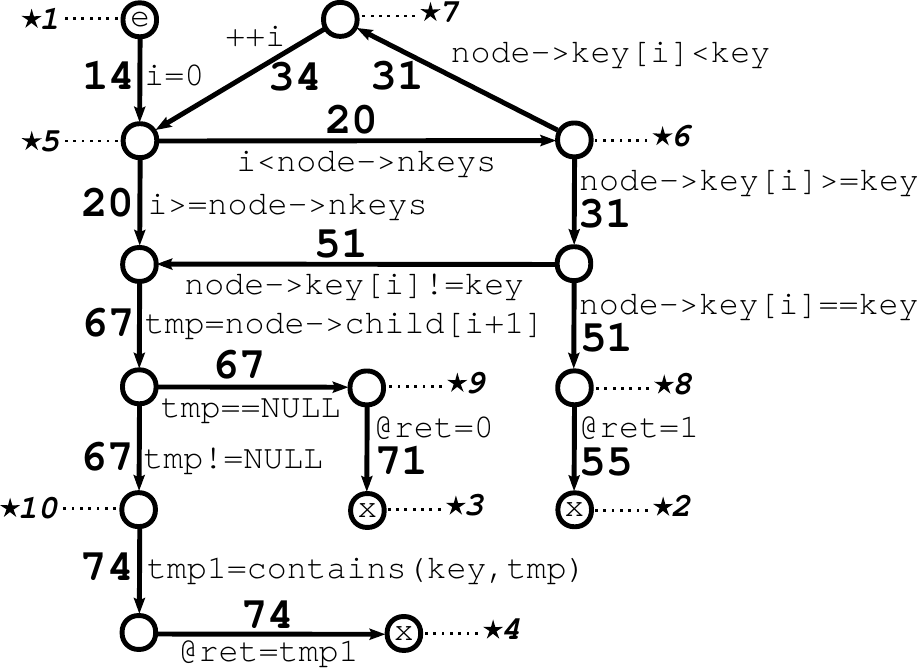}
\\
(b) & & (c)
\end{tabular}
\end{tabular}
\caption{A mapping between a canonical program (b) and an internal program
representation (c) both constructed for a C function (a) with tokens indexed. In
(b) ignore numbers labeling edges and labels \texttt{B} and \texttt{C} stand for
branching and call kinds respectively. (c) references the source code (a) via
indices of root tokens of program expressions (see bolt numbers labeling edges).
Links between nodes of (b) and (c) were computed by the default recipe. Namely,
the link $ \star1 $ was set according to the case 1, links $
\star2,\star3,\star4, $ according to the case 2, and all others according to the
case 3.} %
\label{fig:BTreeContains}
\end{figure}

\noindent
We demonstrate an application of the recipe in an example depicted in Fig.~\ref{fig:BTreeContains}.

\subsubsection{Context}~\\
\label{sec:Context}

A \emph{filter} is a set of kinds of edges of a canonical program (remember that
we distinguish only two kinds of edges in this presentation). A \emph{context}
is a list of edges of a canonical program s.t. each edge has a kind which
belongs to a given filter. For example, any context constructed for the filter $
\{ \mathit{call} \} $ may contain only call edges.

A context constructed for a certain filter represents a set of program paths
starting at the program's entry node. A path belongs to the set if and only if
the context is equal to a list of edges constructed from the path s.t. each edge
with a kind belonging to the filter is preserved and any other is removed.

A context coupled with a node represents the set of its paths restricted to
those terminating at the node.

We demonstrate of a meaning of a pair ``node,context'' on example canonical
program in Fig~\ref{fig:BTreeContains}~(b) constructed for the C function in
Fig~\ref{fig:BTreeContains}~(a). We consider several contexts coupled with the
same node labeled by \{13..15\}. We list the contexts in the following table
together with their creation filters and with description of set of paths they
represent:

\begin{center}
\begin{tabular}{c|c|p{6.8cm}}
Context & Filter & Descritption \\ \hline\hline %
$ [] $ & $ \emptyset $ & All paths reaching the node. \\ %
$ [6,6,6,6] $ & \{\textit{call}\} & All possible paths to the node performing 4
recursive calls \token{btree\_contains}{74}. \\ %
$ [1,2,1,2,3,4,5,6] $ & \{\textit{call}, \textit{branching}\} & The path looping
twice in the \token{while}{17} loop, then taking false branches of both
\token{if}{37} and \token{if}{57} (without evaluation of \token{==}{51}) and
then applying the recusrice call \token{btree\_contains}{74}.
\end{tabular}
\end{center}

Since client's mapping between a canonical program and an internal program
representation may be only partial, it may be impossible for the client to
create some contexts. For example, if the tail node of some branching edge is
not mapped, then the client cannot create a context for a filter with
\textit{branching} edges kind representing program paths passing only through
that edge. Therefore, properties of client's analysis (like call-sensitivity or
path-sensitivity) may not always be  reflected in the communication through
contexts due to incompleteness of the mapping.

Let us return back to the example above. Now we try to translate the node
(labeled by \{13..15\}) and all three contexts in the table to the internal
program representation in Fig.~\ref{fig:BTreeContains}~(c). The link $ \star1 $
allow as to move from the node of the canonical program to the corresponding
node in the internal program representation. In this case we move to the entry
node. Note that in case there is no link for the node then we could not continue
with the example.

The first context $ [] $ can be translated without any loss of accuracy,
i.e.~the set of path represented by the context does not have to be extended.
The reason is simple, the context represents all paths from the programs entry
up to our translated entry node (including any number of recursive calls, etc.).

The second context $ [6,6,6,6] $ can also be translated without a loss of
accuracy. That is because links $ \star10 $ and $ \star4 $ uniquely map the edge
6 to the path consisting of two edges associated the line 74 and just one of
then is the call edge. The call site is thus uniquely identified.

The third context $ [1,2,1,2,3,4,5,6] $ contains edges, namely 3, 4, and 5,
which cannot be translated to the internal program representation, since we do
not have links for tail nodes of the edges 3 and 4. So, we have to reduce the
context into $ [1,2,1,2,6] $. Now we can translate all edges: Edges 1 and 2 of
the loop translate to edges with lines 20 and 31 in the corresponding loop (we
used links $ \star5 $, $ \star6 $, and $ \star7 $ for the translation), and the
edge 6 is translated the same way as for the second context. In terms of the
canonical program (Fig.~\ref{fig:BTreeContains}~(b)) the reduced context $
[1,2,1,2,6] $ extend the set of paths over the original one (as described in the
table) by those paths which go through the node labeled by \{45..52\}.
Nevertheless, if we look into the internal program representation
(Fig.~\ref{fig:BTreeContains}~(c)), then both the original and the reduced
context represent the same set of paths, because of the optimised shape of the
program representation. Namely, there is only one possible path from the loop
(left along the edge with the line 20) to the call site. Therefore, a reduction
of a context does not necessary implied a loss of precision.

Now we look at the opposite direction. We consider the following path in the
internal program representation expressed in term of lines labeling edges:
[14,20,31,34,20,31,34,20,67,67,74]. We now translate this path to the canonical
program in a form of a context constructed for a filter the filter
\{\textit{call}, \textit{branching}\}. The edge 14 can be translated to the
canonical program, since we have links $ \star1 $ and $ \star5 $. Nevertheless,
the corresponding edge is neither call nor branching edge in the canonical
program. So, we skip it. Edges 20 and 31 can be translated to branching edges 1
and 2 respectively. We further skip both edges 34 for the same reason as for the
edge 14. Next, the edge 20 and both edges 67 cannot be mapped because of lack of
links between nodes. And the call edge 74 can be mapped to the call edge 6.
Therefore, we end up with the context $ [1,2,1,2,6] $. This is the reduced
context we discussed above. Observe that we were unable to reconstruct the more
precise context $ [1,2,1,2,3,4,5,6] $ due to lack of mapping between nodes.

Finally, observe that the exit node of the canonical program is mapped to three
exit nodes of the internal program representation. So, if we want to translate a
pair ``the exit node, context'' to the internal program representation, then we
may get three translated pairs ``node$ _{\star2}' $,context$ ' $'', ``node$
_{\star3}' $,context$ ' $'', and ``node$ _{\star4}' $,context$ ' $'', which are
supposed to be considered together (as union).

Now we define a special form of context reduction using a filter. Given a
context and a filter, a \emph{reduced} context by the filter is constructed from
the given context s.t. each edge whose kind does not belong to the intersection
of the filter and the construction filter of the context is removed.

\paragraph{Multi-threading}
\label{sec:Multi-threading}

Different threads may be distinguished through a context: A construct which
causes a creation of a new thread can be modeled in a canonical program by an
artificially introduced branching (e.g.~by two parallel edges). Clients may
decide to use special labels and filters for these edges. We do not introduce
them, because they are not necessary in further presentation.

\subsection{Canonical memory}
\label{sec:CanonicalMemory}

Given a program written in a certain programming language, a \emph{canonical
memory} is a model of program's memory designed and used by clients for purposes
of their communication. Organisation of memory may differ significantly for
different languages, consider for example C and Prolog. Here we provide a
``default'' canonical memory where we model only address space and dereferences
of addresses (which is sufficient for purposes of communication). The model is
thus very low-level.

Addressing of program's memory is based on the common segment-offset style.
Given a program and an ordered list of all its identifiers referencing
memory\footnote{Using the terminology of C language: All those which may be used
as l-values.} a \emph{segment} is any integer between 0 and the number of
identifiers in the list. The segment 0 is reserved for modeling of the standard
invalid memory, represented by the \texttt{NULL} pointer in some languages.
Other segments represent indices of identifiers in the list. We recommend to
order identifiers in the list according to their token-indices (each identifier
is also a lexical token), since there can be several identifiers with the same
name. An \emph{offset} is any non-negative integer.

A sequence of bytes starting at a given segment and offset can have any of the
following \emph{type} interpretations:
\begin{center}
\begin{tabular}{rl}
\texttt{i8}, \texttt{i16}, \texttt{i32}, \texttt{i64} ~~~&~~~
A signed 8,16,32,64-bit integer.  \\
\texttt{u8}, \texttt{u16}, \texttt{u32}, \texttt{u64} ~~~&~~~
An unsigned 8,16,32,64-bit integer.  \\
\texttt{f32}, \texttt{f64} ~~~&~~~
A 32,64-bit floating point number.  \\
\texttt{seg}, \texttt{off} ~~~&~~~
A representation of addresses (pointers). \\
%\texttt{int}, \texttt{real} ~~~&~~~
%An integer and a real number of any magnitude   \\
\end{tabular}
\end{center}
A concrete encoding of these types is given either by the program's deployment
machine, or on a mutual agreement of clients. \texttt{seg} and \texttt{off}
should be encoded as unsigned integers big enough to capture available address
space. Data of composed types can be communicated with other clients only after
their decomposition to elements of basic types above.

A \emph{dereference} is a triple consisting of a segment expression, an offset
expression, and a type. A segment expression is either a segment, or a
dereference of the type \texttt{seg}. An offset expression is any integer
expression over integer constants, interpreted functions for addition,
multiplication, etc., and over dereferences of either any integer type or the
type \texttt{off}. A dereference represents a type-interpreted values of bytes
pointed to by its segment and offset expressions. A dereference is called
\emph{basic} if neither segment nor offset expression contains any dereference.
%We write a dereference with components $ s $, $ o $, and $ t $ as
% \tmb{s}{o}{t}.

A value in the memory referenced by a program identifier is directly denoted by
a basic dereference. A value stored in a non-leaked memory in the program's heap
can always be referenced by a finite sequence of dereferences starting with a
certain program identifier. The sequence can always be expressed by nesting of
dereferences inside segment and offset expressions of a non-basic dereference.
Memory of a record on the top of the program's call stack is directly read by
basic dereferences. Memory of the records deeper in the stack is accessed via
non-basic dereferences exactly the same way as the memory in the program's heap.
\begin{figure}[!htb]
\begin{center}
\begin{tabular}{c}
\begin{tabular}{cc}
\begin{tabular}{l}
\texttt{void foo(int** p) \{\}}\\
\texttt{void main() \{}\\
\texttt{~~int* p=(int*)malloc(100);}\\
\texttt{~~foo(\&p);}\\
\texttt{~~free(p);}\\
\texttt{\}}
\end{tabular}
~~~~~~&~~~~~~
\begin{tabular}{c|c}
Identifier & Segment \\ \hline
\texttt{foo} & 1 \\
\texttt{foo::p} & 2 \\
\texttt{main} & 3 \\
\texttt{main::p} & 4 \\
\end{tabular}
\\
(a) & (b)
\end{tabular}
\\ \\
\begin{tabular}{p{5cm}|p{6cm}}
Dereference & Description \\ \hline\hline
\tmb{2}{0}{seg} & Segment of an address stored in \texttt{foo::p}. \\ \hline %
\tmb{2}{0}{off} & Offset of an address stored in \texttt{foo::p}. \\ \hline %
\tmb{\tmb{2}{0}{seg}}{\tmb{2}{0}{off}}{seg} & Segment of an address stored in
\texttt{main::p}. \\ \hline %
\tmb{\tmb{2}{0}{seg}}{\tmb{2}{0}{off}}{off} & Offset of an address stored in
\texttt{main::p}. \\ \hline %
\tmb{\tmb{\tmb{2}{0}{seg}}{\tmb{2}{0}{off}}{seg}}{\tmb{\tmb{2}{0}{seg}}{\tmb{2}{0}{off}}{off}+12}{i32} & A value of the fourth element of the allocated array (we assume \texttt{sizeof(int)==4}).
\end{tabular}
\\
(c)
\end{tabular}
\end{center}
\caption{Example for dereferences: (a) C program to be considered. (b) An
ordered list of program's identifiers and a mapping to segments. (c) Five
dereferences and their descriptions. It is assumed the instruction counter is
inside the function \texttt{foo}.}
\label{fig:Dereferences}
\end{figure}

Note that threads do not have to be considered here, since they may be
distinguished through a context (see page~\pageref{sec:Multi-threading}). In
Fig.~\ref{fig:Dereferences} there is an example how a canonical memory is used.

\subsection{Client's side of the communication protocol}
\label{sec:ClientProtocol}

Client's communication protocol is a specification of six functions. A client
has to  implement each of them. The specifications are the following: \newline

\noindent\texttt{get\_values(node,context,dereferences,self\_call) -> formula}

A client is queried for a superset of all concrete program states represented
by all his abstract states related to the \texttt{node} and the
\texttt{context}. In other words, the client is queries for an
over-approximation of his current knowledge about program's behaviour regarding
the \texttt{node} and the \texttt{context}.

The passed \texttt{dereferences} can be used to determine what superset will be
computed. Typically, knowledge in the abstract states about memory addressed by
no dereference in \texttt{dereferences} is ignored.
For example, let us suppose that program variables \texttt{a} and \texttt{b} are
presented by basic dereferences \tmb{2}{0}{i32} and \tmb{3}{0}{i32}
respectively, and $ \texttt{dereferences} = \{ \tmb{2}{0}{i32} \} $. Then the
client may ignore his knowledge about \tmb{3}{0}{i32} when computing an answer
to the query. On the other hand, if client's knowledge about \texttt{a} is
expressed in a form of a constraint over several variables, like $ \texttt{a} +
2(\texttt{b} - 3) < 0 $, then \tmb{3}{0}{i32} can be considered for the answer
as well.

\newcommand{\deref}[3]{\texttt{(DEREF\_#3 #1 #2)}}

The superset is then encoded as a quantifier-free first order logic formula over
dereferences in a standard format all client agreed on,
e.g.~SMT-LIB2~\cite{SMTLIBURL}. Any such formula may only contain interpreted
symbols from theories of integers, Peano's arithmetic, and reals. The only
uninterpreted symbols are related to dereferences: A dereference with a segment
expression $ s $, an offset expression $ o $, and a type $ t $ is encoded as an
application of a binary uninterpreted function $ \texttt{DEREF\_}t $ to
arguments $ s $ and $ o $. It is not necessary to provide declarations of these
uninterpreted functions, since all clients know their exact meanings. %
Returning back to the example above, client's answer in SMT-LIB2 format may look
like this: \texttt{(< (+ (+ \deref{2}{0}{i32} (* 2 \deref{3}{0}{i32})) (- 6))
0)}.

It is further highly recommended for clients to agree on additional syntactical
restrictions for formulae. They should not restrict expressivity, they should
ease decoding of answers. Clients may for example consider these: The formula is
in DNF, right hand side of any (in)equality is always a numeric constant,
distributive laws are applied where possible, constant expressions are
simplified, no use of binary subtractions, etc. The last tree restrictions were
applied in the example above.

If the client does not have established the mapping for the passed
\texttt{node}, then he returns a simple tautology, e.g.~\texttt{(= 1 1)}.

The argument \texttt{self\_call} is \texttt{true} if the client is the one who
initiated the query. Otherwise it is \texttt{false}. \newline

\noindent\texttt{get\_coverage(node,context) -> coverage}

A client is supposed to compute a real number between 0 and 1. This number
represents a measure, how much the current client's abstract state space covers
the set of all concrete program states determined by the \texttt{node} and the
\texttt{context}. The value 0 means that the client covers no concrete state and
the value 1 means that all concrete states are covered. Values in-between 0 and
1 can be computed from a progress observed in the abstract state space from
the start of the analysis up to the current time point. We provide more details
in Sec.~\ref{sec:CoverageOracle}.

The client returns 0, if he does not have established the mapping for the passed
\texttt{node}. \newline

\noindent\texttt{is\_relevant\_coverage(node,context,coverage,self\_coverage) -> bool}

A client is asked whether \texttt{coverage} computed by some other client is big
enough so that this client would accept result of a call to the function
\texttt{get\_values} of the other client. Clearly, the client returns
\texttt{false} whenever \texttt{coverage} is 0. The argument
\texttt{self\_coverage} is the result of the function \texttt{get\_coverage} of
this client. This values may be useful for the client to make the decision. We
provide more details to the decision in Sec.~\ref{sec:CoverageOracle}.

The client returns \texttt{false}, if he does not have established the mapping
for the passed \texttt{node}.
\newline

\noindent\texttt{is\_memory\_over\_approximated() -> bool}

A client is supposed to return \texttt{false} if any of the following cases:
\begin{enumerate}
\item The client has updated an abstract state but he has not send a notification
\texttt{on\_location\_outdated} to the server about the update yet.
\item Client's function \texttt{can\_improve\_memory\_over\_approximation} returns \texttt{true}.
\end{enumerate}
Otherwise, the client returns \texttt{true}. Observe that the function may
return \texttt{true} even though client's abstract state space currently does
not expresses all possible behaviour of the analysed program. This may happen
when client's computation used values from other clients (via
\texttt{get\_values}) for coverages less then 1. \newline

\noindent\texttt{can\_improve\_memory\_over\_approximation() -> bool}

A client returns \texttt{true} if he sees a chance in making a progress
towards expressing all possible behaviour of the analysed program. Otherwise, he
returns \texttt{false}. Note that the answer \texttt{true} does not imply the
client will necessarily ever do such a progress. \newline

\noindent\texttt{on\_location\_outdated(node,context,coverage,self\_call) -> none}

A client is notified through this function that abstract state space of some
client has been updated. The \texttt{node} and the \texttt{context} provide a
closer description what set of concrete program states the update was relevant
for.

The argument \texttt{coverage} is the result of the function
\texttt{get\_coverage} of the client who issued this notification on the updated
version of his abstract state space. The argument \texttt{self\_call} is
\texttt{true} if the client is the one who issued the query. Otherwise it is
\texttt{false}.

If the client does not have established the mapping for the passed
\texttt{node}, he should perform a graph search in the canonical program to find
all nearest reachable nodes with mappings and apply the notification for each of
them. Note that the \texttt{context} has to be extended/reduced for each node
separately using client filter by considering all paths between the passed and
the searched node. \newline

A passed \texttt{context} to some of the protocol functions may contain edges
whose nodes do not appear in the mapping between the canonical program and
client's internal program representation. Such edges has to be removed. This
removal may produce a context representing a larger set of program paths.
Nevertheless, clients computation in protocol functions must be based on such
larger contexts.

All protocol functions above are supposed to terminate quickly. We cannot give
exact complexities, but each of them should terminate (much) sooner than
client's algorithm which transforms an abstract state along a program's edge.

If an evaluation of any function above does not finish before the time-out, then
the client may cancel its evaluation without sending any response to the server.
Also, any queries from the server issued after the time-out may be ignored.

A client may assume correctness of data received from other clients in
individual queries. He may also assume other clients meet all requirements given
in this section. And he may further assume the server meets all requirements
given in the section below. But he cannot make any other assumptions about the
communication. It means that based on the mentioned assumptions he has to check
by himself for the correctness of implementations of his communication related
code in isolation from others by considering all possible scenarios how the
communication may affects his results and performance.

\subsection{Server's side of the communication protocol}
\label{sec:ServerProtocol}

Server's communication protocol is a specification of four functions. A server
has to  implement each of them. The specifications are the following: \newline

\noindent\texttt{get\_values(node,context,trace,dereferences,client) -> formula}

A client \texttt{client} calls this function in order to receive a knowledge
about memory content from all clients. The server handles the query as follows:
\begin{verbatim}
   1 formula := client.get_values(node,context,dereferences,true)
   2 cov0 := client.get_coverage(node,context)
   3 for every other client C do
   4   cov := C.get_coverage(node,context)
   5   if client.is_relevant_coverage(node,context,cov,cov0)) then
   6     F := C.get_values(node,context,dereferences,false)
   8     formula := make_conjunction(formula,F)
   9 return formula
\end{verbatim}
The use of conjunction as the connective between formulae has the meaning of
intersection of sets of concrete program states represented by the formulae.
\newline

\noindent\texttt{is\_memory\_over\_approximated() -> bool}

A client calls this function in order to check whether abstract state spaces of
all clients caver all possible behaviour of the analysed program. The server
handles the query by invoking functions \texttt{is\_memory\_over\_approximated}
of all clients and returns \texttt{true} if all the clients respond by
\texttt{true} and no \texttt{on\_location\_outdated} query is being evaluated
during evaluation of this query. Otherwise, \texttt{false} is returned. \newline

\noindent\texttt{can\_improve\_memory\_over\_approximation() -> bool}

A client calls this function in order to check whether at least one client can
hopefully get his abstract state spaces closer to complete coverage of all possible
behaviour of the analysed program. The server handles the query by invoking
functions \texttt{can\_improve\_memory\_over\_approximation} of all clients and
returns \texttt{true} if at least one client respond by \texttt{true}.
Otherwise, \texttt{false} is returned. \newline

\noindent\texttt{on\_location\_outdated(node,context,coverage,client) -> none}

A client has to calls this function in order to notify others about the fact
that his abstract state space has been updated. The passed \texttt{client} is
the one who issues the notification. The \texttt{node} and the \texttt{context}
provide a closer description what set of concrete program states the update is
relevant for. The server is responsible for delivering the notification to all
clients. So, the server invokes functions \texttt{on\_location\_outdated} of all
clients with the last argument \texttt{true} in case of the \texttt{client} and
with \texttt{false} for all other clients. \newline

A client may call any of these function any time before the time-out. If an
evaluation of a query does not finish before the time-out the server may cancel
the evaluation without any response to any client. Also, queries issued after
the time-out may be ignored by the server.

\subsection{Issuing {\tt get\_values} for highly over-approximated operators}
\label{sec:IssueingGetValues}

Although it is completely up to a client to decide where and when to issue
\texttt{get\_values} queries, there is rather general scenario for recommended
application of such queries: It is quite common that a client highly
over-approximates effects of some language operators. Consider for example C
language and bit-operators (like conjunction, negation, etc.), pointer
arithmetic, or some math functions (like \texttt{sqrt}, \texttt{pow},
\texttt{sin}, etc.). The client may take advantage of the communication to
improve the precision. We demonstrate this on a the following simple example.

Let us consider a client who does not support pointers and who is about to
execute a C assignment \texttt{i=*p}, where \texttt{i} and \texttt{p} have types
\texttt{int} and \texttt{int*} respectively. Although the client cannot evaluate
the  sub-expression \texttt{p} (since it is a pointer), he can issue a query to
other clients. Let us say the answer is $ \texttt{(and (= \deref{2}{0}{seg} 3)
(= \deref{2}{0}{off} 4))}$, where we assume that \texttt{p} is mapped to the
segment 2. If the basic dereference \tmb{3}{4}{i32} corresponds to a program
variable, \texttt{j} say, than the client may evaluate the assignment as it was
\texttt{i=j}. Otherwise, the client may issue another query for the dereference
\tmb{3}{4}{i32}. A received answer represents (an over-approximation of) an
\texttt{i32} value to be assigned to \texttt{i}.

\subsection{Decoding answers from {\tt get\_values} can loose precision}
\label{sec:DecodingGetValues}

A formula returned from a \texttt{get\_values} query may encode an information
whose precision is beyond expressivity of client's abstract state space. In any
such case a lose of precision is inevitable. Another source of the lose can lie
in an incomplete implementation of client's decoding procedure. Indeed, a
returned formula can be arbitrarily complex. A useful information may be encoded
in several transitively dependent constraints. Client's decoding procedure may
simply ignore them. It is worth noting that a complete implementation of the
decoding may lead to a non-trivial programming effort.

We show this on the example from the previous sub-section, where we suppose the
client performs the classic interval analysis and we consider the situation when
the client receives an answer for the second query, i.e.~for the dereference
\tmb{3}{4}{i32}. Possible answers for the query may look like these:
\begin{itemize}
\item \texttt{(= \deref{3}{4}{i32} 3)}
\item \texttt{(and (< (- \deref{3}{4}{i32}) (- 3)) (< \deref{3}{4}{i32} 10))}
\item \texttt{(or (< \deref{3}{4}{i32} 3) (= \deref{3}{4}{i32} 10))}
\item \texttt{(= (+ (* 2 \deref{3}{4}{i32}) \deref{4}{0}{i32}) 10)}
\end{itemize}
Although all of these formulae can be converted to intervals, the client may
decide to restrict implementation of his decoding procedure to first two
formulae, since decoding of others would lead to a loss of precision anyway.
Therefore, the last two formulae are decoded as they are tautologies (i.e.~any
possible value).

\subsection{Coverage oracle}
\label{sec:CoverageOracle}

Client's protocol functions \texttt{get\_coverage} and
\texttt{is\_relevant\_coverage} require an estimation of how many concrete
program states are represented by client's current abstract state space.

The estimation can be based on observation of progress of client's abstract
state space from the beginning of the analysis up to the current time. For a
given node and context the client may for example observe changes in counts of
abstract states, sizes of abstract states, or counts of updates of abstract
states in time. In order to compute an estimate, the client needs to
compare these ``progress properties'' with similar ones in client's knowledge
base. We can build a knowledge base for a client s.t. we run the client in
isolation (i.e.~without any communication) on a sufficiently large training set
of benchmarks. For each benchmark the client observes a progress of his progress
properties. Then we relate properties of analysed benchmarks (like number of
tokens, number of loops, recursive functions, pointer dereferences, etc.) with
functions from analysis time (scaled into the interval $ [0,1) $) to the
corresponding values of progress properties. In order to use the knowledge base
the client need three functions. A distance function between program properties,
a distance function between progress properties, and a ``progress'' function
which for a given program properties and a passed (and scaled) analysis time
finds a closed progress properties in the knowledge base.

For \texttt{get\_coverage} the client should extend the knowledge base s.t.
progress properties are annotated by numbers between 0 and 1 representing
coverages. This is possible, since the client has a complete information for
each benchmark. The estimate (i.e.~the resulting coverage) then corresponds to
annotation of the closest progress properties in the knowledge base to the
current ones.

In case of \texttt{is\_relevant\_coverage} the client uses his progress function
to find a time for which the function returns closest properties to the current
ones. The client tweaks the protocol function according to the time he found
subtracted by the current time (scaled to $ [0,1) $) in a way: bigger the
result, higher coverages are accepted. The tweaking can ideally be sensitive to
a passed node and context.

\subsection{Safety outdate}
\label{sec:SafetyOutdate}

There are situations which cannot be resolved only by functions
\texttt{get\_coverage} and \texttt{is\_relevant\_coverage}. Let us consider a
client whose goal is to compute an over-approximation of program's behaviour and
there is a part of his abstract state space (for some node and context) whose
abstract states were computed from an information of other clients with
coverages less than 1. The client cannot be satisfied with these states, if
server's query \texttt{is\_memory\_over\_approximated} returns \texttt{false}.
It may happen the query remain \texttt{false} till the time-out. It may also
happen that the client does not receive notifications
\texttt{on\_location\_outdated} for that part of his abstract state space.

In order to succeed in the computation of an over-approximation the client may
force a re-computation of that part of the state space in sufficiently long time
before the time-out. To prevent repetition of this situation in the subsequent
computation the client may tweak his \texttt{is\_relevant\_coverage} function
s.t. for the node and context of the problematic part of the state space only
coverages 1 are accepted.

\section{Case Study: Box, Polka, Symbolic execution}
\label{sec:CaseStudy}

In this case study we attempt to experimentally evaluate how much clients may
ideally improve their results due to the communication. We thus investigate the
limit case, where client are offered a maximal opportunities for the
communication: communication at each node, no overhead of message delivery, etc.

We embedded three clients called Box, Polka, and Symbolic execution into a
single tool~\cite{StatorURL}. The tool itself stands for the server.
Clients are separate and completely independent modules in the tool, they can
communicate only with the server through his protocol functions. The tool builds
a single read-only internal program representation, which is referenced by all
clients. The internal program representation simultaneously serves as a
canonical program. Clients may thus communicate at each node. Since all clients
run on a single main thread, they perform their computations in small regularly
interleaved steps. A step corresponds to an update of an abstract state space by
taking one or more edges which all always share either head or tail node. A
client determines by himself, i.e.~independently to other clients, which edges
he will take in what computation step. A client may issue communication queries
to the server only during his step. Responses from other clients are also
computed and returned in that step. This process is purely sequential.

\subsection{Clients: Box and Polka}
\label{sec:BoxPolka}

These two clients have many things in common. First of all, implementations of
their communication-related code is very similar. Also, they represent two
instances of the abstract interpretation framework~\cite{AI_CousotCousot77}.
While Box uses intervals to represent a memory content, Polka uses polyhedrons.
Moreover, both are implemented in the same library called Apron~\cite{ApronURL}.
Therefore, we join their description here, but all differences are explicitly
stated.

Both clients are integrated into the tool in the default setting: Join operators
are applied at each node with an in-degree greater then 1, widening operators
are applied at each loop head and function entry node (because of recursion),
and all heap allocations performed along the same edge are indistinguishable for
the same context. Both clients perform call-sensitive analysis, and they
construct contexts for filters consisting of call edges kind.

Both clients issue queries to the server in same situations. Whenever a part of
an abstract state space is updated the query \texttt{on\_location\_outdated} is
issued. The query \texttt{get\_values} is issued in two situations during a
computational step of a client, when a source abstract state is supposed to be
transformed along a program edge. First, instead of taking the source abstract
state directly from the abstract state space there is used a result of
\texttt{get\_values} (after its translation to an abstract state). Second,
during evaluation of a program expression associated with the edge: When the
client does not have enough information for an evaluation of some sub-expression
(e.g.~a pointer dereference), then the client asks other (e.g.~for possible
values of the pointer) and continues the evaluation with that information.

In the following subsections we discuss those modules of these clients, which
are related to the communication.

\subsubsection{Coverage oracle}~\\
\label{sec:CoverageSracleBoxPolka}

This module records statistical data about a progress of client's abstract state
space and it plays an important role in the implementation of the protocol
function \texttt{get\_coverage}.

The oracle maintains a map from nodes and contexts to ordinal numbers of
client's computational steps in which the corresponding parts of the abstract
state space were updated for the last.

Given a node, a context, and a count $ n $ of all passed computation steps
(including the current one) the oracle computes the corresponding coverage as a
number $ 1 - (0.25 * u + 0.75 * v) / n $, where $ u $ is an average of all
values in the map mapped to by the context (and any node), and $ v $ is the
value in the map for the node and the context. The term was inferred
experimentally.

\subsubsection{Safety outdate}~\\
\label{sec:SafetyOutdateBoxPolka}

\noindent
There are three actions which may be applied during the safety outdate:
\begin{enumerate}
%\item Remembering that the safety update was applied. %
\item Tweaking of client's function \texttt{is\_relevant\_coverage} s.t. it
returns \texttt{false} for any coverage less then 1. %
\item Marking of all abstract states appearing in a list extended by the
function \texttt{on\_location\_outdated} (see the description of the protocol
function below) as to be recomputed. %
\item Marking of all abstract states in the current abstract state space to be
recomputed.
\end{enumerate}
Until the safety outdate is applied, it can be triggered in the following cases:
\begin{enumerate}
\item[(a)] If server's function \texttt{is\_memory\_over\_approximated} returns
\texttt{true}, then actions 1. and 2. are applied. %
\item[(b)] If server's function
\texttt{can\_improve\_memory\_over\_approximation} returns \texttt{false}, then
actions 1. and 3. are applied. %
\item[(c)] Let $ m $ be a number of communicating clients, $ n $ be a number of
nodes in the program, $ t $ be the time point of the time-out in milliseconds,
and $ c $ be 250 for Box and 500 for Polka. If the current time point is greater
or equal to $ \max\{0, t - c \cdot m \cdot n\} $, then actions 1. and 3. are
applied. Both ``$ \max $'' terms were adjusted experimentally.
\end{enumerate}

\subsubsection{Communication protocol}~\\
\label{sec:CommunicationBoxPolka}

Box and Polka implement the mandatory functions of the protocol as follows:

\paragraph{\tt get\_values: }

The function first collects all abstract states relevant to the passed node and
a reduced context computed from the passed context and the client's filter. An
abstract state is a list of Apron's linear constraints over dereferences
(represented via strings in Apron). Each state is converted into a conjunction
of predicates, where each predicate is a direct conversion of a constraint from
Apron's internal data structures. Subsequently, from each conjunction there are
removed all predicates which are irrelevant to the passed set of dereferences.
The function then returns a disjunction of the resulting conjunctions.

\paragraph{\tt get\_coverage: }

For each abstract states relevant to the passed node and a reduced context
computed from the passed context and the client's filter it calls the coverage
oracle with the node and the reduced context. When $ n $ is a number of such
states and $ c_i $ is a value returned by the oracle for the $ i $-th state,
then the function returns the value $ (\sum_{i=1}^{n} c_i) / \max\{ 1,n \} $.
The term was adjusted experimentally.

\paragraph{\tt is\_relevant\_coverage: }

If the safety outdate was already applied, then it returns \texttt{true} only if
the passed coverage is equal to 1. Otherwise, there is computed a scale for the
passed self-coverage.  The scale is initialised to 1 and according to each
property the of the passed node it is multiplied by coefficients in this table:

\hspace{5mm}
\begin{tabular}{c||c|c|c|c}
\textbf{Client} &
    \textbf{Loop head} &
    \textbf{Function entry} &
    \textbf{Return node} &
    \textbf{Join node}
    \\
\hline
Box & 0.25 & 0.25 & 0.5 & 0.25 \\
Polka & 0.5 & 0.5 & 0.75 & 0.9
\end{tabular}

\noindent The values were adjusted experimentally. The function returns
\texttt{true} only if the passed coverage is greater or equal to the
self-coverage multiplied by the scale.

\paragraph{\tt is\_memory\_over\_approximated: }

Returns \texttt{true} when a fixed-point is reached, i.e.~when there is no
abstract state (identified by a node and context) to be updated. Otherwise, it
returns \texttt{false}.

\paragraph{\tt can\_improve\_memory\_over\_approximation: }

Returns a negated value from the function \texttt{is\_memory\_over\_approximated}.

\paragraph{\tt on\_location\_outdated: }

If this is not the self-call and a call to the function
\texttt{is\_relevant\_coverage} for the passed coverage returns \texttt{false},
then the function adds all abstract states corresponding to the passed node and
context to a list of ``states for safety outdate'' and then terminates.
Otherwise, it marking all abstract states corresponding the passed node and
context for a re-computation.

\subsubsection{Correctness}~\\
\label{sec:CorrectnessBoxPolka}

The client (Box or Polka) claims that program's behaviour was successfully
over-approximated only if his function \texttt{is\_memory\_over\_approximated}
return \texttt{true} and the safety outdate was applied.

After application of the safety outdate only values with coverage equal to 1 are
accepted. If the safety outdate was triggered by the case (b) or (c), each
abstract state was recomputed. Otherwise, the case (a) was triggered and so the
server returned \texttt{true} from his function
\texttt{is\_memory\_over\_approximated}. The computation performed so far was
thus based either on final memory over-approximations from other clients or on
re-computed abstract states where the client rejected final values of others due
to low coverages.

\subsection{Client: Symbolic Execution}
\label{sec:SymbolicExecution}

Symbolic execution~\cite{SE_King76} is similar to standard program execution.
The key difference is that while the standard program execution runs the program
on concrete input data (numbers), the symbolic execution runs the program on
symbols. Each symbol represents a concrete, but yet unknown, input value.
Executed program paths are recorded in a tree, where the root corresponds to the
entry node and a leaf node either represents a termination of the execution or
it belongs into the exploration frontier, i.e.~it remains to compute its
successors.

From the communication point of view it is important to mention that this
technique attempts to explore all feasible program paths. This goal is achieved
very rarely in practice, only for programs with finite and quite low number of
such paths. Therefore, the technique primarily focuses on searching for defects
and typically end up with an under-approximation of program's behaviour.

The client is integrated into the tool in the default setting: There is fixed the
breadth-first exploration strategy of feasible program paths, the variable
storage-referencing problem is resolved by branching, there is only one SMT
solver used for SAT queries, and there is no cache in front of the solver. The
filter of the client consist of both call and branching kinds of edges.

In the following subsections we discuss only those modules of the client, which
are related to the communication with others.

\subsubsection{Coverage oracle}~\\
\label{sec:CoverageSracleSE}

This module records statistical data about both a progress of client's abstract
state space and queries issued to the client. It serves as the key part in
implementations of protocol functions \texttt{get\_values} and
\texttt{get\_coverage}.

The main part of the oracle is a map from node and context to triples
(\texttt{formula}, \texttt{num\_state\_updates},
\texttt{num\_formula\_updates}). The component \texttt{formula} is a formula
over dereferences, \texttt{num\_state\_updates} is a counter of updates of the
triple, and \texttt{num\_formula\_updates} counts number of updates of the
\texttt{formula}. Each triple is initialised as (\texttt{true},0,0).

The oracle is called at the end of each client's step in order to reflect
updates he made. Given a node and context related to some of client's updates,
the oracle performs the following three steps: First, he expresses all abstract
states corresponding to the node and the context in a form of a formula over
dereferences according to a process described in the next sub-section. Then, he
searches in the map for a triple using the node and the context. Lastly, the
triple is updated s.t. \texttt{num\_state\_updates} is incremented, and if the
newly computed formula is not logically equal to \texttt{formula} in the triple,
then \texttt{formula} is overwritten by the new one and the counter
\texttt{num\_formula\_updates} is incremented. In the case the \texttt{formula}
was updated in the triple whole the record is registered into a special queue,
if it is not already there. The client extracts records of the queue regularly
in each step with a delay 10s and issues the query
\texttt{on\_location\_outdated} to the server. The delay prevents frequent
notifications for nodes inside loops, where symbolic execution tends to cycle
very quickly for long time.

The oracle is called from function of client's communication protocol. These
functions query the oracle fro triples stored in the map. For each such query,
if the searched triple has \texttt{num\_state\_updates} equal to 0, then the
oracle automatically performs its update according to the procedure above,
before the triple is returned.

\subsubsection{Building of response formula}~\\
\label{sec:BuildingResponseFormulaSE}

Given a node, a context, and the filter of the client, the procedure first
collects all abstract states attached to nodes of the client's symbolic
execution tree which are relevant to the given program node and a reduced
context computed from the given context and the filter. Note that the client's
filter consists of both branching and call edge kinds, since symbolic execution
is fully path and context sensitive. Also note that each collected abstract
state consists of a mapping from dereferences to expressions over input symbols
and a quantifier-free conjunction of predicates over input symbols.

For each abstract state there is constructed a formula, which is initialised as
the conjunction of predicates of the state. Next, for each pair in the map of
the state a new conjunct is added into the formula. This conjunct is an equality
between the dereference and the expression of the pair. Then, each equality in
the formula, which consists of a dereference and an input symbols, all
occurrences of the symbol in the formula is replaced by the dereferences.
Finally, each predicate in the formula which still contains any input symbol is
removed from the formula.

From formulae we computed for each collected abstract state we now build a
resulting formula. The formula is a conjunction of predicates which are common
to all computed formulae (according to logical equality) and the strongest
inequalities between dereferences and numerical constants deducible from all
computed formulae. Although we attempt to find as many common predicates and
inequalities as possible, we may early terminate the search in order to preserve
performance of the whole query.

\subsubsection{Communication protocol}~\\
\label{sec:CommunicationSymbolicExecution}

The client implements the mandatory functions of the protocol as follows:

\paragraph{\tt get\_values: }

The function queries the coverage oracle for a triple for the passed node and
context and returns the component \texttt{formula} of the triple.

\paragraph{\tt get\_coverage: }

If \texttt{is\_memory\_over\_approximated} returns 1, the also this function
returns 1. Otherwise, it queries the coverage oracle for a triple for the passed
node and context. The function then returns a coverage value computed according
to the following experimentally established term $ 1 - \frac{1}{(s - f + 1)
\cdot s + 1} $, where $ s $ and $ f $ stands for \texttt{num\_state\_updates}
and \texttt{num\_formula\_updates} respectively.

\paragraph{\tt is\_relevant\_coverage: }

This function always returns \texttt{false}, since no information is taken from
other clients. This client only provides information to others.

\paragraph{\tt is\_memory\_over\_approximated: }

Returns \texttt{true} when the exploration frontier is empty and no state was
force-terminated due to a failure. A majority of failures come from an SMT
solver, which fails to decide satisfiability of a path condition of a program
state. Another important sources are unsupported bit operators and presence of
inline assembly.

\paragraph{\tt can\_improve\_memory\_over\_approximation: }

Returns \texttt{true} if the exploration frontier is not empty. Otherwise,
returns \texttt{false}.

\paragraph{\tt on\_location\_outdated: }

This function does nothing, since this client does not take any information from
others. The client only provides information.

\subsubsection{Correctness}~\\
\label{sec:CorrectnessSE}

The client does not accept values from others.

\subsection{Server: The tool}
\label{sec:ServerSTATOR}

The server implements all his protocol functions (see
section~\ref{sec:ServerProtocol}) in purely sequential manner. It means that
clients are queried for responses one by one and when all responses are
collected the result is returned back the the client initiating the query.

The server and all clients run of the same main thread and all queries from
clients are issued also from that thread. It means that all queries are issued
sequentially: A next query may be issues only after the previous one is
completed.

\subsection{Evaluation}
\label{sec:Evaluation}

We evaluated clients of the case study in five different configurations. Each
configuration specifies what clients are used and whether they can communicate
or not. We denote configurations using the following abbreviations: b*p*s,
b+p+s, b+p, b+s, p+s. Symbols `b', `p', and `s' stand for Box, Polka, and
Symbolic execution respectively, and `+' and `*' stand for communication enabled
and disabled respectively. For each configuration we assume that either all
clients communicate with each other (the use of `+') or none of them (the use of
`*').

We performed the evaluation on SV-COMP~2015~\cite{SVCOMPURL} benchmark suite,
revision 571. The revision consists of 5861 benchmarks in 48 directories. In
order to make the evaluation manageable for us, we put a requirement that whole
the evaluation should finish within one week of continuous
computation\footnote{The used a server: 32xIntel Xenon E5-2650 @ 2GHz, 64GB RAM,
Debian~4.6.3.}. Therefore, we picked 10 randomly chosen benchmarks from each
directory (or less if there was not enough) and so we got 473 benchmarks in 48
directories, see Appendix~\ref{sec:BenchList}. Since this was still not enough
we set a time-out 2.5 minutes and a memory-out 512MB for each client in each
configuration. It means, for example, that b+p had the time-out 5 minutes and
the memory-out 1024MB, b*p*s had the time-out 7.5 minutes and the memory-out
1536MB, etc. Remember that clients share time (steps are interleaved) and memory
(all run on a single thread) within a configuration, see
page~\pageref{sec:CaseStudy}.

%Surely, this way one client may consume resources at the expense of others.
%Since clients are implemented in external libraries (Apron) the prevention of
%such behaviour would be difficult. Besides, this behaviour is allowed whenever
%the communication is enabled and disabled.

We compared results of each combination of configurations. The comparison was
always done separately per client: given two configurations and a client
appearing in both of them we only consider results of that client in both
configurations for the comparison.

\begin{figure}[!htb]
\begin{center}
\begin{tabular}{c}
\begin{tabular}{|cc||ccccc|ccccccc|}
\hline
\multicolumn{2}{|c||}{Configuration} &
\multicolumn{5}{c|}{Comparison per node} &
\multicolumn{7}{c|}{Comparison per benchmark}
\\
1st & 2nd &
fail & neq & eq & 1st & 2nd &
fail & neq & 1st & 2nd & 1st! & 2nd! & eq!
\\ \hline
b*p*s & ~b+p+s &
13 & 246 & 13021 & 627 & 9563 & 
1 & 20 & 56 & 191 & 23 & 148 & 35
\\
b*p*s & b+p &
13 & 342 & 15042 & 613 & 10043 & 
1 & 37 & 46 & 243 & 7 & 186 & 34
\\
b*p*s & b+s &
13 & 309 & 12444 & 450 & 8202 & 
1 & 10 & 36 & 128 & 28 & 115 & 71
\\
b+p+s & b+p &
13 & 1167 & 25025 & 3707 & 1851 & 
1 & 42 & 120 & 123 & 43 & 48 & 97
\\
b+p+s & b+s &
13 & 142 & 24753 & 1170 & 708 & 
1 & 14 & 66 & 34 & 43 & 17 & 155
\\
b+p & b+s &
13 & 889 & 21311 & 1798 & 2899 & 
1 & 41 & 106 & 76 & 56 & 28 & 99
\\
\hline
\end{tabular}

\\ \\
\begin{tabular}{|cc||ccccc|ccccccc|}
\hline
\multicolumn{2}{|c||}{Configuration} &
\multicolumn{5}{c|}{Comparison per node} &
\multicolumn{7}{c|}{Comparison per benchmark}
\\
1st & 2nd &
%Fails & Incmp & Equal & 1st & 2nd &
%Fails & Incmp & 1st & 2nd & 1st* & 2nd* & Equal*
fail & neq & eq & 1st & 2nd &
fail & neq & 1st & 2nd & 1st! & 2nd! & eq!
\\ \hline
b*p*s & ~b+p+s &
20 & 487 & 10616 & 401 & 12672 & 
1 & 42 & 38 & 199 & 16 & 157 & 43
\\
b*p*s & b+p &
26 & 319 & 11691 & 168 & 12923 & 
1 & 39 & 23 & 224 & 5 & 183 & 51
\\
b*p*s & p+s &
20 & 609 & 10689 & 455 & 12119 & 
1 & 57 & 44 & 196 & 12 & 136 & 50
\\
b+p+s & b+p &
14 & 663 & 30462 & 2217 & 1489 & 
1 & 29 & 93 & 87 & 50 & 53 & 137
\\
b+p+s & p+s &
0 & 253 & 32268 & 1307 & 446 & 
0 & 19 & 74 & 40 & 53 & 24 & 182
\\
b+p & p+s & 
14 & 691 & 30187 & 1960 & 1903 & 
1 & 46 & 115 & 88 & 58 & 33 & 129
\\
\hline
\end{tabular}

\end{tabular}
\end{center}
\caption{Comparison of invariants for clients Box (top) and Polka (bottom).
Meaning of columns from left: ``1st'',``2nd'' - 1st and 2nd compared
configuration, ``fail'' - failures of Z3, ``neq'' - incomparable (neither is
stronger), ``eq'' - logically equal, ``1st'',``2nd'' - 1st,2nd configuration has
stronger invariant ; ``fail'' - at least one Z3 failure, ``neq'' - contains
incomparable invariants, ``1st'',``2nd'' - has at least one stronger invariant
than in 2nd,1st configuration, ``1st!'',``2nd!'' - at least one stronger but no
weaker invariant than in 2st,1nd configuration, ``eq!'' - all invariant are
logically equal.} \label{fig:Invariants}
\end{figure}

We focused on two kinds of measurements. First, we compared a precision of
invariants computed by clients Box and Polka. Symbolic execution does not
provide this kind of information. The clients attempt to compute for each node a
strongest invariant over-approximating all concrete states which can be seen at
the node. We used Z3~\cite{Z3URL} SMT solver to compare invariants. Results are
presented in Fig.~\ref{fig:Invariants}. The numbers for ``Comparison per node''
are summary counts of nodes of all considered benchmarks together. And the
numbers for ``Comparison per benchmark'' are simply counts of considered
benchmarks. Note that for each client there were only considered those
benchmarks for which the client terminated with the state ``Success'' in both
compared configurations.

\noindent
We can observe the following facts in data in Fig.~\ref{fig:Invariants}:
\begin{itemize}
\item \emph{Each configuration may bring us improvements over others}: We can
clearly see this phenomena for all pairs of configurations in both kinds of
comparisons in tables of both clients. We can thus expect that for any chosen
combination of clients we get some new results in terms of strengthened
invariants. %
\item \emph{There is no configuration strictly dominating all others}: We can
only read patterns in the data, like:
\begin{itemize}
\item A configurations with communicating clients gives us at least one order of
magnitude more precise invariants than isolated clients. %
\item More communicating clients, more strengthened invariants. %
\item Count of incomparable invariants and lower count of strengthened
invariants can be expected in the same order of magnitude. %
\item More improved invariants typically yields more improved benchmarks,
i.e.~improvements are rather regularly distributed than highly concentrated in
few benchmarks. Nevertheless, a degree of correlation is sensitive to kinds of
clients appearing in configurations, cf.~fourth and sixth rows for both Box and
Polka. %
\end{itemize}
Observations made for invariants can easily be adopted to very similar
observations for benchmarks.
\end{itemize}

\begin{figure}[!htb]
\begin{center}
\begin{tabular}{c}
\begin{tabular}{|cc||ccc|ccc|ccc|ccc|}
\hline
\multicolumn{2}{|c||}{Configuration} &
\multicolumn{3}{c|}{Success} &
\multicolumn{3}{c|}{Time-out} &
\multicolumn{3}{c|}{Memory-out} &
\multicolumn{3}{c|}{Crash}
\\
1st & 2nd &
eq & 1st & 2nd &
eq & 1st & 2nd &
eq & 1st & 2nd &
eq & 1st & 2nd
\\ \hline
b*p*s & ~b+p+s &
251 & 35 & 24 & 
10 & 0 & 38 & 
137 & 19 & 0 & 
7 & 14 & 6
\\
b*p*s & b+p &
285 & 1 & 25 & 
10 & 0 & 3 & 
143 & 13 & 0 & 
7 & 14 & 0
\\
b*p*s & b+s &
230 & 56 & 23 & 
10 & 0 & 200 & 
0 & 156 & 0 & 
7 & 14 & 3
\\
b+p+s & b+p &
275 & 0 & 35 & 
13 & 35 & 0 & 
137 & 0 & 6 & 
7 & 6 & 0
\\
b+p+s & b+s &
241 & 34 & 12 & 
45 & 3 & 165 & 
0 & 137 & 0 & 
10 & 3 & 0
\\
b+p & b+s &
245 & 65 & 8 & 
12 & 1 & 198 & 
0 & 143 & 0 & 
7 & 0 & 3
\\
\hline
\end{tabular}

\\ \\
\begin{tabular}{|cc||ccc|ccc|ccc|ccc|}
\hline
\multicolumn{2}{|c||}{Configuration} &
\multicolumn{3}{c|}{Success} &
\multicolumn{3}{c|}{Time-out} &
\multicolumn{3}{c|}{Memory-out} &
\multicolumn{3}{c|}{Crash}
\\
1st & 2nd &
eq & 1st & 2nd &
eq & 1st & 2nd &
eq & 1st & 2nd &
eq & 1st & 2nd
\\ \hline
b*p*s & ~b+p+s &
263 & 21 & 30 & 
10 & 0 & 20 & 
137 & 21 & 0 & 
7 & 14 & 6
\\
b*p*s & b+p &
283 & 1 & 28 & 
10 & 0 & 2 & 
143 & 15 & 0 & 
7 & 14 & 0
\\
b*p*s & p+s &
267 & 17 & 27 & 
10 & 0 & 15 & 
143 & 15 & 0 & 
7 & 14 & 4
\\
b+p+s & b+p &
289 & 4 & 22 & 
10 & 20 & 2 & 
137 & 0 & 6 & 
7 & 6 & 0
\\
b+p+s & p+s &
287 & 6 & 7 & 
21 & 9 & 4 & 
137 & 0 & 6 & 
11 & 2 & 0
\\
b+p & p+s &
294 & 17 & 0 & 
12 & 0 & 13 & 
143 & 0 & 0 & 
7 & 0 & 4
\\
\hline
\end{tabular}

\\ \\
\begin{tabular}{|cc||ccc|ccc|ccc|ccc|}
\hline
\multicolumn{2}{|c||}{Configuration} &
\multicolumn{3}{c|}{Success} &
\multicolumn{3}{c|}{Time-out} &
\multicolumn{3}{c|}{Memory-out} &
\multicolumn{3}{c|}{Crash}
\\
1st & 2nd &
eq & 1st & 2nd &
eq & 1st & 2nd &
eq & 1st & 2nd &
eq & 1st & 2nd
\\ \hline
b*p*s & ~b+p+s & 
151 & 3 & 0 & 
155 & 0 & 35 & 
119 & 16 & 0 & 
12 & 17 & 1
\\
b*p*s & b+s & 
148 & 6 & 7 & 
155 & 0 & 153 & 
0 & 136 & 0 & 
10 & 19 & 0
\\
b*p*s & p+s & 
149 & 5 & 0 & 
155 & 0 & 34 & 
124 & 11 & 0 & 
11 & 18 & 0
\\
b+p+s & b+s & 
147 & 4 & 8 & 
190 & 0 & 118 & 
0 & 119 & 0 & 
10 & 3 & 0
\\
b+p+s & p+s & 
148 & 3 & 1 & 
185 & 5 & 4 & 
119 & 0 & 5 & 
11 & 2 & 0
\\
b+s & p+s & 
148 & 7 & 1 & 
189 & 119 & 0 & 
0 & 0 & 124 & 
10 & 0 & 1
\\
\hline
\end{tabular}

\end{tabular}
\end{center}
\caption{Comparison of termination states for clients Box (top), Polka (middle),
and Symbolic execution (bottom). Columns of ``Configuration'': ``1st'',``2nd'' -
1st and 2nd compared configuration~;~All other columns: ``eq'' - equal state,
``1st'',``2nd'' - 1st,2nd configuration has the state while 2nd,1st has
some other.} \label{fig:Termination}
\end{figure}
% crashes Box: 27 + 21 + 24 + 13 + 13 + 10
% crashes Polka: 27 + 20 + 25 + 13 + 13 + 11
% crashes SE: 30 + 29 + 29 + 13 + 13 + 11
% avrg: (27+21+24+13+13+10 + 27+20+25+13+13+11 + 30+29+29+13+13+11)/18=19
% avrg percentage of crashes: 19/473=0.040169133

In the second measurement we focused on comparison of termination states of
individual clients as they are used in different configurations. We distinguish
termination states ``Success'', ``Time-out'', ``Memory-out'', and ``Crash'', all
with obvious meanings. Results are presented in Fig.~\ref{fig:Termination}.
Numbers in each table represent counts of benchmarks.

%We present counts of crashes only for completeness. They represent less than 5\%
%of all benchmarks, so we can safely exclude them from the consideration for
%making conclusions.

%Majority of crashes came from Apron library\footnote{It is likely they are
%caused by a misuse of the library.} and from translating long complex
%initialiser lists.

\noindent
We can observe the following facts in data in Fig.~\ref{fig:Termination}:
\begin{itemize}
\item \emph{Consumption of resources via communication does not imply a decrease
of successful termination}: Considering ``Success'' data for all configurations
comparing with b*p*s for all clients, the communication caused a lose of success
termination states in the following percentages:

\hspace{25mm}
\begin{tabular}{cc||ccc}
1st & 2nd & Box & Polka & Sym.exec.
\\ \hline
b*p*s & ~b+p+s &
3.8 &  % 1-(251+24)/(251+35)=0.038461538
-3.2 & % 1-(263+30)/(263+21)=-0.031690141
1.9    % 1-(151+0)/(151+3)=0.019480519
\\
b*p*s & b+p &
-8.4 & % 1-(285+25)/(285+1)=-0.083916084
-9.5 & % 1-(283+28)/(283+1)=-0.095070423
-
\\
b*p*s & b+s &
11.5 &  % 1-(230+23)/(230+56)=0.115384615
- &
-0.6    % 1-(148+7)/(148+6)=-0.006493506
\\
b*p*s & p+s &
- &
-3.5 & % 1-(267+27)/(267+17)=-0.035211268
3.2   % 1-(149+0)/(149+5)=0.032467532
\end{tabular}
% avrg=(3.8 -3.2 +1.9 -8.4 -9.5 +11.5 -0.6 -3.5 +3.2)/9=-0.533333333
% sorted= -9.5 -8.4 -3.5 -3.2 -0.6 +1.9 +3.2 +3.8 +11.5
% crashes Box: 27 + 21 + 24
% crashes Polka: 27 + 20 + 25
% crashes SE: 30 + 29 + 29
% avrg: (27+21+24 + 27+20+25 + 30+29+29)/9=25.777777778
% avrg percentage of crashes: 26/473=0.054968288

In 5 of 9 cases we see an increase of ``Success'' termination states. The
average of these numbers is -0.53\%. We may thus expect about 0.5\% increase of
``Success'' termination states on average per client due to reduced overall time
and memory consumption (the influence of crashes is less than 6\%). %
\item \emph{Resources consumption via communication heavily depends on kinds of
clients}: This statement is based on the following patterns which dominate
data:
\begin{itemize}
\item Symbolic execution is a major source of ``Time-out'' termination states.
We can see this is tables of all clients: Whenever the client is present, there
is a high count of time-outs. %
\item Polka is a major source of ``Memory-out'' termination states.
We can see this is tables of all clients: Whenever the client is present, there
is a high count of memory-outs. %
\end{itemize}
\end{itemize}

\noindent
The facts we observed in data of both figures lead as to a conclusion:
\vspace{2mm}

\emph{In order to get the best result we should run as many configurations as
possible and then merge their results. Since resources are always limited (like
a number of available computers or threads), we should use patterns (similar to
those we observed) in order to express preferences for certain configurations.}

\vspace{2mm}
It is important to add that the presented anonymous communication approach
provides a cheap way for building combinations of clients. Indeed, a
construction of a new client mostly involves an implementation of the
communication interface and choosing right locations in client's code, where to
issue communication queries. Now, having the best result for a certain benchmark
for a group of $ n $ clients, an addition of a single new one will automatically
gives as $ 1 + \frac{1}{n+1} \sum_{k=1}^{n} k\binom{n+1}{k} $ new configurations
each with a potential to improve the old result. In the case of our evaluation
we would immediately get eight new such configurations ready for execution.

An interested reader may have a look to Appendix~\ref{sec:Plots}, where we put
the presented data in a form of histograms. A package with sources and binaries
of the tool used in the evaluation together with the computed results is
freely available here~\cite{StatorURL}. Details about the
download, installation, and the use can be found in
Appendix~\ref{sec:AccessToStatorEvaluation}.

\section{Related Work}
\label{sec:RelatedWork}

% Searched conferences: 
%   ASE,ATVA,CAV,ESA,ESOP,FSE,ICSE,ISSTA,PLDI,POPL,SAS,TACAS,VMCAI   
%   years: 2015,2014,2013,2012,2011
% Other search:
%   kdws: combine,combining,composing,communication,communicating,analyses,analyzes
%   engines: Google (scholar), Yahoo!, http://academic.research.microsoft.com/
% Depth of transitive search in references of collected papers: at most 1

There is a broad class of approaches dedicated to combining of lattice-based
analyses. They all are based either on a direct or a reduced
product~\cite{DirectReducedProduct}.

A direct product is fully automatic, it requires neither writing of resulting
operations nor modifications of the input ones. Nevertheless, composed analyses
do not interact and the product represents the least precise Cartesian product
of the analyses. No check for correctness is required.

A reduced product is in contrary the most precise Cartesian product, but it is
based on (non-computable) concretisation functions used in so called ``reduce''
function. In practice, the issue is typically resolved by providing an
approximation of the reduce function. But this implies a dedicated
implementation for each given combination of analyses. It is also necessary to
check for correctness of the approximation.

A logical product~\cite{LogicalProduct} automatically construct the most precise
reduced product of analyses whose elements are conjunctions of atomic facts over
theories that are convex, stably infinite, and disjoint. No check for
correctness is required. 

In~\cite{ProductShapes} authors combine abstract domains for shape analyses
using reduced product. Each component reasons independently about different
aspects of the data structure invariant and then separately exchange information
via a reduction operator.

Granger's product~\cite{GrangerProduct} provides an elegant approximation of the
reduce function of a reduced product. Each input analysis is extended by its own
reduce function from domains of all analyses to domain of the analysis owning
the function. Each analysis must only check whether its reduce function
satisfies necessary requirements. Note that each combination of analysis
requires new dedicated implementations of reduce functions.

An open product~\cite{OpenProduct} substantially improves the Granger's product,
since it removes the dependence between analyses. The only common property is an
a priory given set of queries (e.g.~in the logic paradigm: ``is a variable
\texttt{x} surely bound to a ground term?''). Each operation of each analysis is
extended s.t. it is parametrised by all possible valuations of the queries.
Analyses are thus also extended by boolean functions capable to compute
individual parametrisations. Each analysis has to check for itself for
correctness of its extension.

The requirement of a set of predefined queries in an open product was later
relaxed in~\cite{InteractingPlugins} by replacing them by a language of the
first order logic. Operations of all analyses can then be parametrised by any
formula of the language.

Composition of configurable program analyses~\cite{CPA} is based on a direct
product, whose precision can be improved via relations ``transfer'', ``merge'',
``stop'', ``compare'', and ``strengthen''. Each of these relations is defined
over domains of all composed analyses. Implementations of relations transfer,
merge, and stop cannot access abstract states of individual analyses, they can
only apply their operators and relations. There is no such restriction for
implementations of compare and strengthen. These two relations can be used in
implementations of the previous three. The composition requires neither
modification nor extension of implementations of input analyses. Nevertheless,
the five relations mentioned above require dedicated implementations for each
combination of analyses. An improved concept by a dynamic precision
adjustment~\cite{CPAplus} introduced a relation ``prec''. This relation is also
defined over domains of all composed analyses, so its implementations also has
to be provided for each combination of analyses. In both the original and the
improved concept one needs to check whether his implementations of all the
relations satisfy necessary requirements. It is further important to mention
that in~\cite{CPA} there was introduced a ``location analysis'' modeling
instruction counter. It can be composed with other analyses.

An advanced combination of lattice-based analyses can be found in
Astr{\'e}e~\cite{Astree07}. It is based on the idea of an open product with
several extensions. The set of fixed queries was replaced by an extensible set
of kinds of constraints. An extension of the set be a new kind implies
extensions and checks for correctness of only those analyses which want to use
constraints of that kind. A constraint represents an information of its kind
about an analysed program. Analyses may exchange information through input and
output channels. Input channels provide information on both the postcondition
being computed and the precondition computed in the last computation step.
Output channels are used when an analysis wants to send a message to others.
Messages are elements of a separate abstract domain. Messages are not always
exchanged freely between analyses. An order of analyses in a computational step
matters. Typically, an analysis may freely communicate with any predecessor.

An obvious difference between our approach and approaches above is that our
approach is not restricted to lattice-based analyses. Further important
differences can be expressed in terms of three features we mentioned in the
introduction:
\begin{itemize}
\item \emph{Independence of clients}: This feature provide only approaches based
on the open product, i.e.~\cite{OpenProduct,InteractingPlugins,Astree07}. %
\item \emph{Asynchronous execution of clients}: Approaches above do not natively
provide this feature. Computational steps of all analyses are synchronised,
i.e.~they all run in the same speed. An asynchronous execution can be emulated
in some extent in~\cite{CPA} through the use of several location analyses in one
combination: some location analysis stays on a certain location in several
subsequents steps in order to simulate slow-down of dependent analyses.
Nevertheless, by a use of more than one location analysis in a combination we
may expect enormous increase of complexity in designing relations transfer,
merge, etc. %
\item \emph{Reuse of current implementations}: Approaches based on the open
product are suited for the reuse. Nevertheless,
approaches~\cite{OpenProduct,InteractingPlugins,Astree07} require that all
composed analyses (synchronously) run on the same internal program
representation. And in~\cite{CPA} the reuse requires to look inside
implementation of each analysis and discover how the program is internally
represented. Then corresponding location analyses can be designed and
implemented together with all relations transfer, merge, etc. Nevertheless, this
may imply enormous amount of programing effort (which has to be repeated for
different combinations of analyses). %
\end{itemize}

There is another broad class of approaches devoted to combining of program
analyses. They are focused on combining specific kinds of analyses. Typically,
two or more particular analyses are considered, e.g.~predicate abstraction with
dynamic test generation~\cite{SMASH}, static checking and
testing~\cite{CheckNCrash,StaticDynamic}, different testing
techniques~\cite{CombiningTesting}, symbolic and concrete
execution~\cite{Concolic}, static and dynamic analyses via program
partitioning~\cite{ProgramPartitioning}, data-flow with predicate
lattices~\cite{DataflowPredicates}, data-flow analyses in a
compiler~\cite{CombiningOptimisations,CompilerTranformations}, etc., and a
result is a new program analysis possessing advantages of individual analyses.
Although these approaches give us interesting new algorithms and ideas, they are
orthogonal to our approach: Our approach does not build ``a new analysis'', we
proposed an analysis-independent communication model. In other words, we took an
alternative path.

%abstract interpretation with symbolic execution~\cite{},
%predicate abstraction with testing~\cite{},
%shape analysis with a model-checker~\cite{},

\section{Conclusion}
\label{sec:Conclusion}

We presented an approach for a light-weight anonymous online communication of
existing implementations of program analyses (i.e.~clients). In order to
communicate with others a client has to implements the proposed client's
communication protocol and to identify places in his implementation where to
issue queries s.t. the received information may subsequently help to improve
client's results. These steps are completely independent on other clients. The
client also has to check by himself for correctness of his analysis for all
possible communication scenarios which may occur during analysis of a given
program. Each client performs his work on a private data and an internal program
representation. Nevertheless, the communication is performed in terms of common
model of instruction counter (canonical program) and common memory addressing
(canonical memory). Therefore, communication queries are are typically coupled
with conversions from client's internal data to common terms and back.

We also presented a case study with three communicating clients: two abstract
interpreters (intervals, polyhedrons) and one classic symbolic execution. We
evaluated five their configurations (how many and what client will communicate)
and we pairwise compared their results per client. We measured strengthening of
invariant (for abstract interpreters) and termination states (for all). The data
shows that each combination may bring us new improvements in both strengthened
invariants and increased count of ``success'' termination states. Therefore,
since there are $ 2^n - 1 $ configurations for $ n $ clients, we can achieve
substantial gain compared to results of $ n $ isolated clients.

\bibliographystyle{plain}
\bibliography{aocbpa}

\begin{thebibliography}{10}

\bibitem{CPA}
D.~Beyer, T.~A. Henzinger, and G.~Th{\'e}oduloz.
\newblock Configurable software verification: Concretizing the convergence of
  model checking and program analysis.
\newblock In {\em Proceedings of CAV}, pages 504--518. Springer-Verlag, 2007.

\bibitem{CPAplus}
D.~Beyer, T.~A. Henzinger, and G.~Theoduloz.
\newblock Program analysis with dynamic precision adjustment.
\newblock In {\em Proceedings of ASE}, pages 29--38. IEEE, 2008.

\bibitem{InteractingPlugins}
N.~Charlton.
\newblock Verification of java programs with interacting analysis plugins.
\newblock {\em Electron. Notes Theor. Comput. Sci.}, 145:131--150, 2006.

\bibitem{CombiningOptimisations}
C.~Click and K.~D. Cooper.
\newblock Combining analyses, combining optimizations.
\newblock {\em ACM Trans. Program. Lang. Syst.}, 17(2):181--196, 1995.

\bibitem{OpenProduct}
A.~Cortesi, B.~Le~Charlier, and P.~Van~Hentenryck.
\newblock Combinations of abstract domains for logic programming: Open product
  and generic pattern construction.
\newblock {\em Sci. Comput. Program.}, 38(1-3):27--71, 2000.

\bibitem{CombiningTesting}
D.~Cotroneo, R.~Pietrantuono, and S.~Russo.
\newblock A learning-based method for combining testing techniques.
\newblock In {\em Proceedings of ICSE}, pages 142--151. IEEE, 2013.

\bibitem{AI_CousotCousot77}
P.~Cousot and R.~Cousot.
\newblock Abstract interpretation: A unified lattice model for static analysis
  of programs by construction or approximation of fixpoints.
\newblock In {\em Proceedings of the POPL}, pages 238--252. ACM, 1977.

\bibitem{DirectReducedProduct}
P.~Cousot and R.~Cousot.
\newblock Systematic design of program analysis frameworks.
\newblock In {\em Proceedings of POPL}, pages 269--282. ACM, 1979.

\bibitem{Astree07}
P.~Cousot, R.~Cousot, J.~Feret, L.~Mauborgne, A.~Min{\'e}, D.~Monniaux, and
  X.~Rival.
\newblock Combination of abstractions in the {ASTR\'EE} static analyzer.
\newblock In {\em Proceedings of ASIAN}, pages 272--300. Springer-Verlag, 2007.

\bibitem{CheckNCrash}
Ch. Csallner and Y.~Smaragdakis.
\newblock Check 'n' crash: Combining static checking and testing.
\newblock In {\em Proceedings of ICSE}, pages 422--431. ACM, 2005.

\bibitem{DataflowPredicates}
J.~Fischer, R.~Jhala, and R.~Majumdar.
\newblock Joining dataflow with predicates.
\newblock {\em SIGSOFT Softw. Eng. Notes}, 30(5):227--236, 2005.

\bibitem{SMASH}
P.~Godefroid, A.~V. Nori, S.~K. Rajamani, and S.~D. Tetali.
\newblock Compositional may-must program analysis: Unleashing the power of
  alternation.
\newblock {\em SIGPLAN Not.}, 45(1):43--56, 2010.

\bibitem{GrangerProduct}
P.~Granger.
\newblock Improving the results of static analyses programs by local decreasing
  iteration.
\newblock In {\em Proceedings of FSTTCS}, pages 68--79. Springer-Verlag, 1992.

\bibitem{LogicalProduct}
S.~Gulwani and A.~Tiwari.
\newblock Combining abstract interpreters.
\newblock {\em SIGPLAN Not.}, 41(6):376--386, 2006.

\bibitem{ProgramPartitioning}
P.~Jalote, V.~Vangala, T.~Singh, and P.~Jain.
\newblock Program partitioning: A framework for combining static and dynamic
  analysis.
\newblock In {\em Proceedings of WODA}, pages 11--16. ACM, 2006.

\bibitem{SE_King76}
J.~C. King.
\newblock Symbolic execution and program testing.
\newblock {\em Commun. ACM}, 19(7):385--394, 1976.

\bibitem{CompilerTranformations}
S.~Lerner, D.~Grove, and C.~Chambers.
\newblock Composing dataflow analyses and transformations.
\newblock {\em SIGPLAN Not.}, 37(1):270--282, 2002.

\bibitem{Concolic}
K.~Sen, D.~Marinov, and G.~Agha.
\newblock Cute: A concolic unit testing engine for c.
\newblock {\em SIGSOFT Softw. Eng. Notes}, 30(5):263--272, 2005.

\bibitem{StaticDynamic}
Y.~Smaragdakis and Ch. Csallner.
\newblock Combining static and dynamic reasoning for bug detection.
\newblock In {\em Proceedings of TAP}, pages 1--16. Springer-Verlag, 2007.

\bibitem{ProductShapes}
A.~Toubhans, B.-Y. Chang, and X.~Rival.
\newblock Reduced product combination of abstract domains for shapes.
\newblock In {\em Proceedings of VMCAI}, volume 7737, pages 375--395. Springer
  Berlin Heidelberg, 2013.

\bibitem{ApronURL}
\textsc{Apron}.
\newblock \url{http://apron.cri.ensmp.fr/library}.

\bibitem{BugstURL}
\textsc{Bugst}.
\newblock \url{git://git.code.sf.net/p/bugst/src}.

\bibitem{CPAcheckerURL}
\textsc{CPAchecker}.
\newblock \url{http://cpachecker.sosy-lab.org}.

\bibitem{StatorURL}
\textsc{Evaluation package}.
\newblock \url{https://github.com/trtikm/aocbpa/releases/tag/v1.0}.

\bibitem{SMTLIBURL}
\textsc{SMT-LIB}.
\newblock \url{http://www.smt-lib.org}.

\bibitem{SVCOMPURL}
\textsc{SV-COMP}.
\newblock \url{http://sv-comp.sosy-lab.org}.

\bibitem{Z3URL}
\textsc{Z3}.
\newblock \url{https://github.com/Z3Prover/z3}.

\end{thebibliography}

\clearpage

\appendix
\section{Plots}
\label{sec:Plots}

Here we present data summarised in Fig.~\ref{fig:Invariants} and
Fig~\ref{fig:Termination} in a form of histograms. They are automatically
produced by the tool after evaluation (and then prettify by our utility
'\texttt{STATOR-tool/plotcopy.py}'). Data in the mentioned figures were
collected from the histograms.

\begin{figure}
\begin{tabular}{cc}
\includegraphics[scale=0.25]{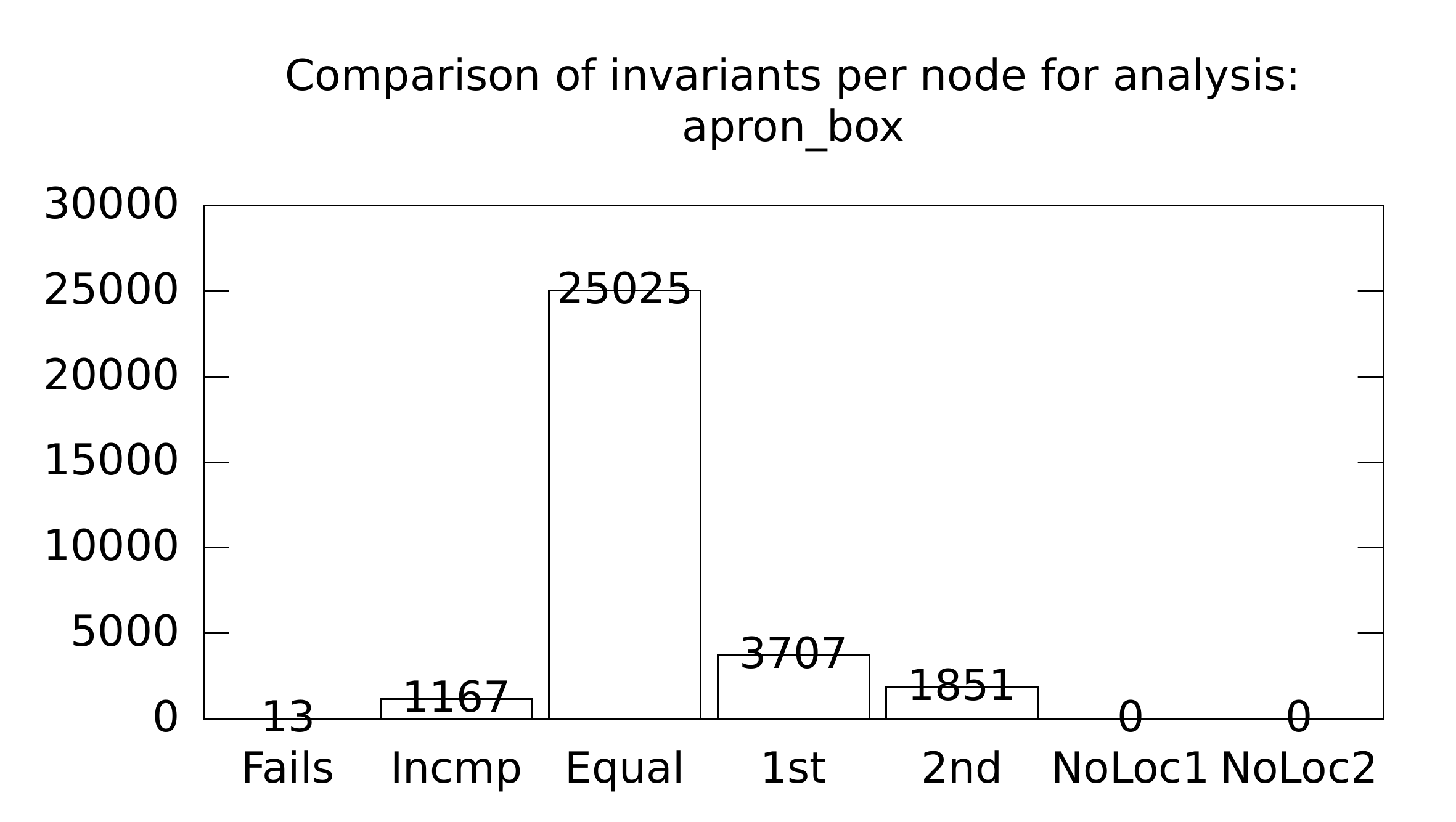}
&
\includegraphics[scale=0.25]{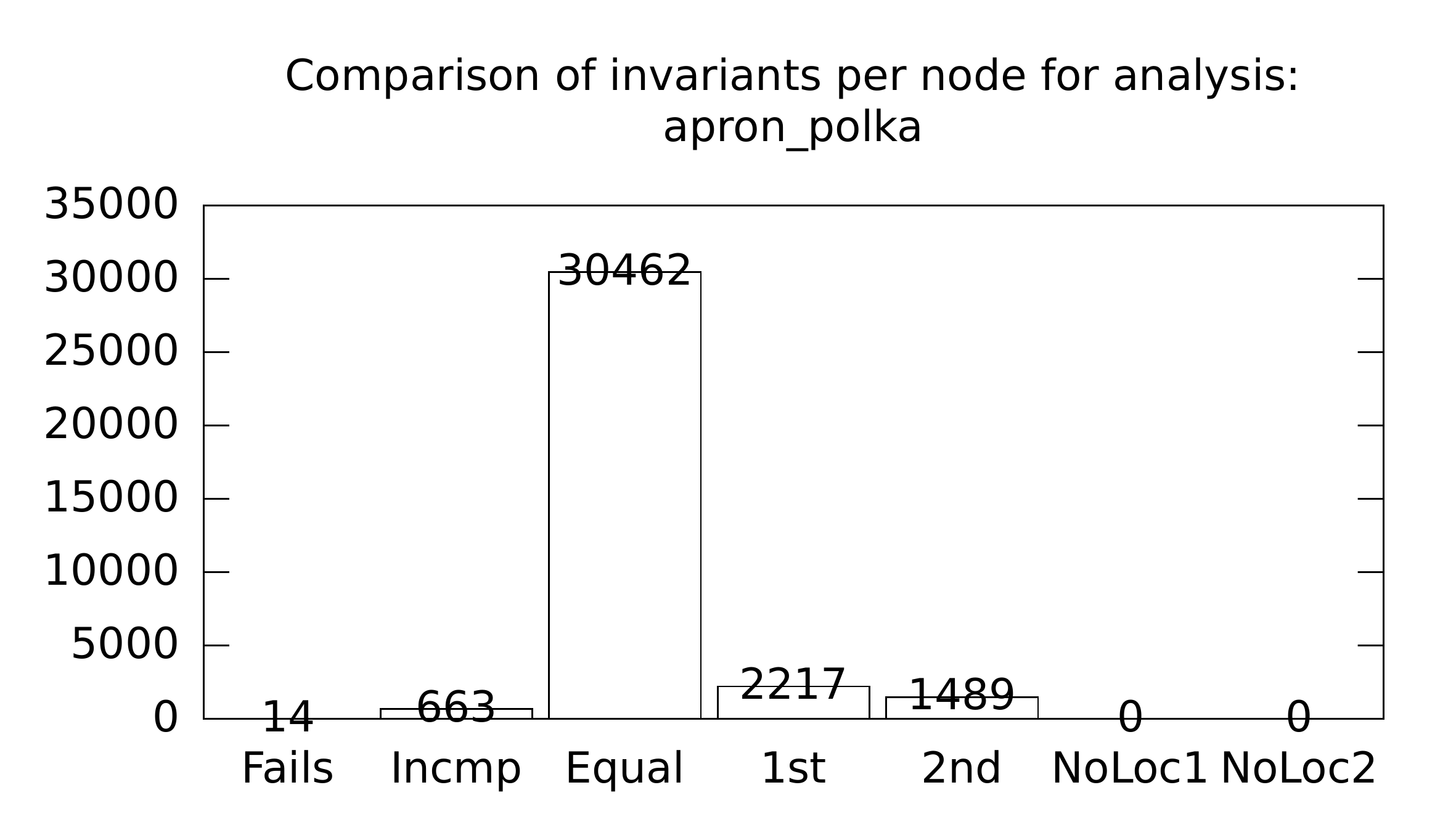}
\\
\includegraphics[scale=0.25]{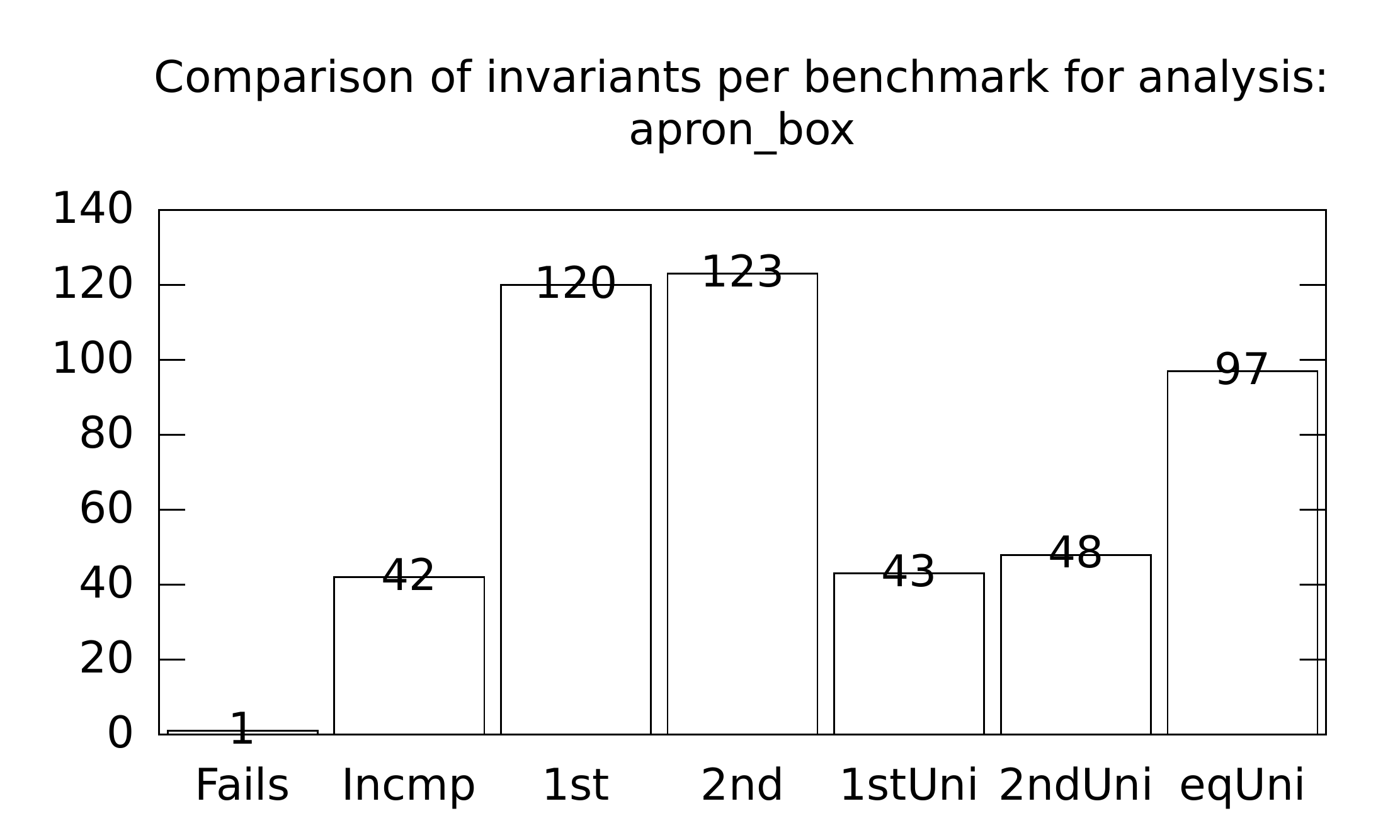}
&
\includegraphics[scale=0.25]{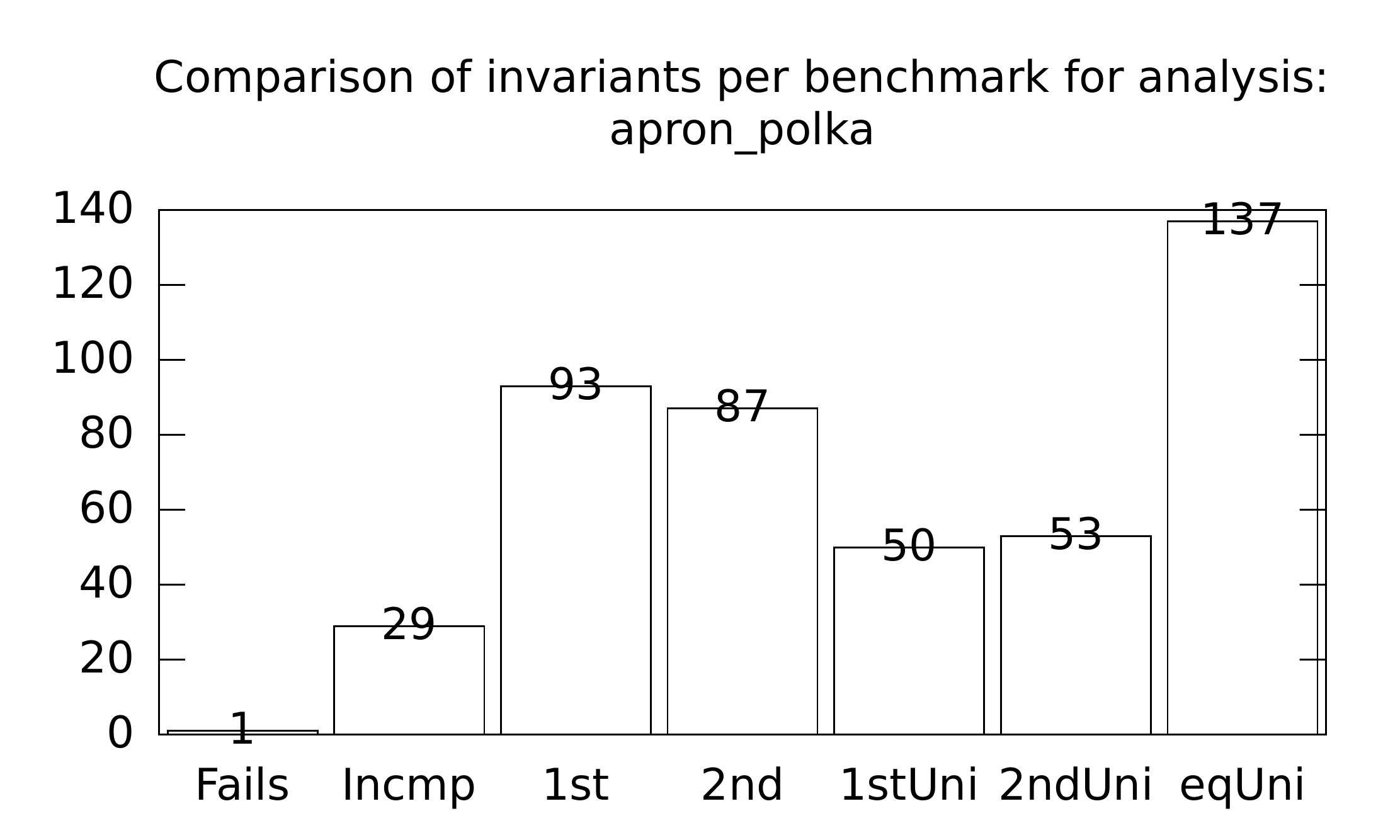}
\\
\includegraphics[scale=0.25]{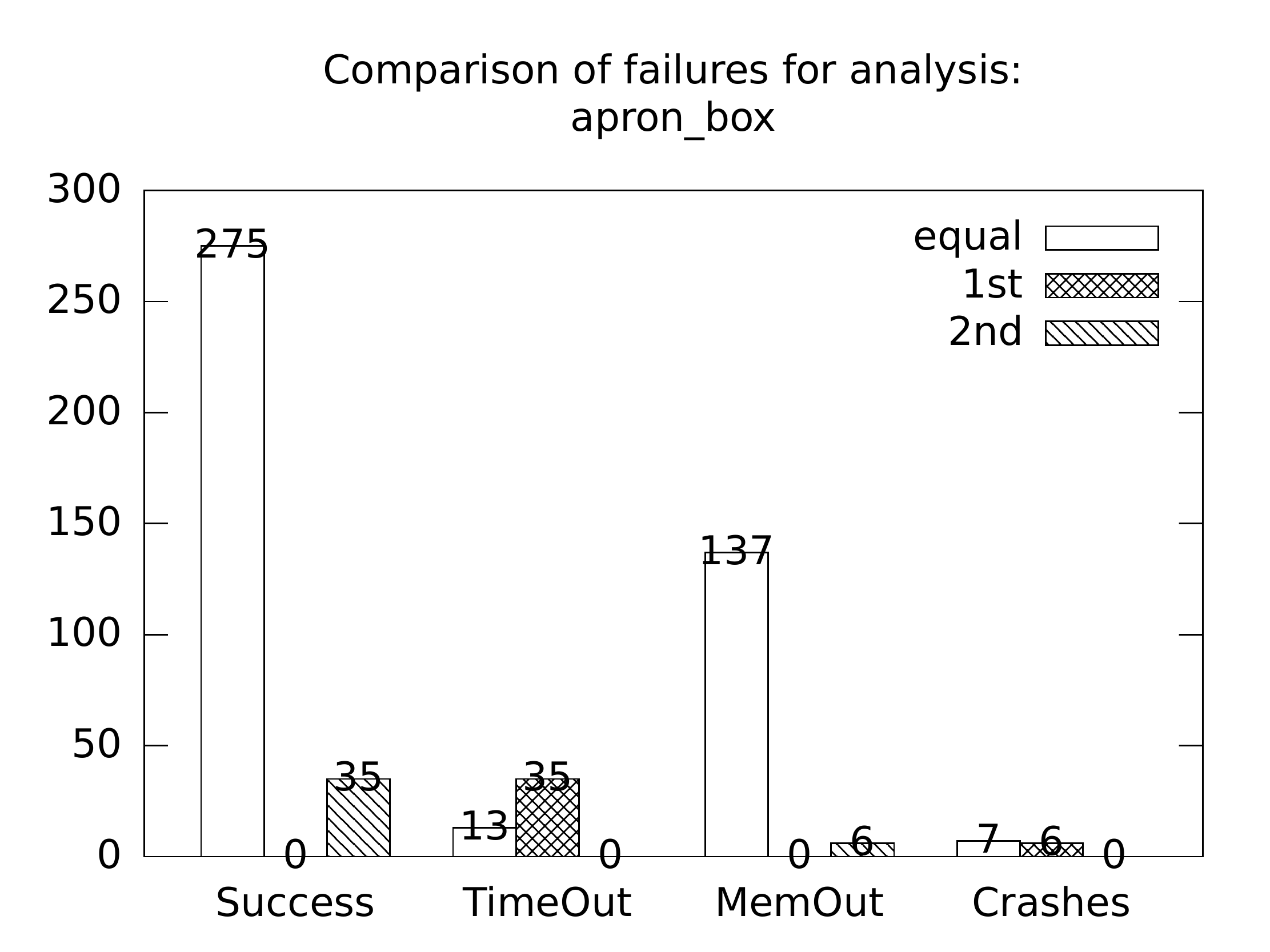}
&
\includegraphics[scale=0.25]{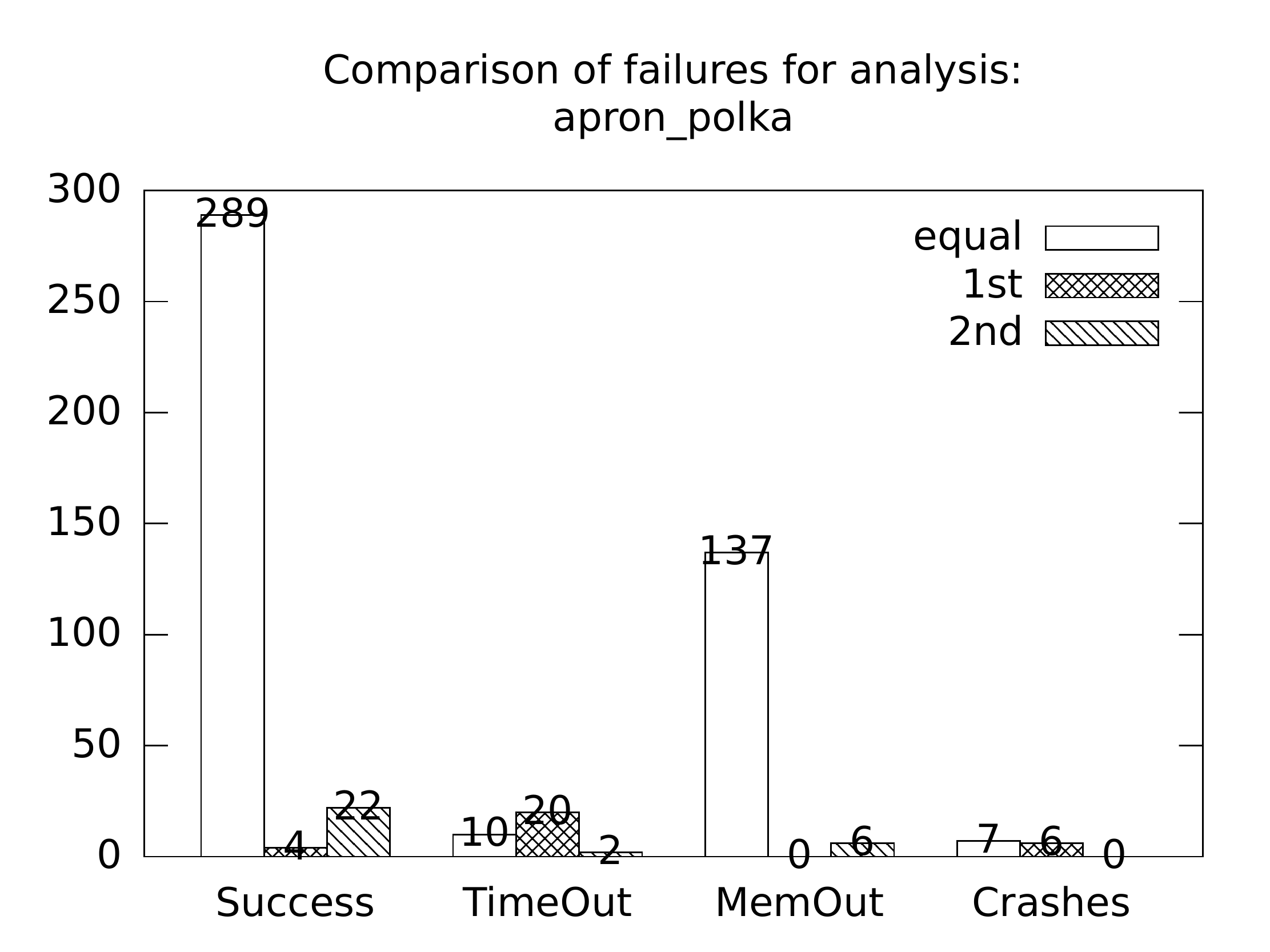}
\end{tabular}
\caption{Comparison:~~b+p+s~~vs.~~b+p.}
\end{figure}

\begin{figure}
\begin{tabular}{cc}
\includegraphics[scale=0.25]{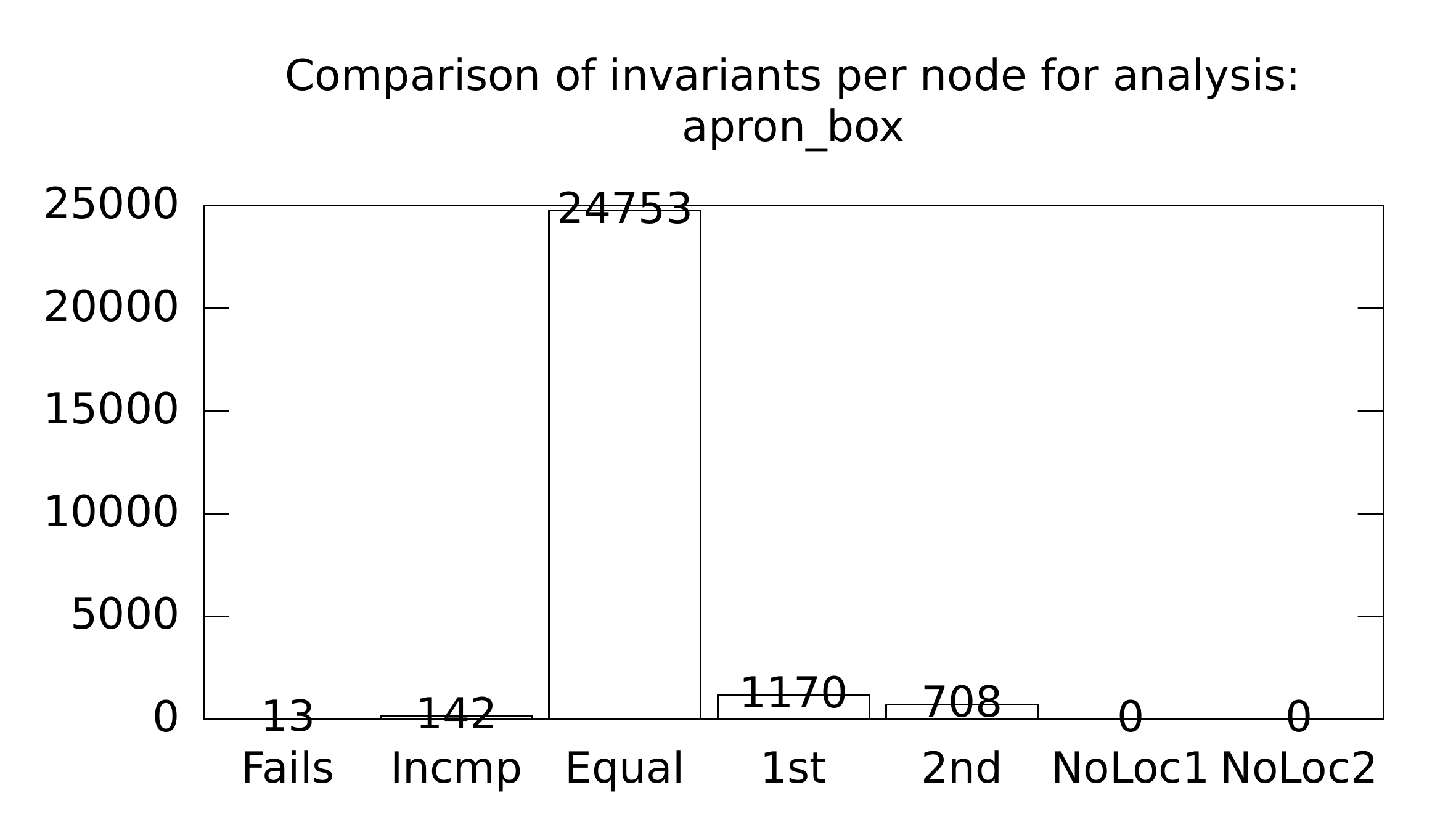}
&
\includegraphics[scale=0.25]{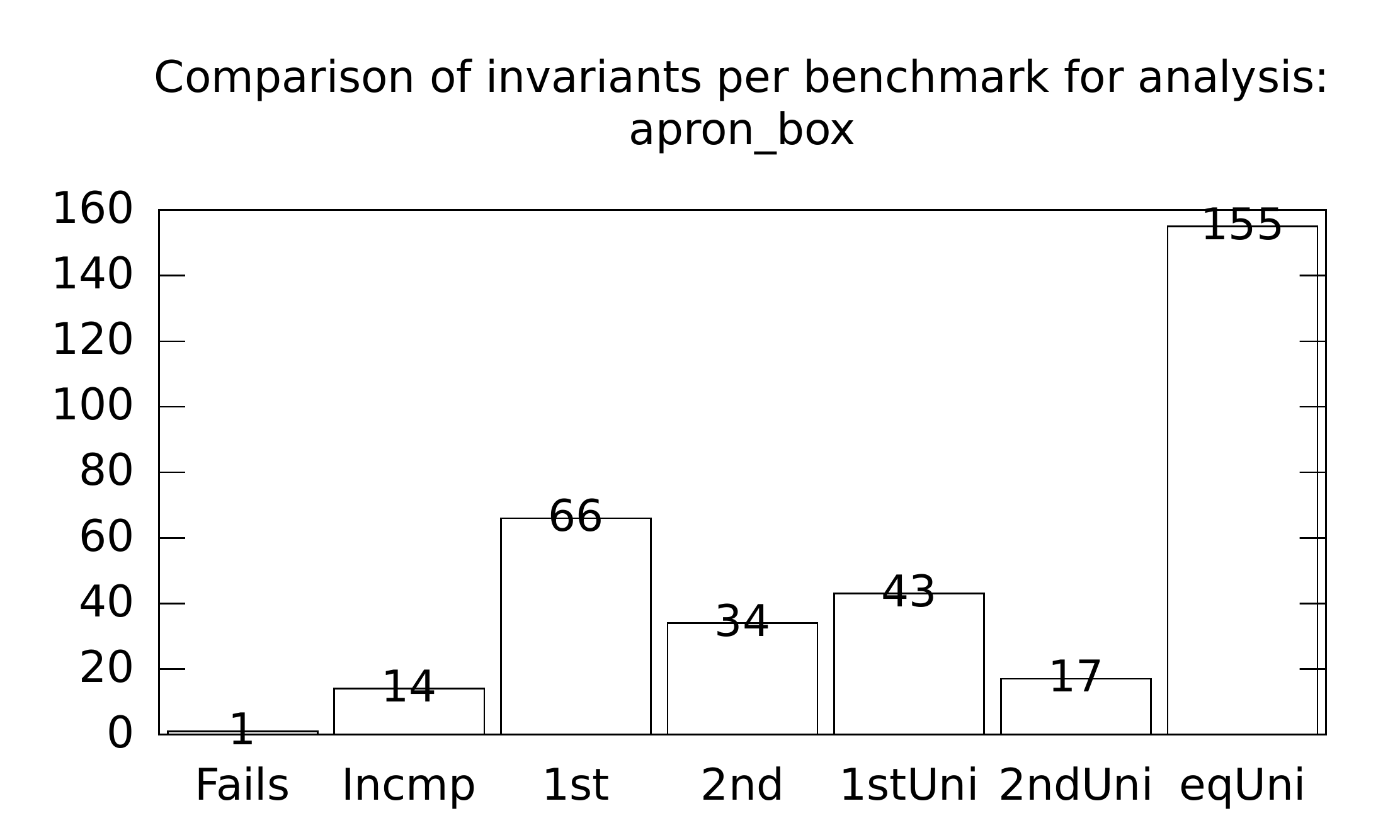}
\\
\includegraphics[scale=0.25]{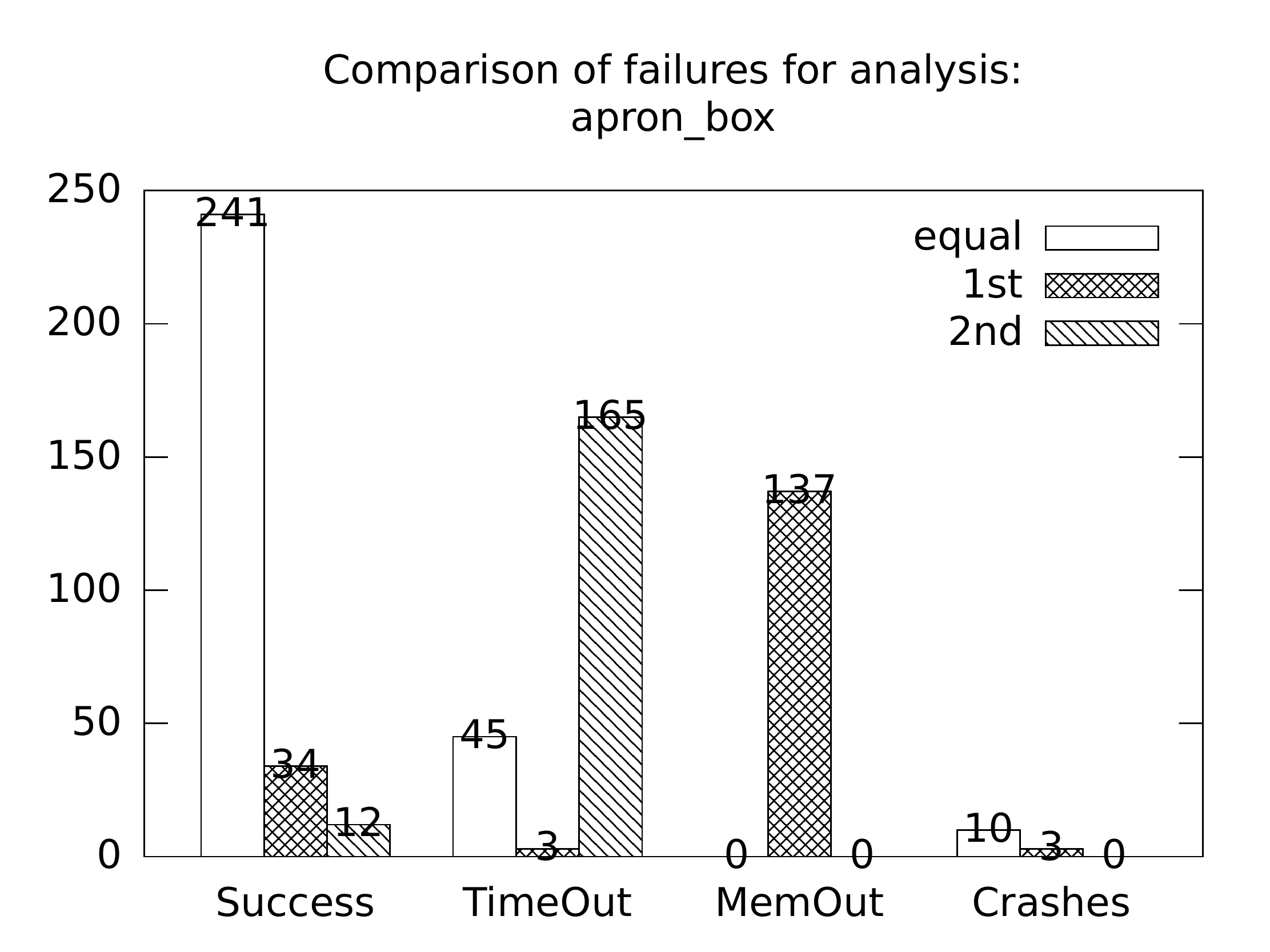}
&
\includegraphics[scale=0.25]{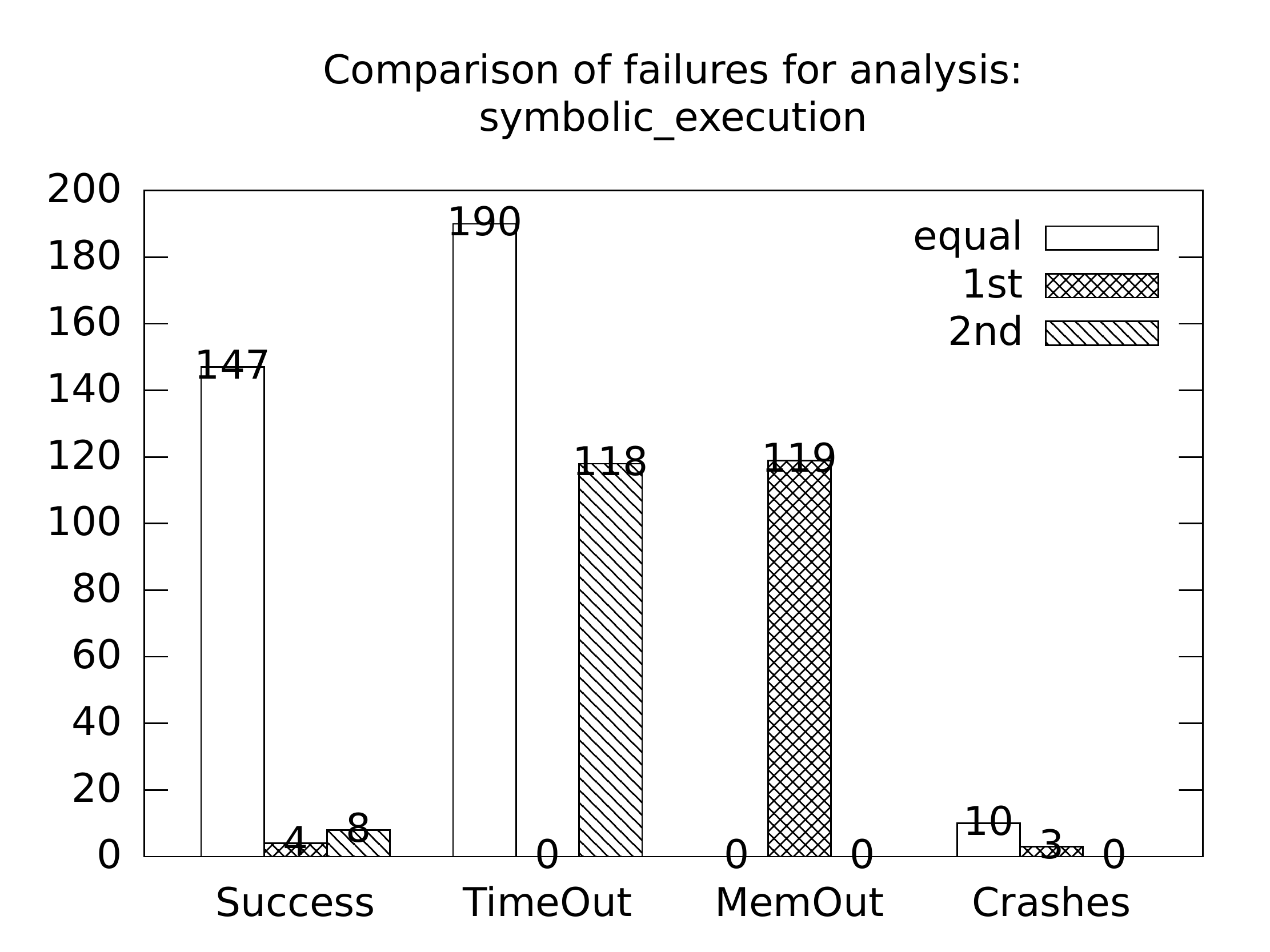}
\end{tabular}
\caption{Comparison:~~b+p+s~~vs.~~b+s.}
\end{figure}

\begin{figure}
\begin{tabular}{cc}
\includegraphics[scale=0.25]{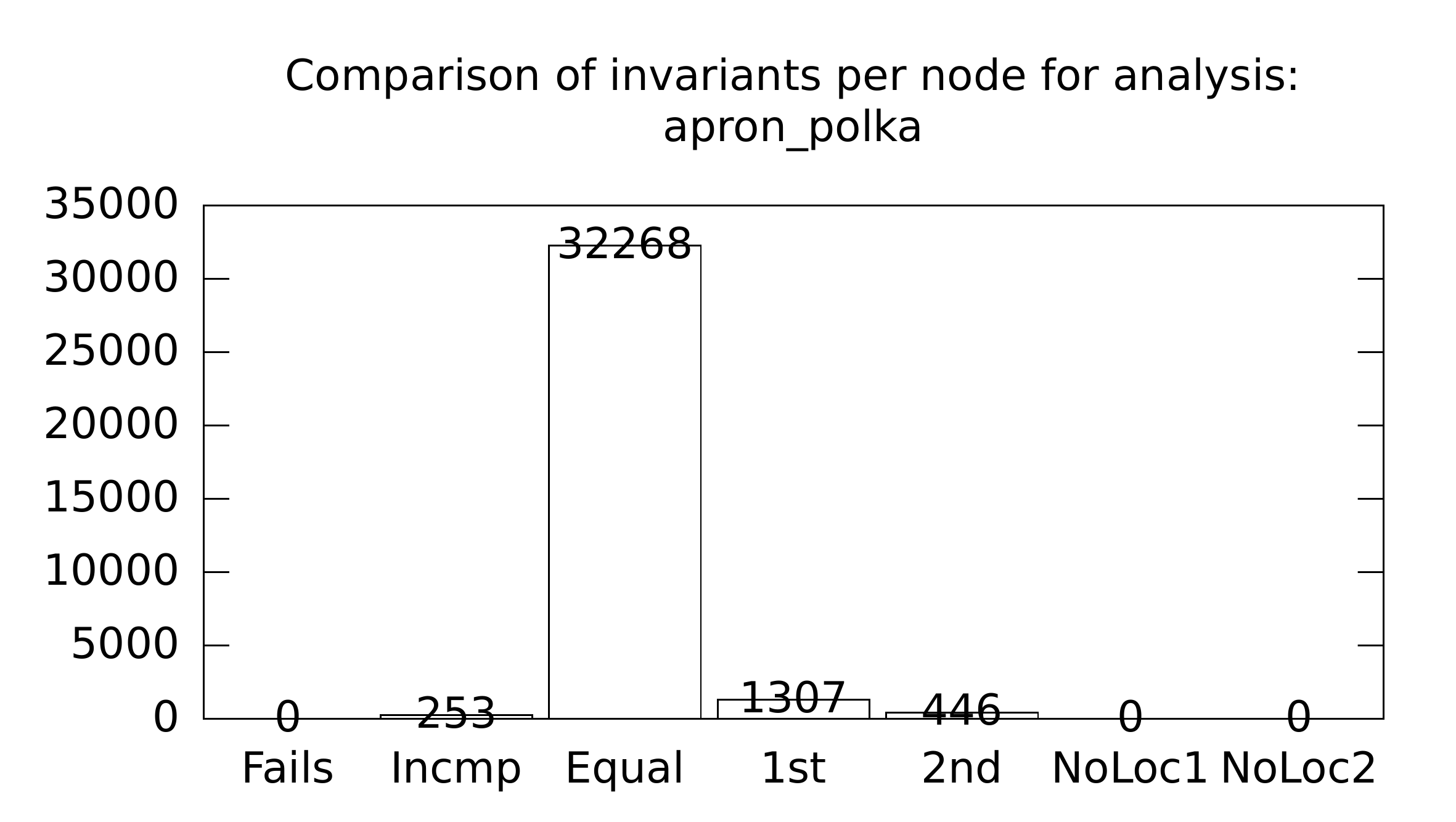}
&
\includegraphics[scale=0.25]{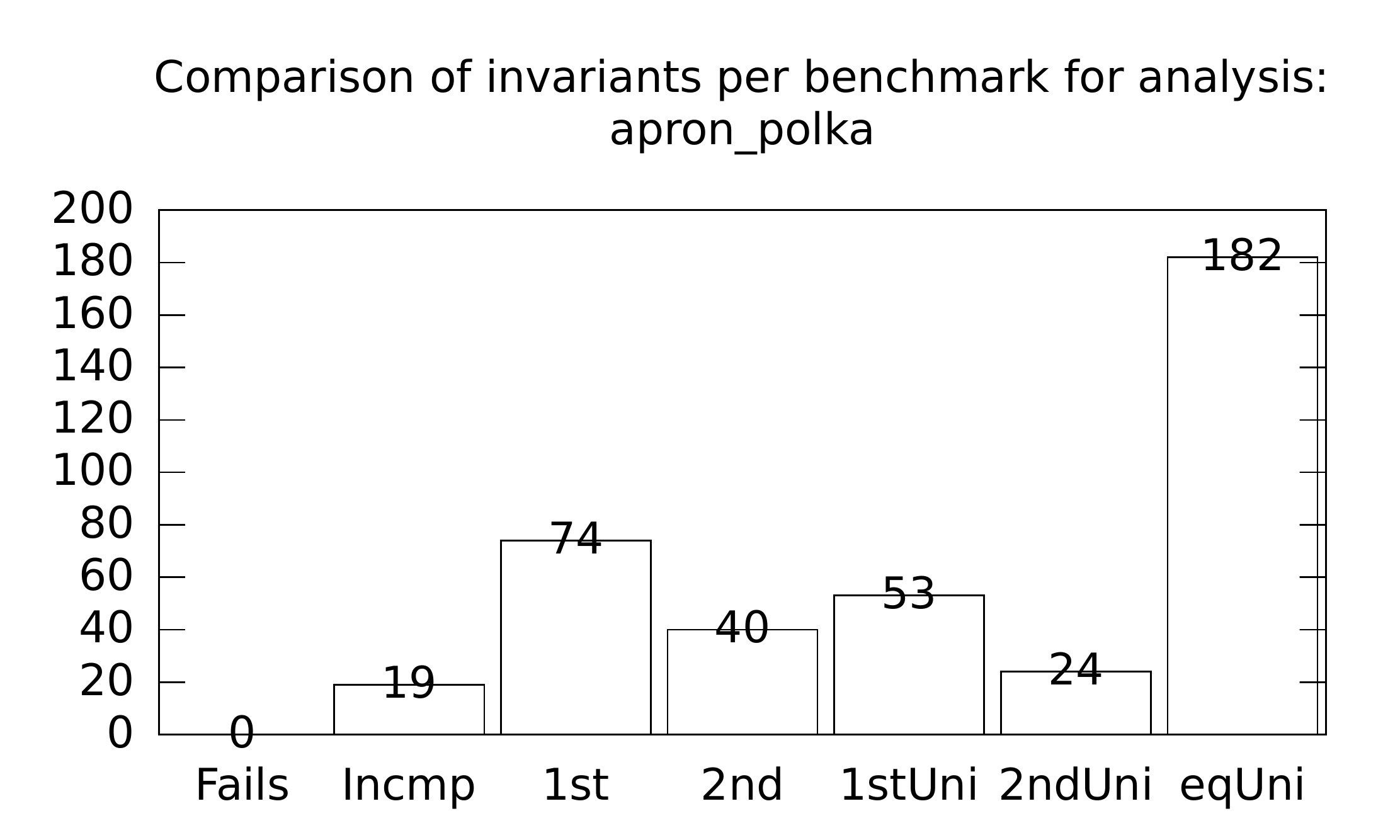}
\\
\includegraphics[scale=0.25]{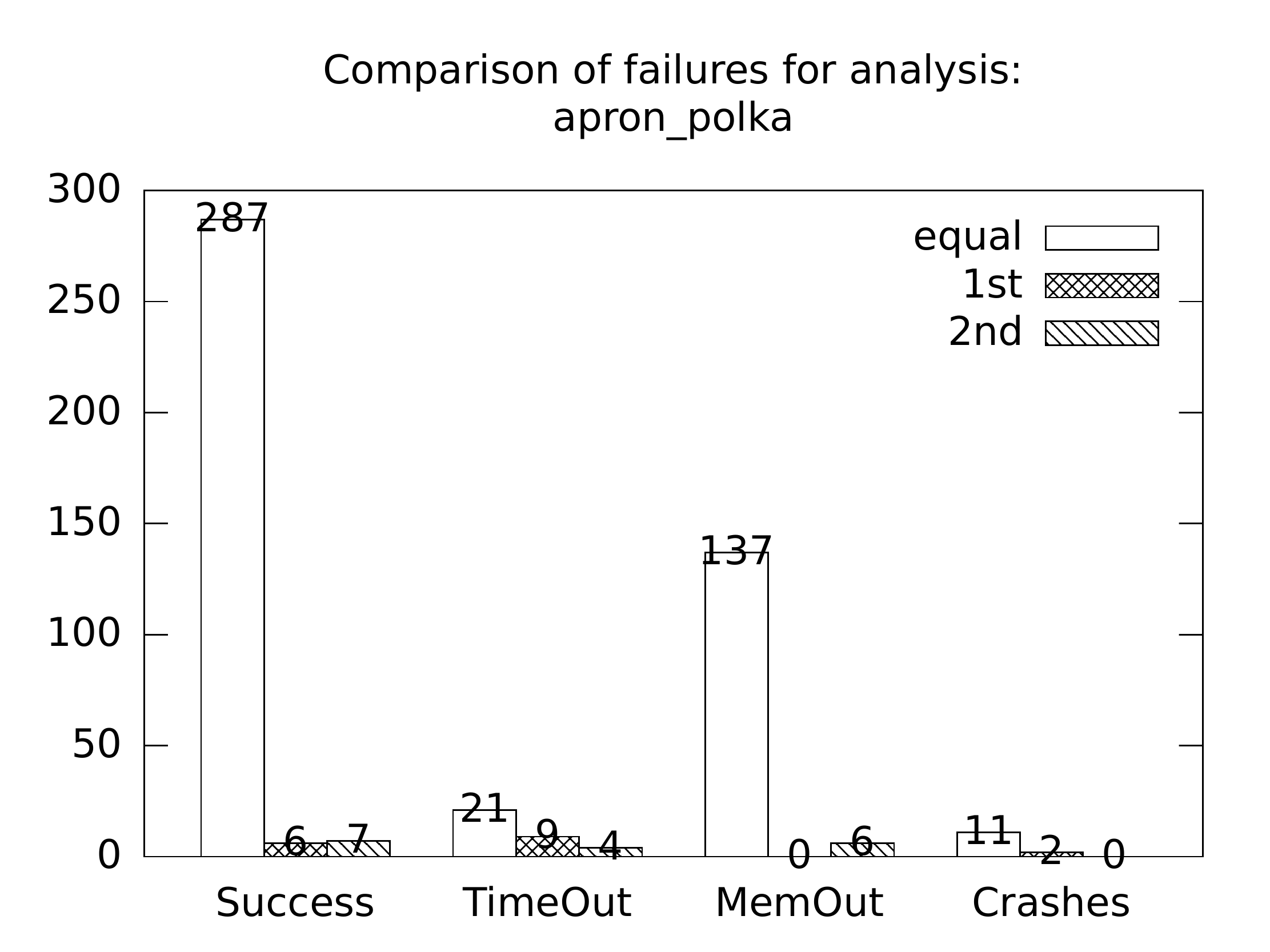}
&
\includegraphics[scale=0.25]{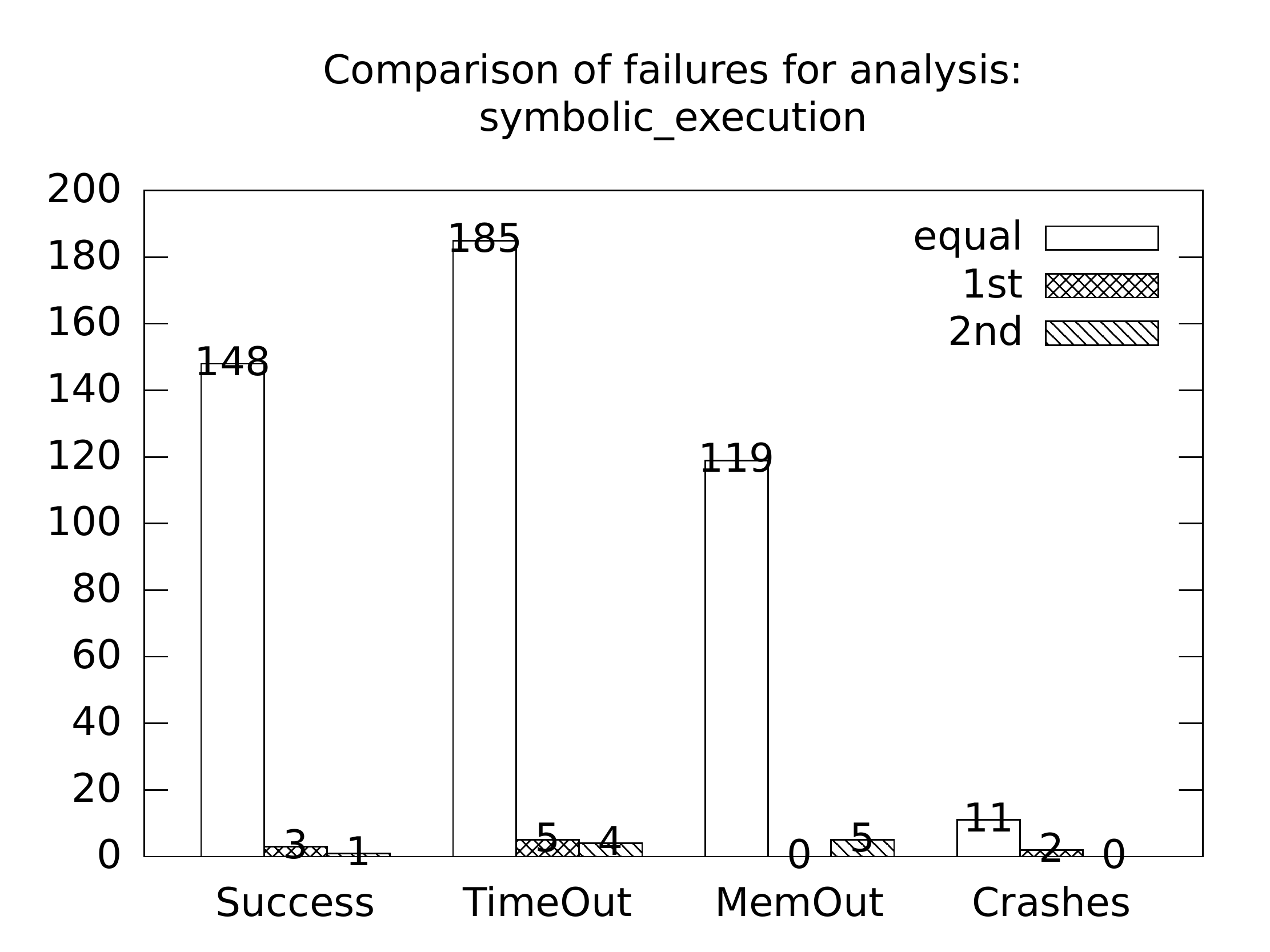}
\end{tabular}
\caption{Comparison:~~b+p+s~~vs.~~p+s.}
\end{figure}

\begin{figure}
\begin{tabular}{cc}
\includegraphics[scale=0.25]{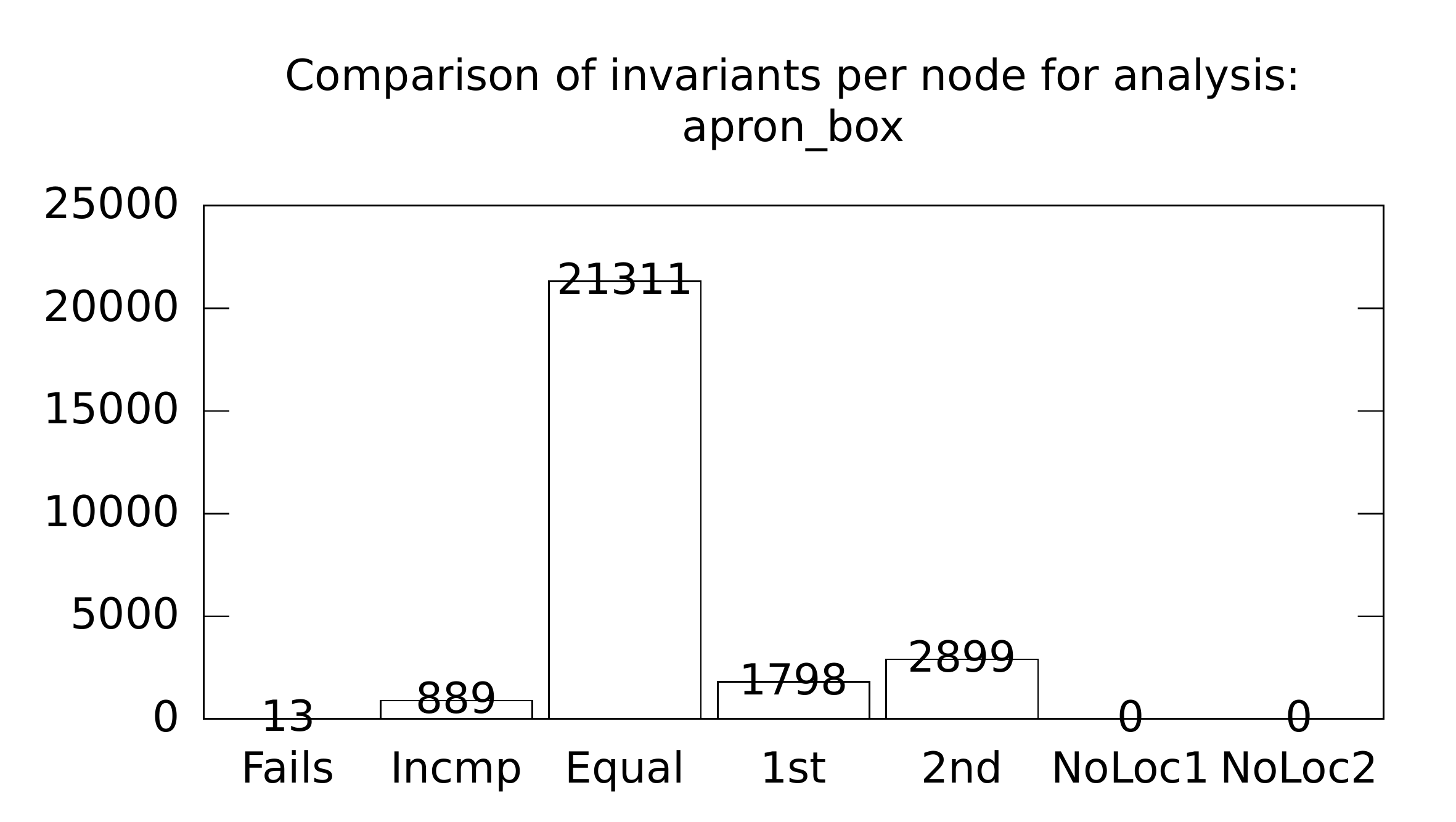}
&
\includegraphics[scale=0.25]{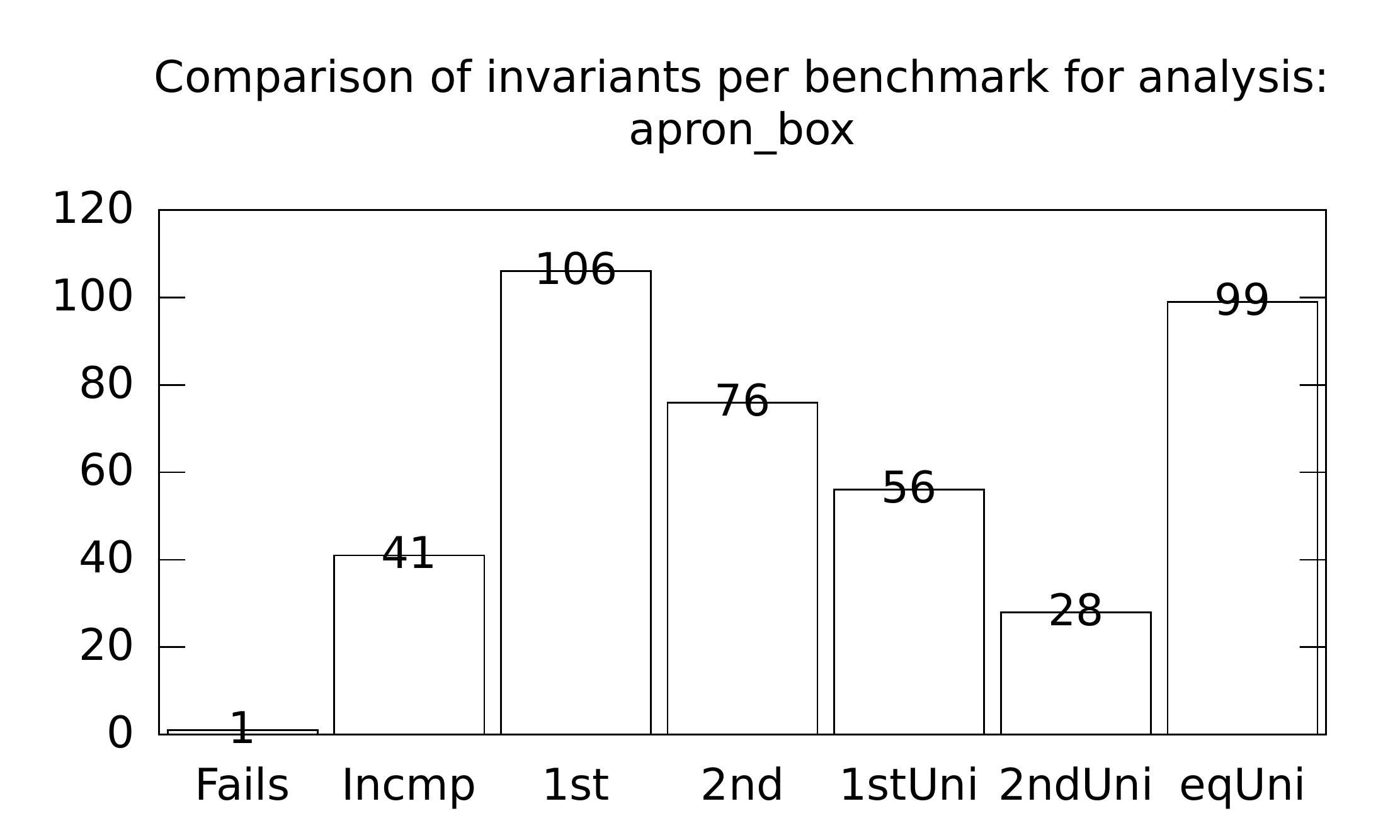}
\\
\includegraphics[scale=0.25]{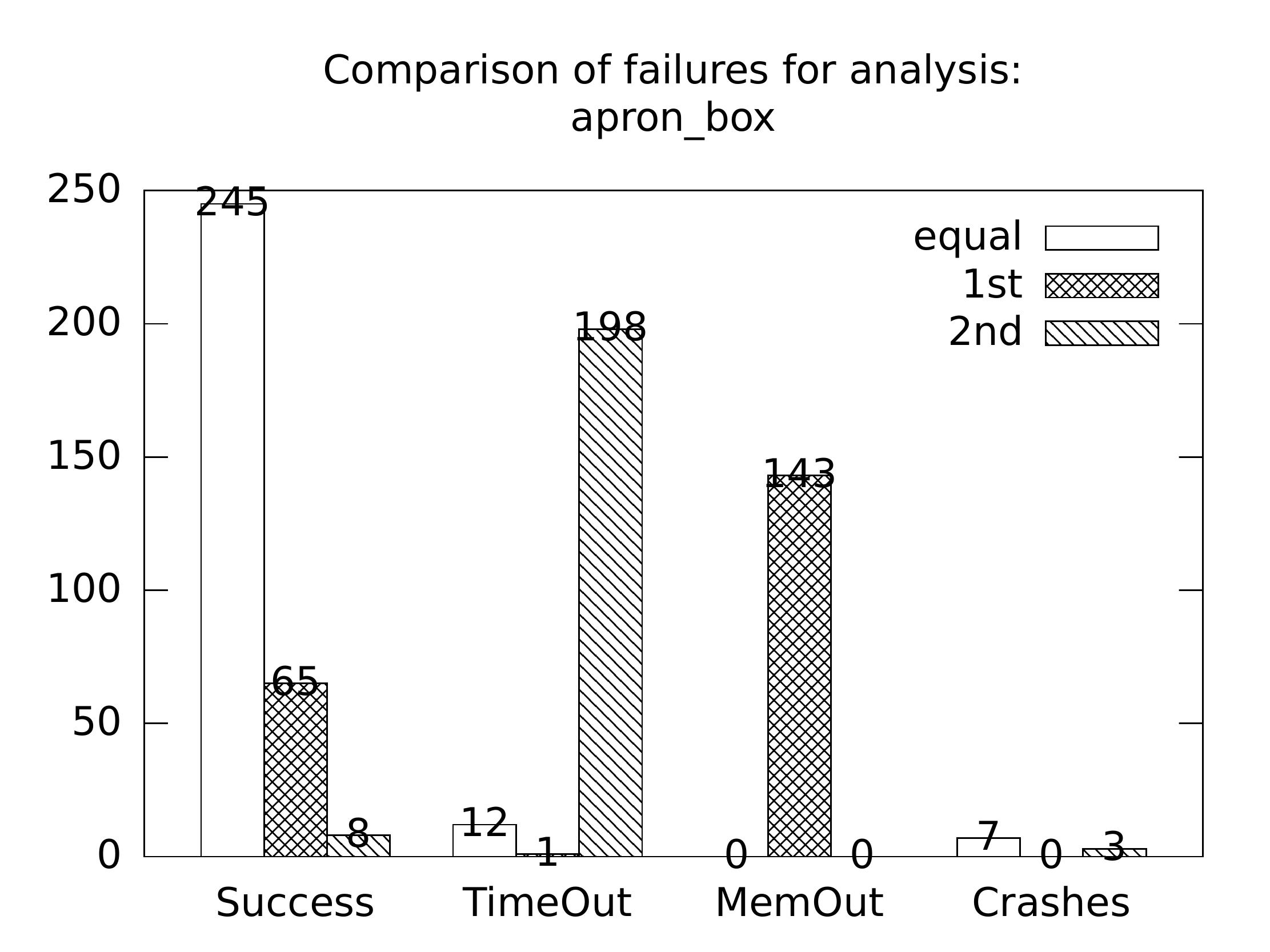}
\end{tabular}
\caption{Comparison:~~b+p~~vs.~~b+s.}
\end{figure}

\begin{figure}
\begin{tabular}{cc}
\includegraphics[scale=0.25]{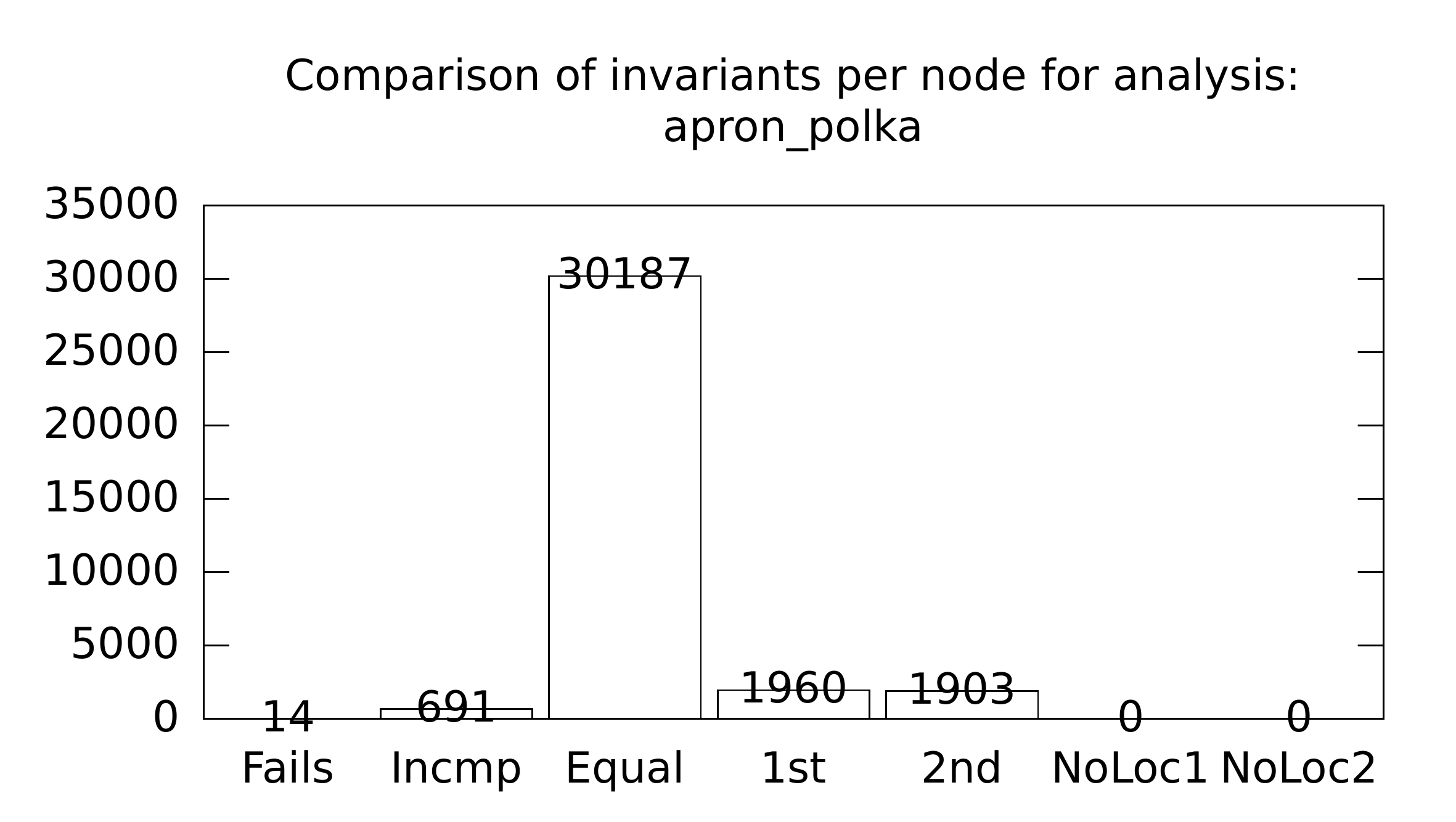}
&
\includegraphics[scale=0.25]{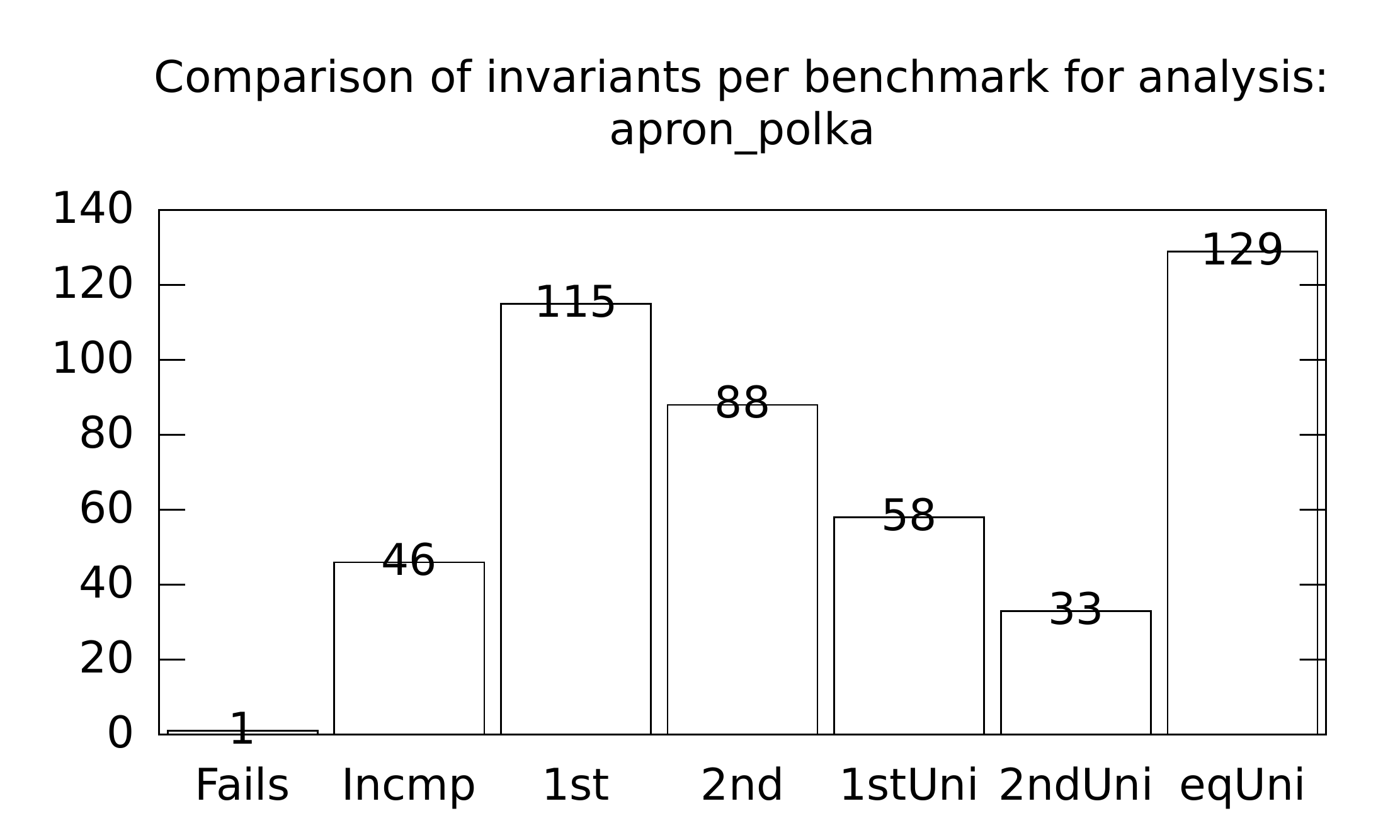}
\\
\includegraphics[scale=0.25]{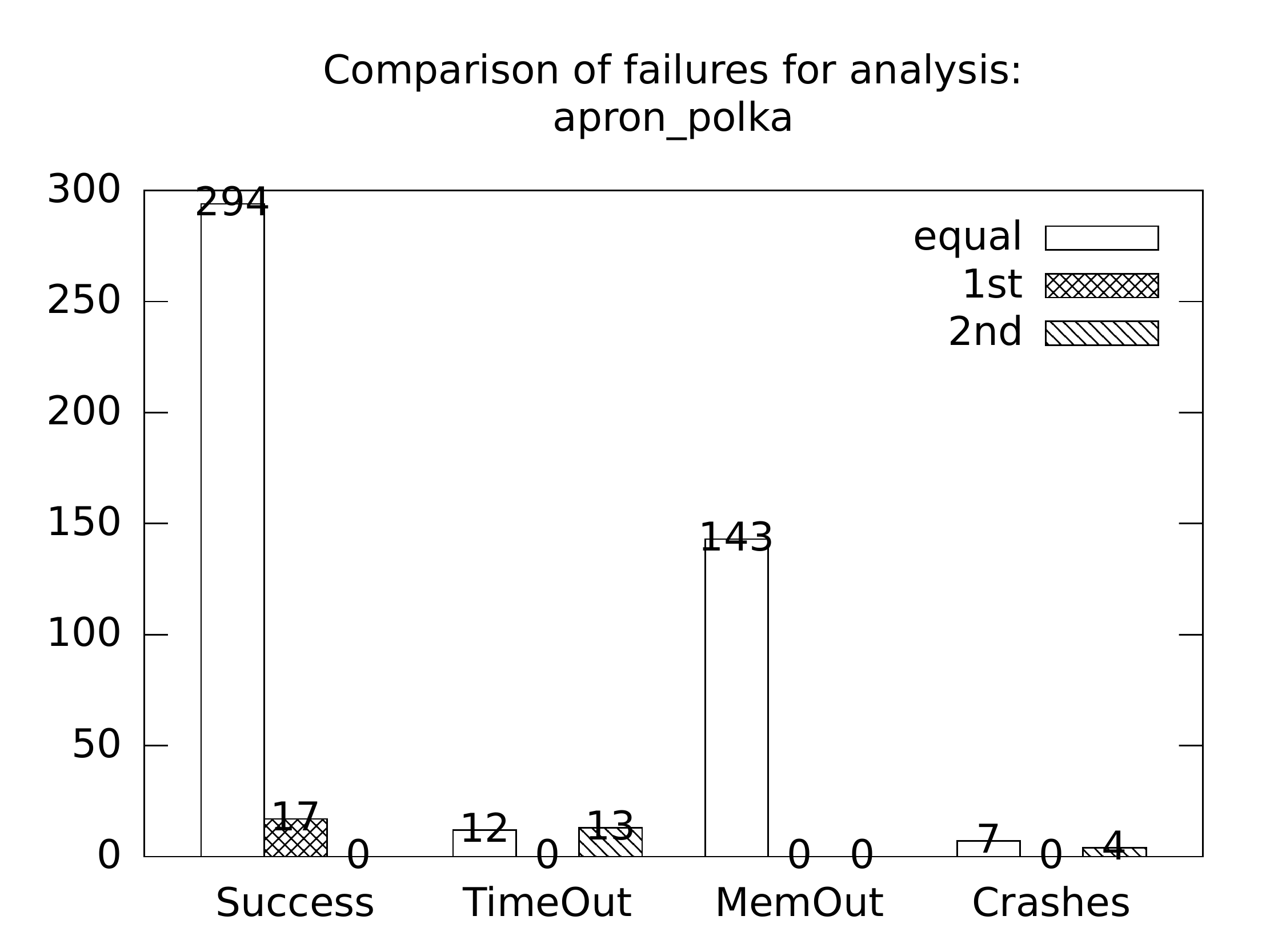}
\end{tabular}
\caption{Comparison:~~b+p~~vs.~~p+s.}
\end{figure}

\begin{figure}
\begin{tabular}{cc}
\includegraphics[scale=0.25]{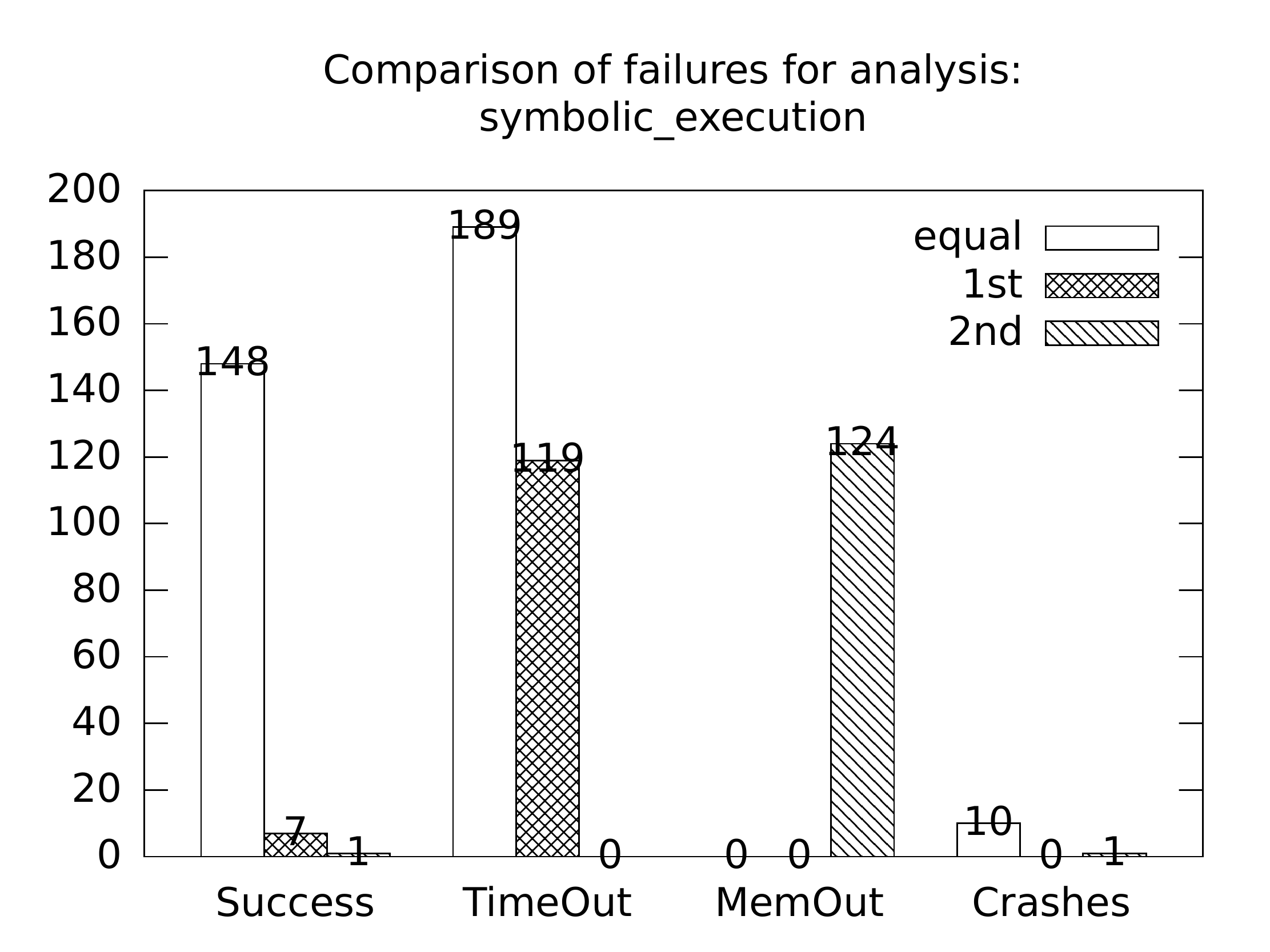}
\end{tabular}
\caption{Comparison:~~b+s~~vs.~~p+s.}
\end{figure}

\begin{figure}
\begin{tabular}{cc}
\includegraphics[scale=0.25]{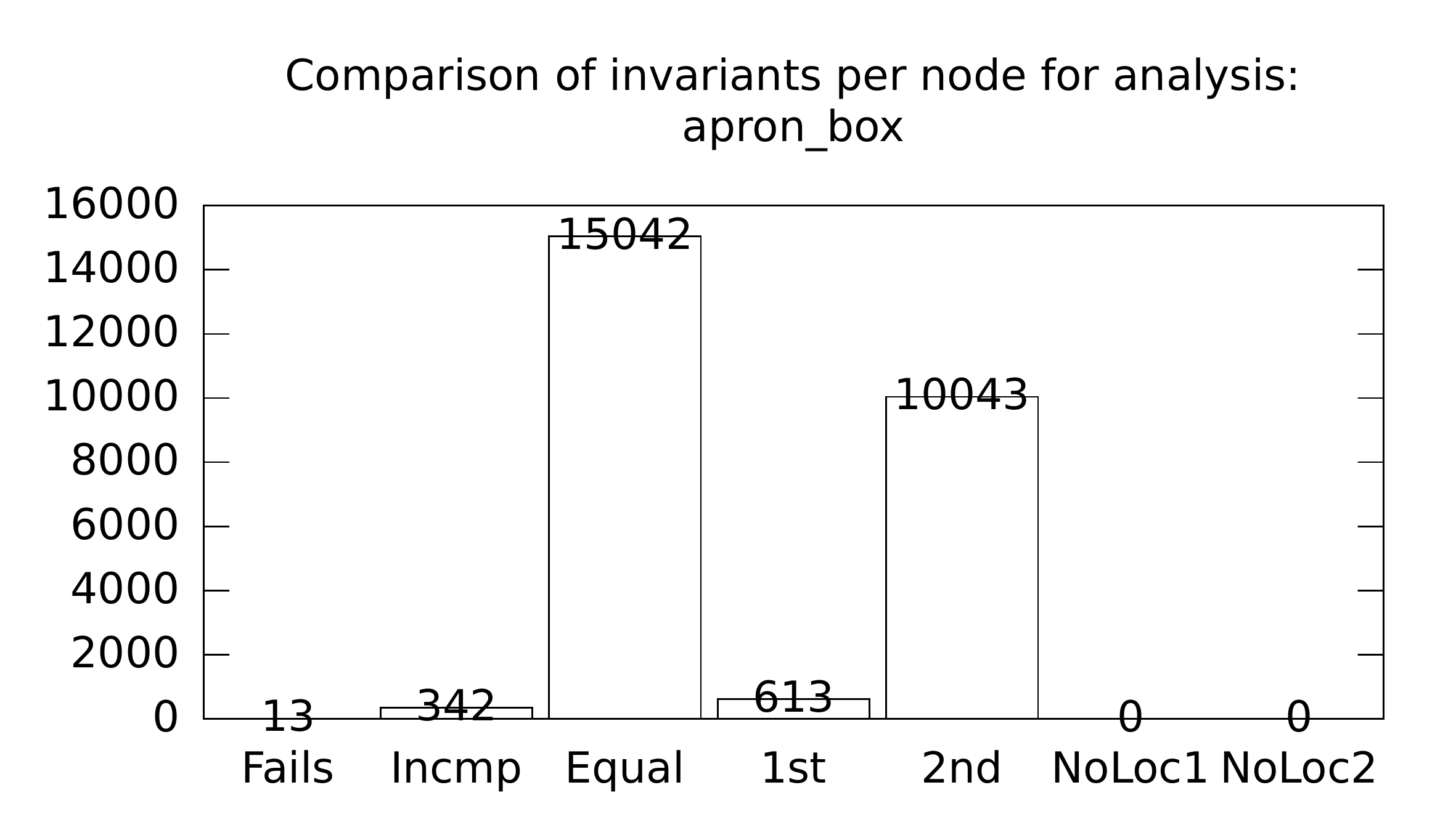}
&
\includegraphics[scale=0.25]{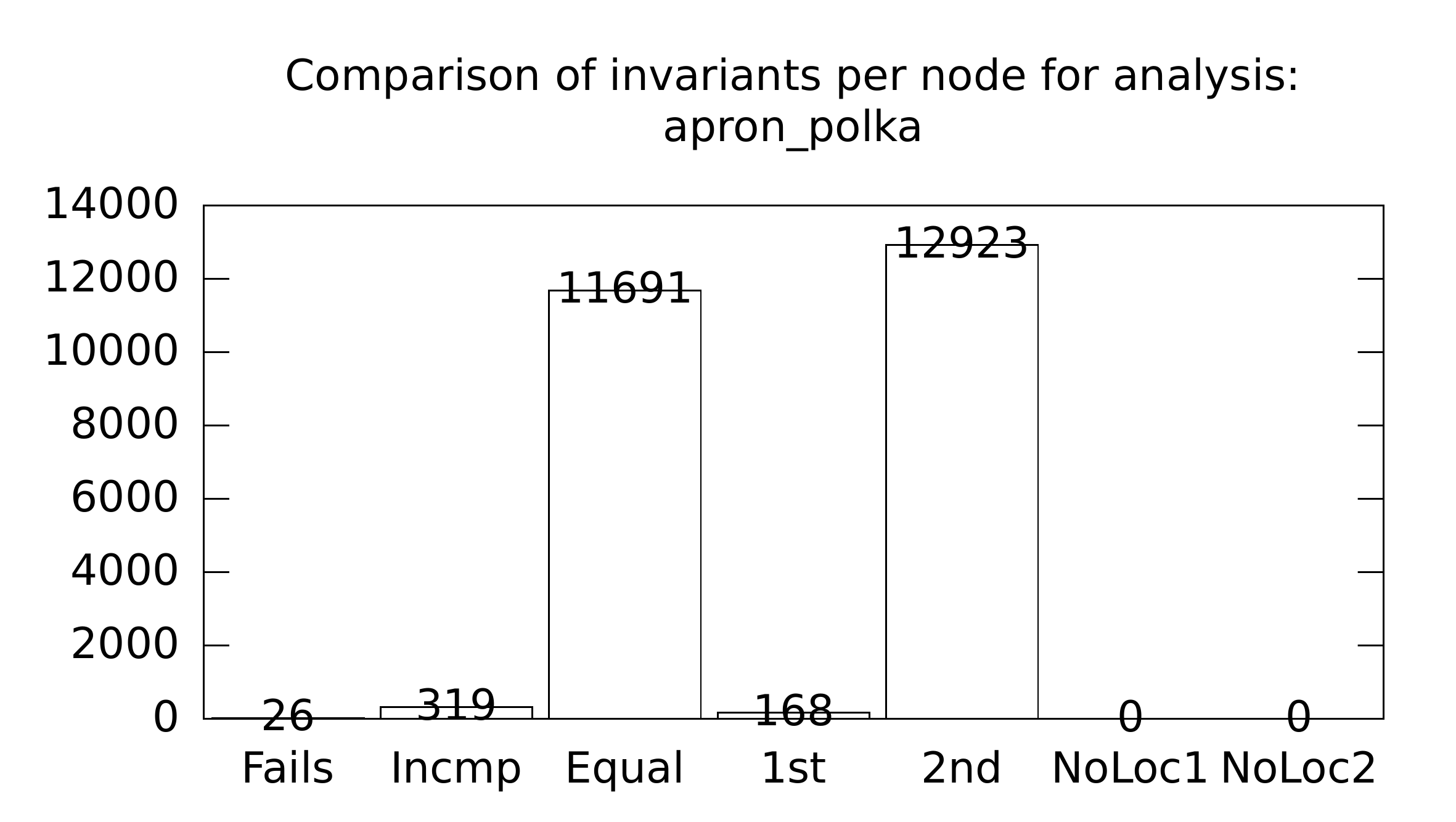}
\\
\includegraphics[scale=0.25]{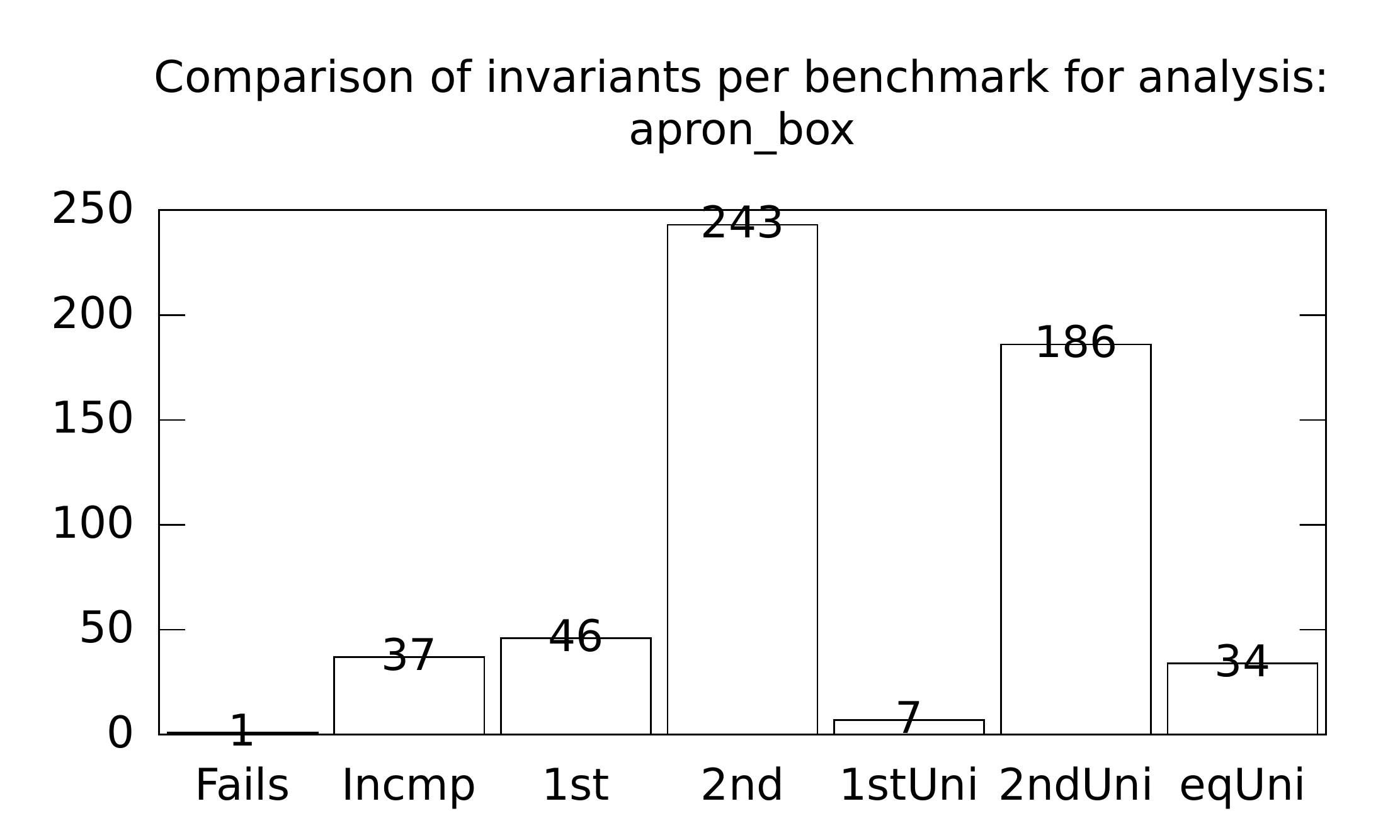}
&
\includegraphics[scale=0.25]{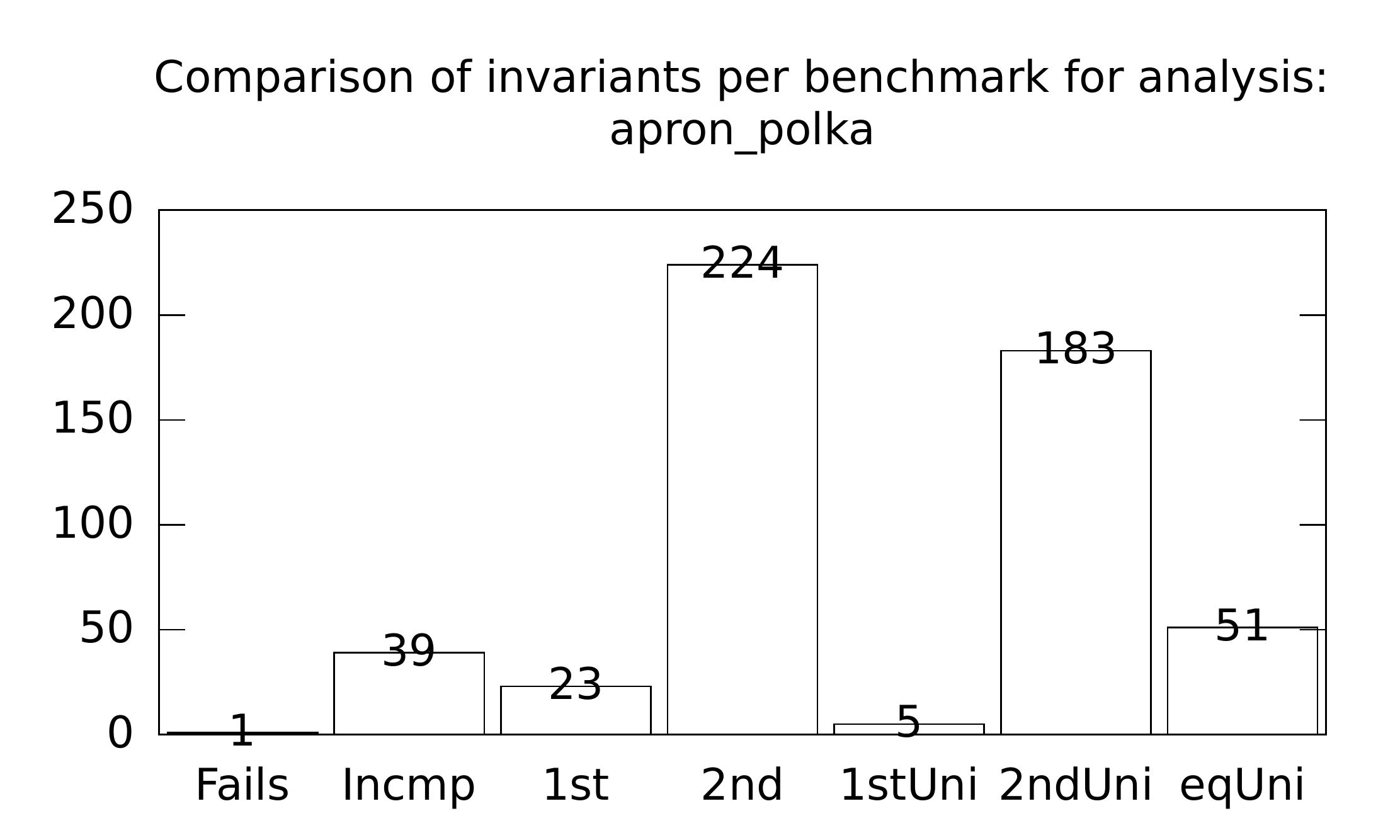}
\\
\includegraphics[scale=0.25]{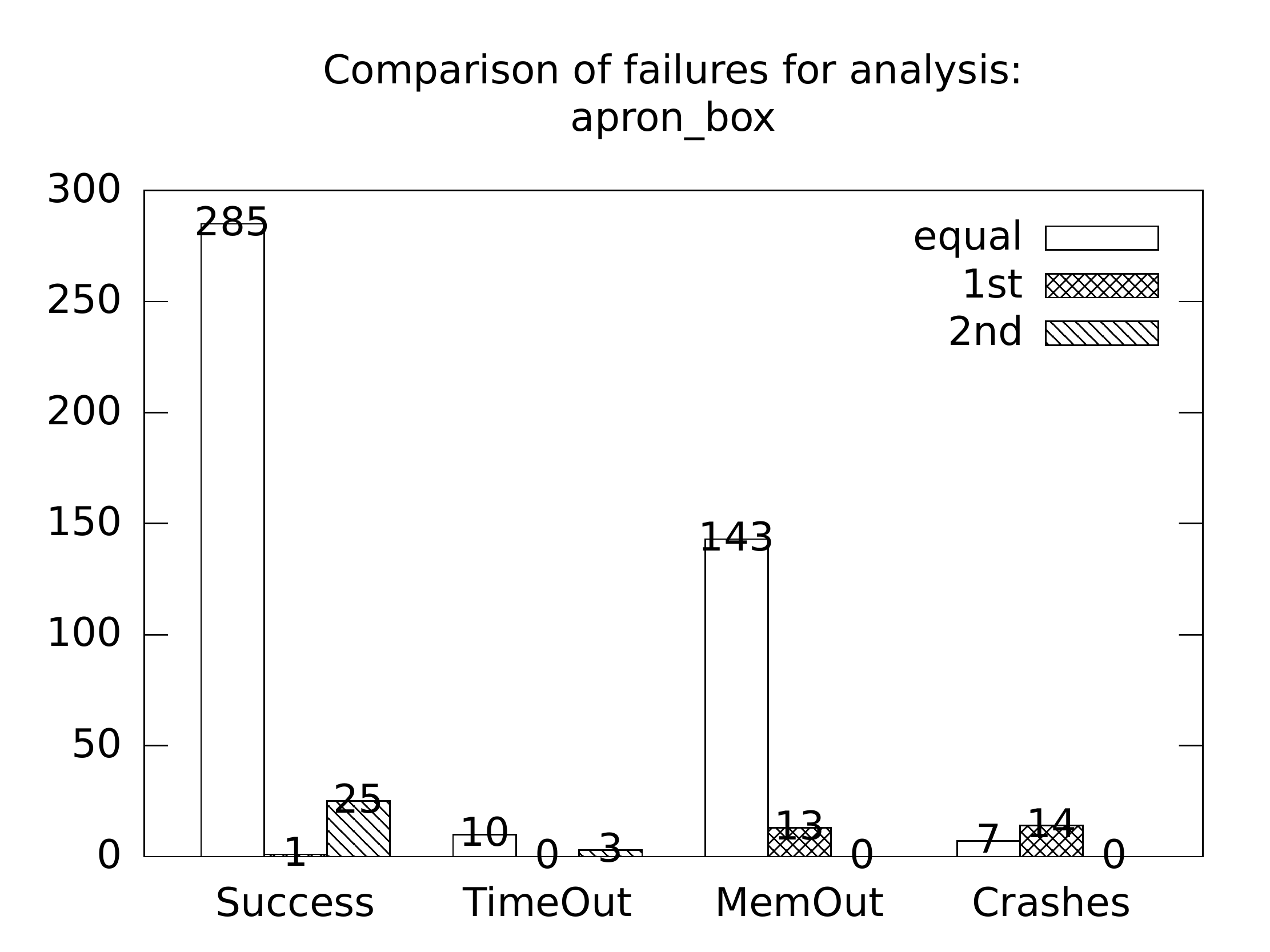}
&
\includegraphics[scale=0.25]{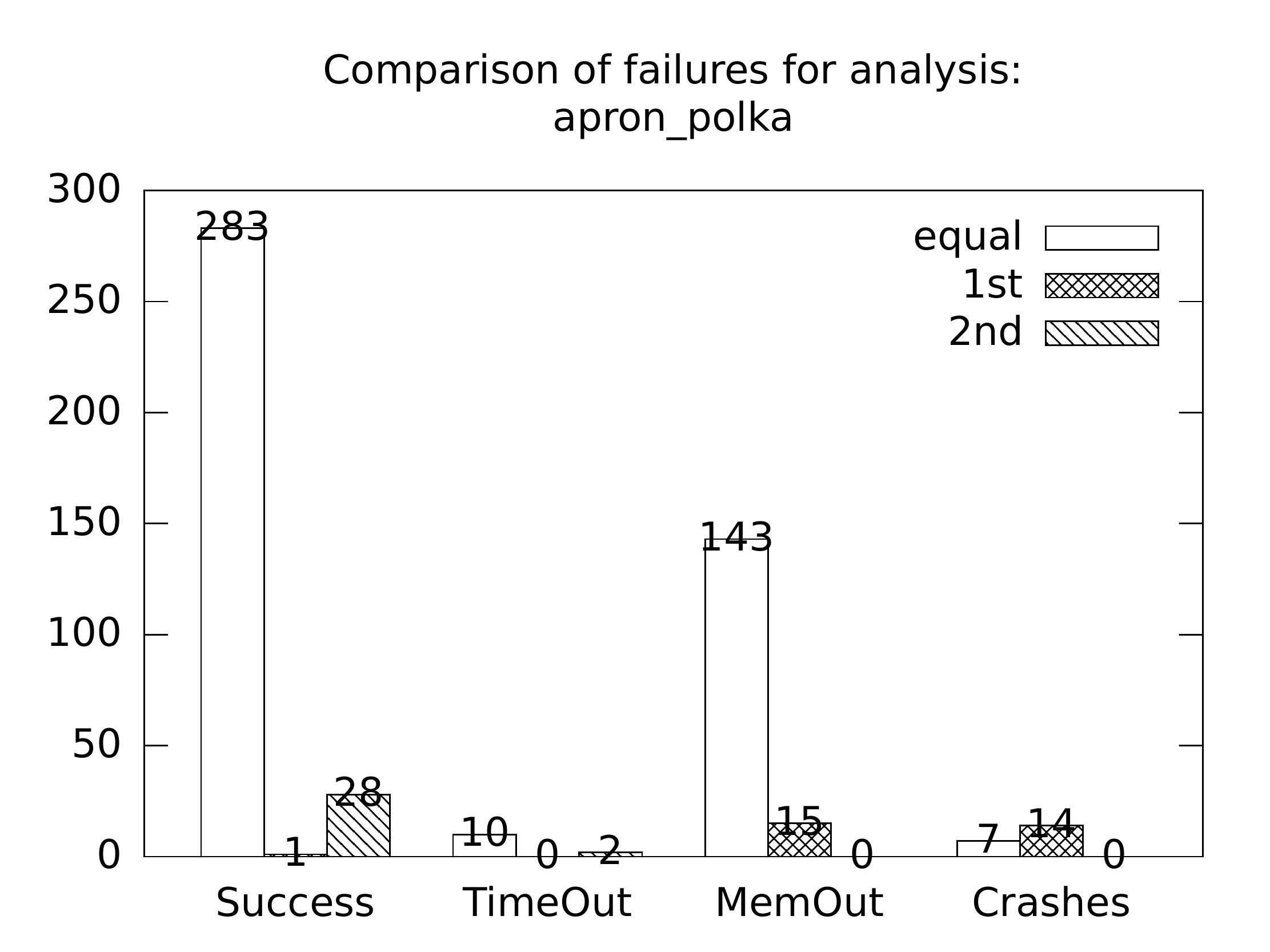}
\end{tabular}
\caption{Comparison:~~b*p*s~~vs.~~b+p.}
\end{figure}

\begin{figure}
\begin{tabular}{cc}
\includegraphics[scale=0.25]{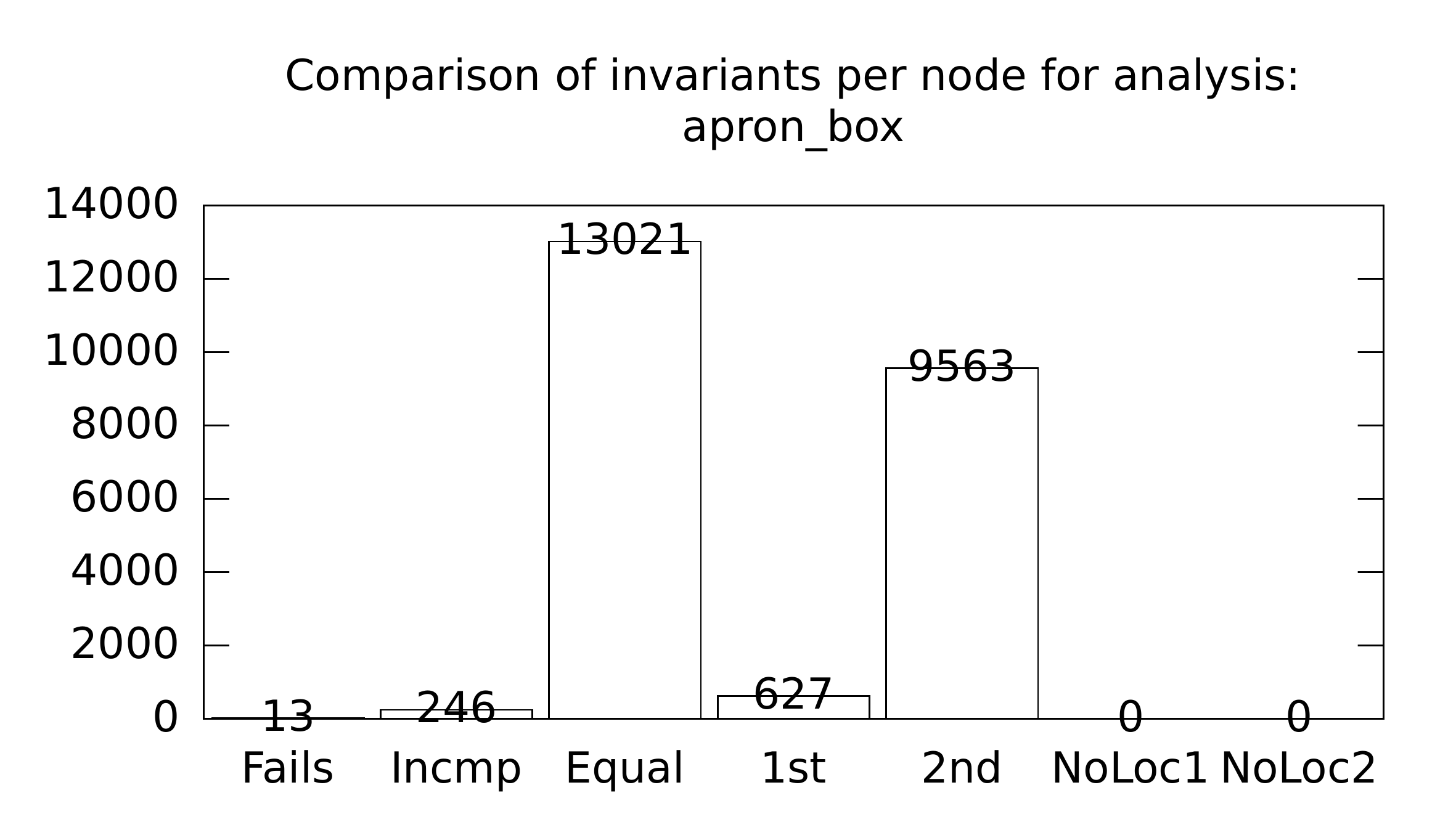}
&
\includegraphics[scale=0.25]{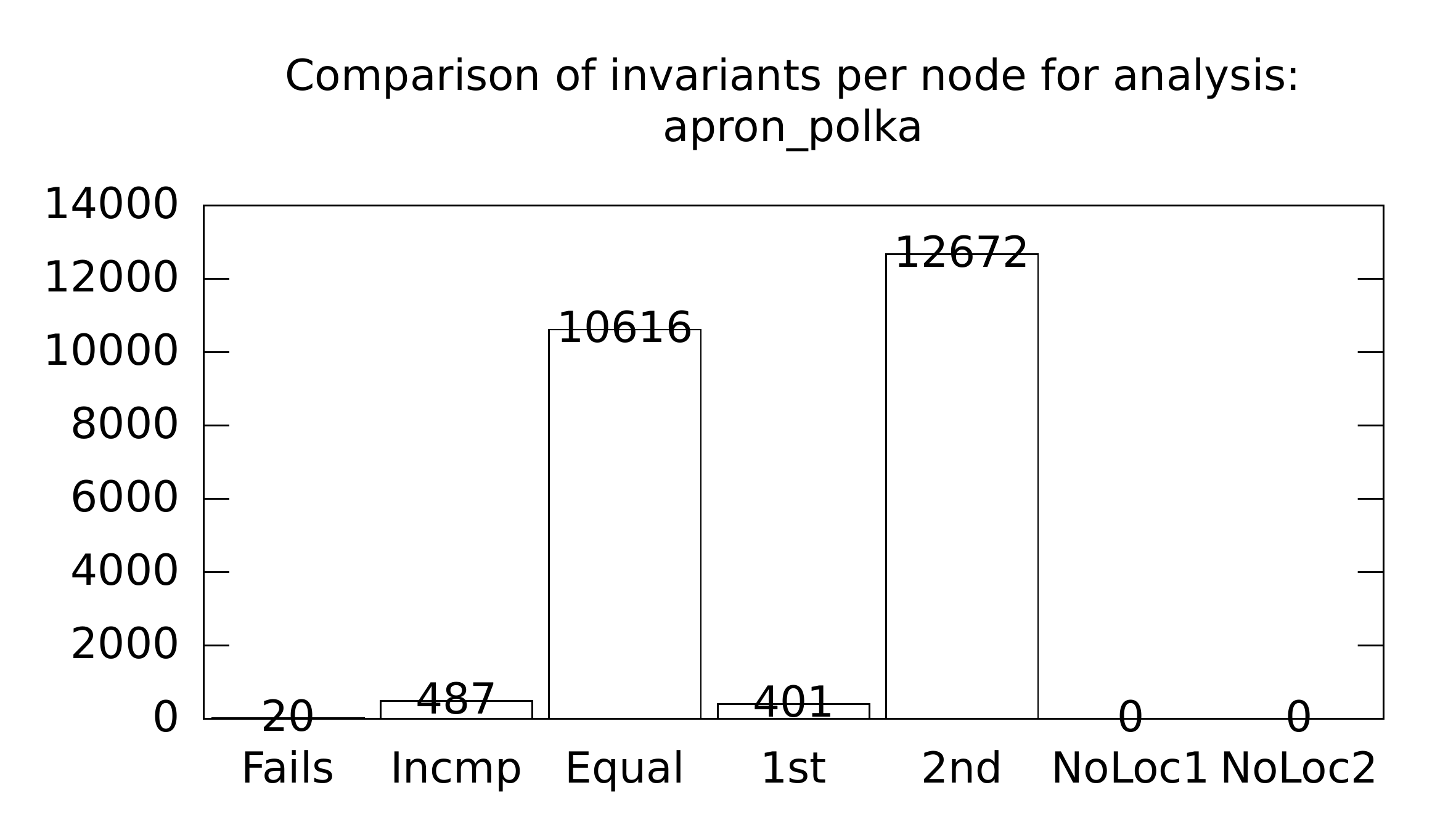}
\\
\includegraphics[scale=0.25]{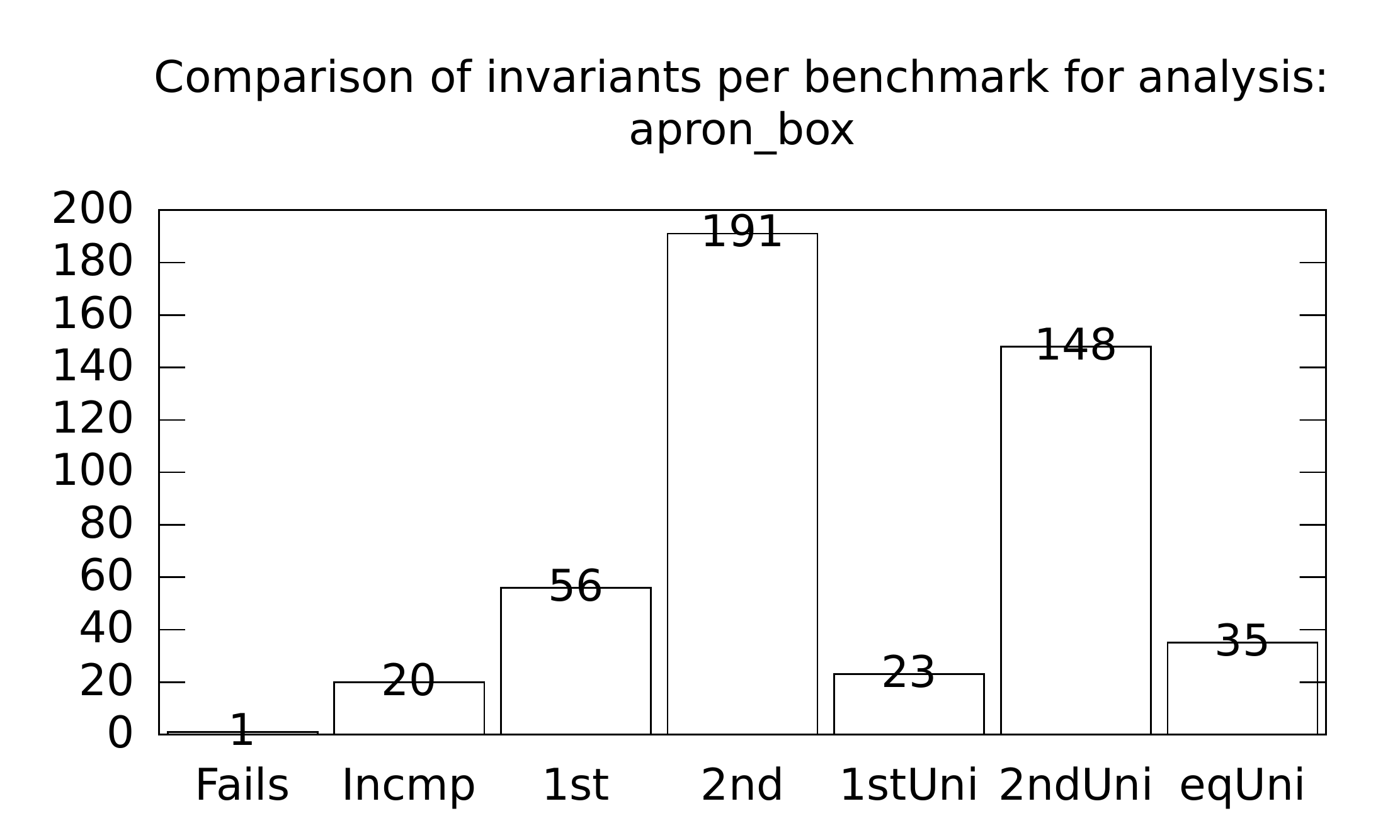}
&
\includegraphics[scale=0.25]{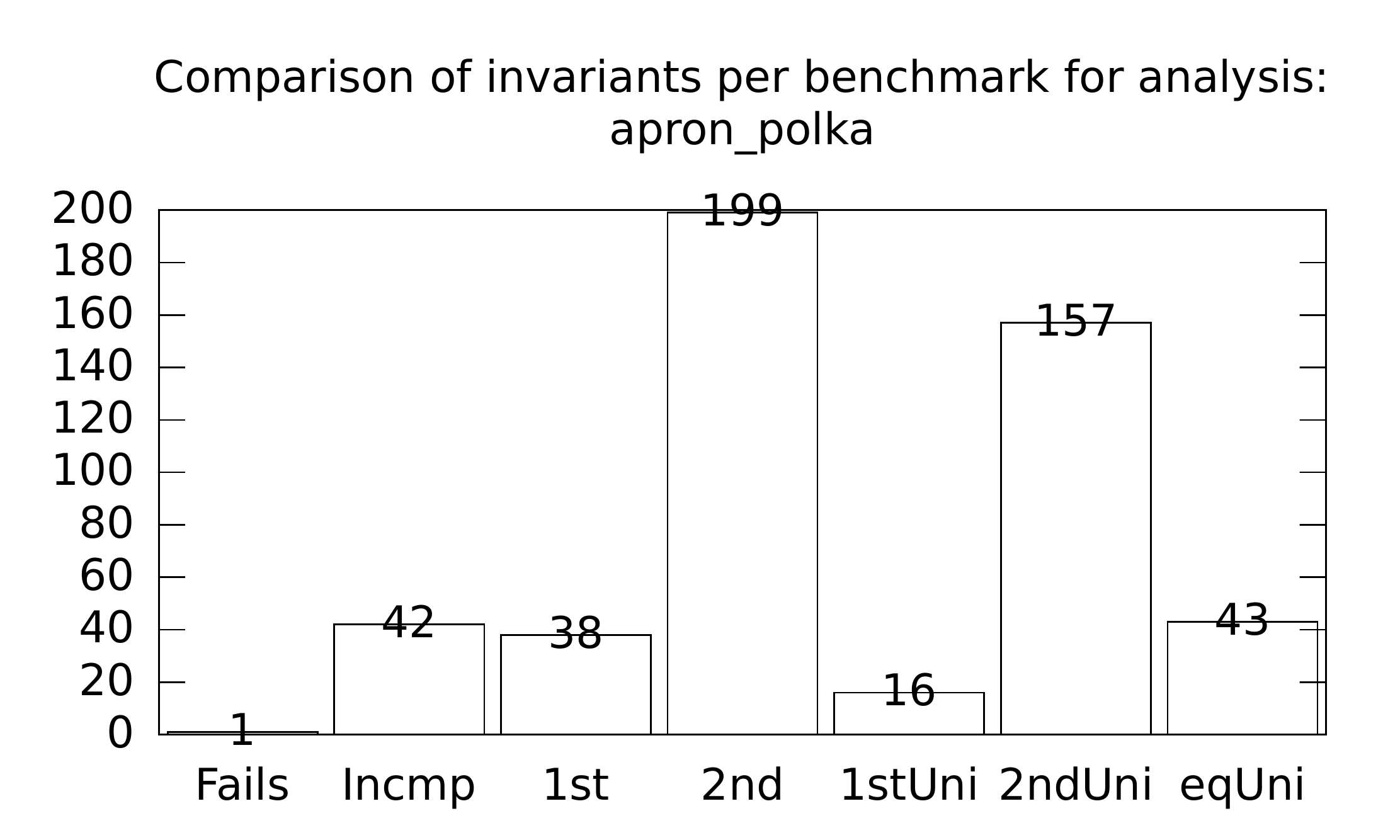}
\\
\includegraphics[scale=0.25]{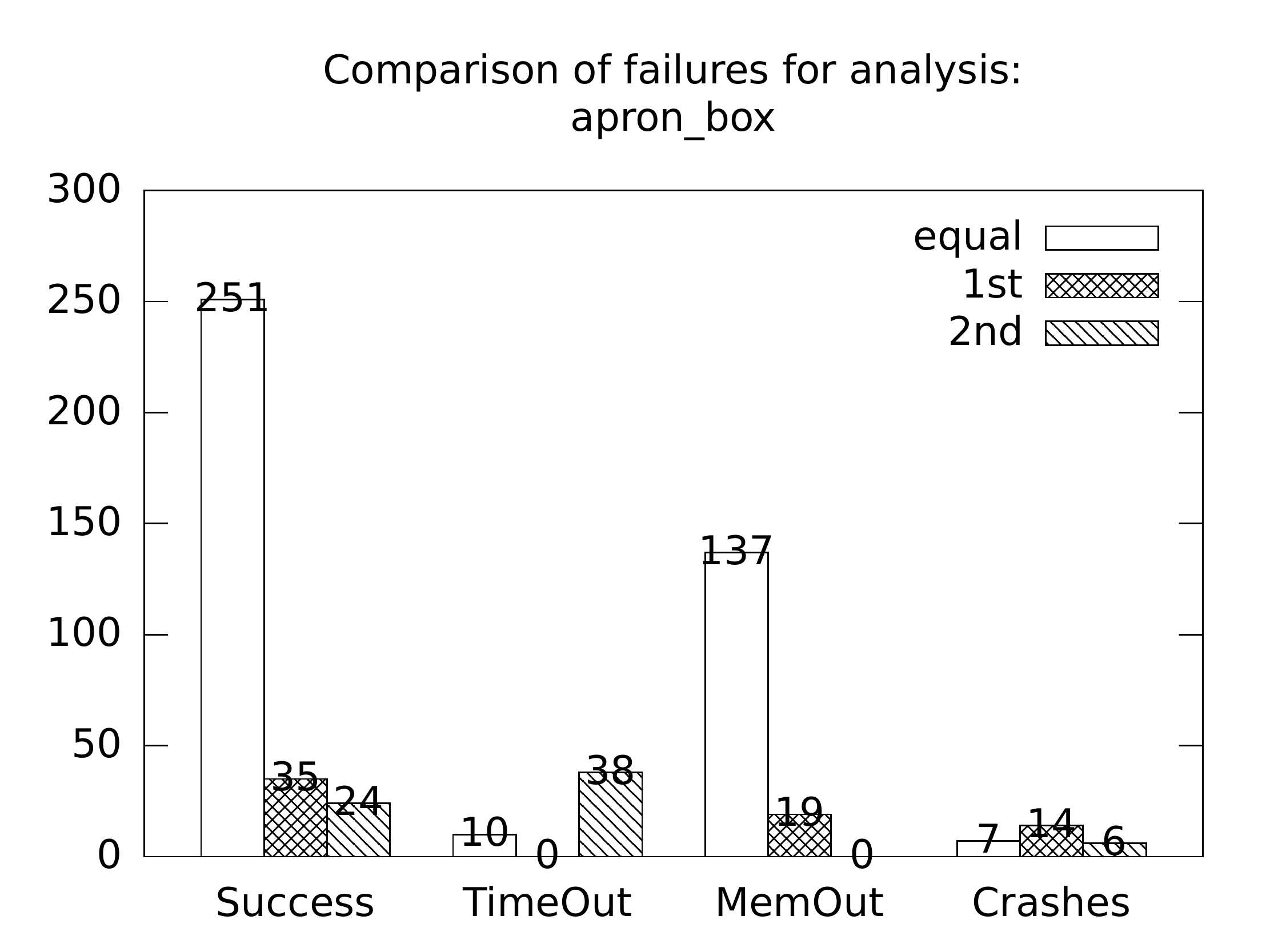}
&
\includegraphics[scale=0.25]{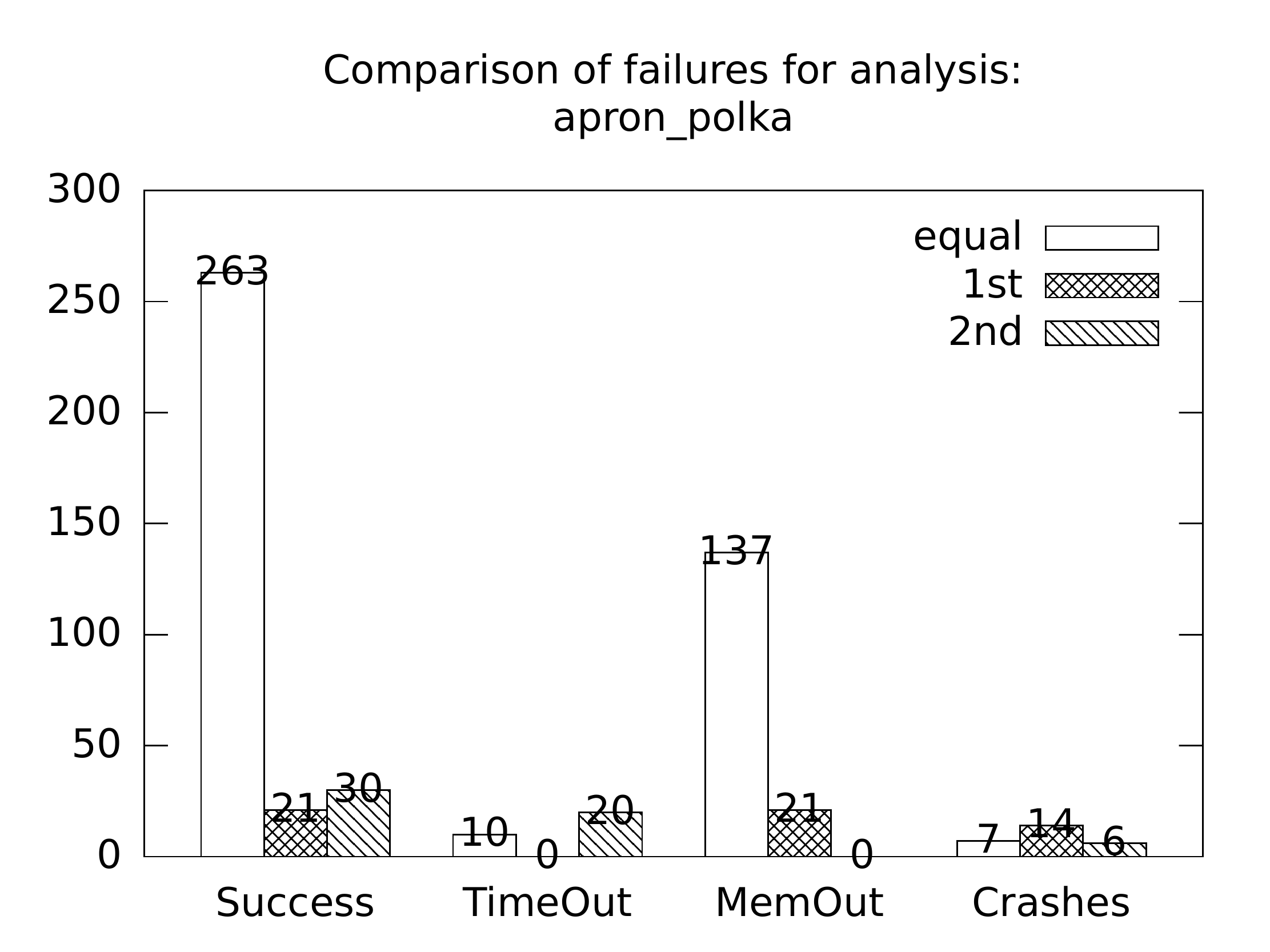}
\\
\includegraphics[scale=0.25]{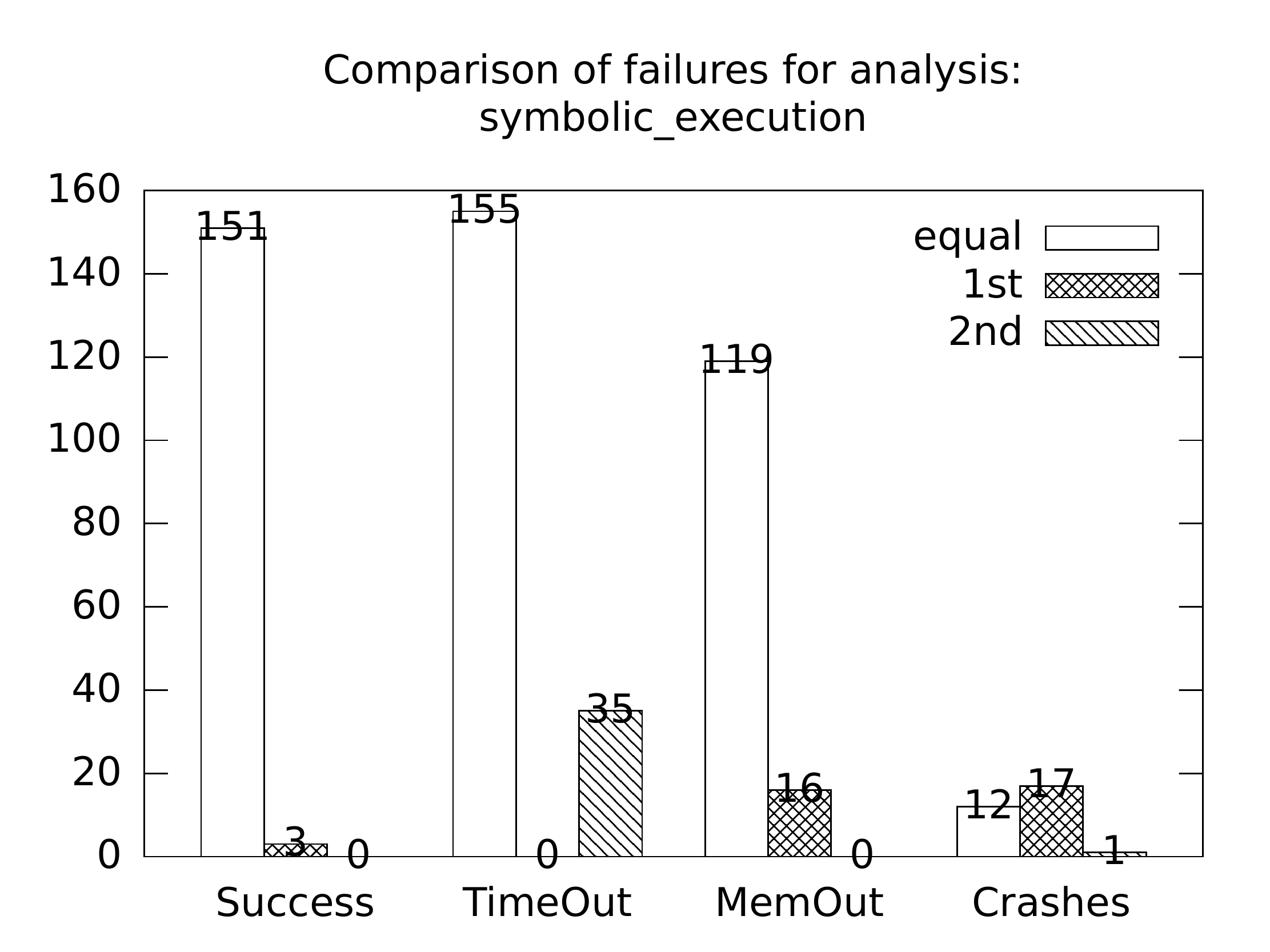}
\end{tabular}
\caption{Comparison:~~b*p*s~~vs.~~b+p+s.}
\end{figure}

\begin{figure}
\begin{tabular}{cc}
\includegraphics[scale=0.25]{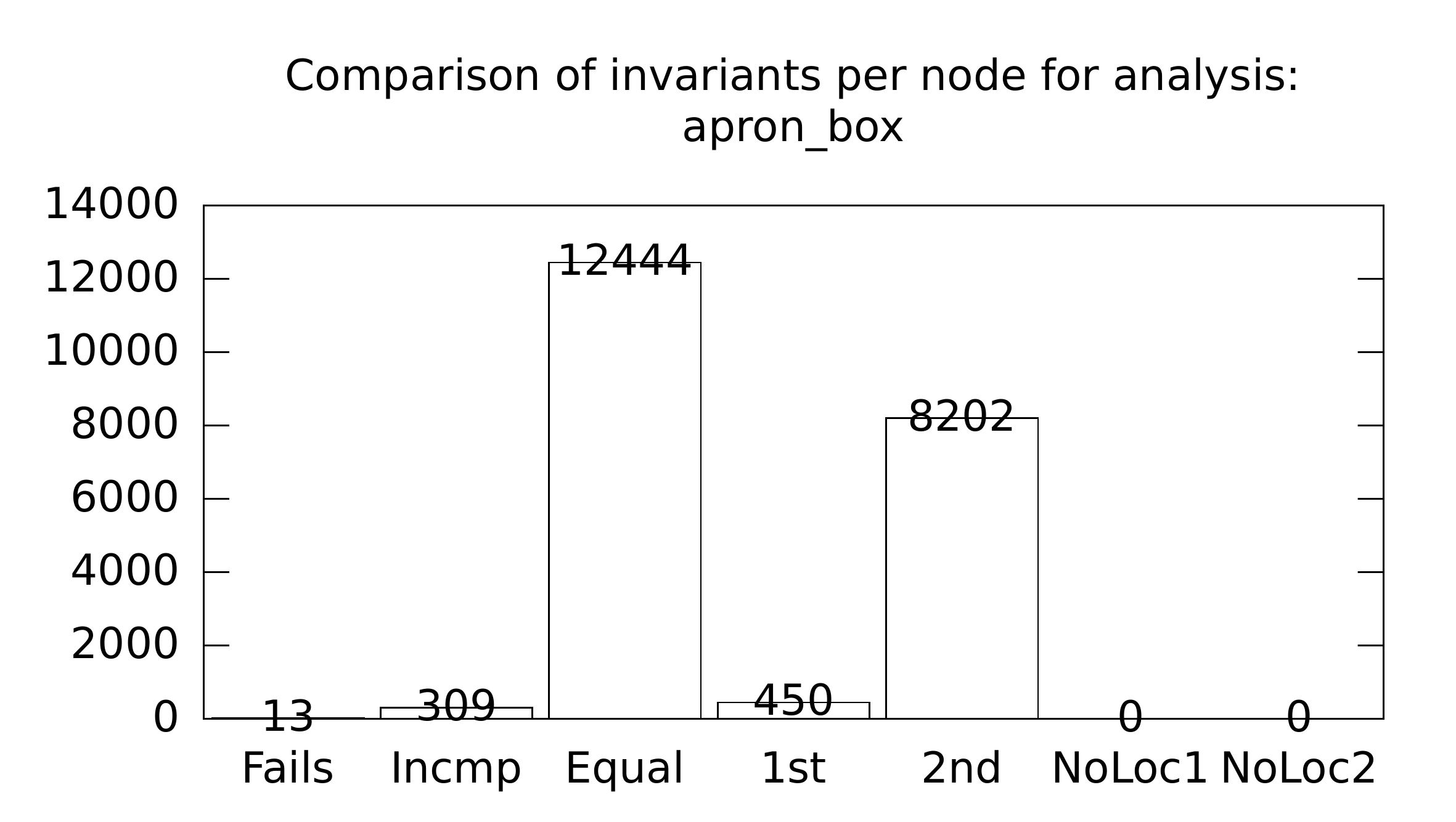}
&
\includegraphics[scale=0.25]{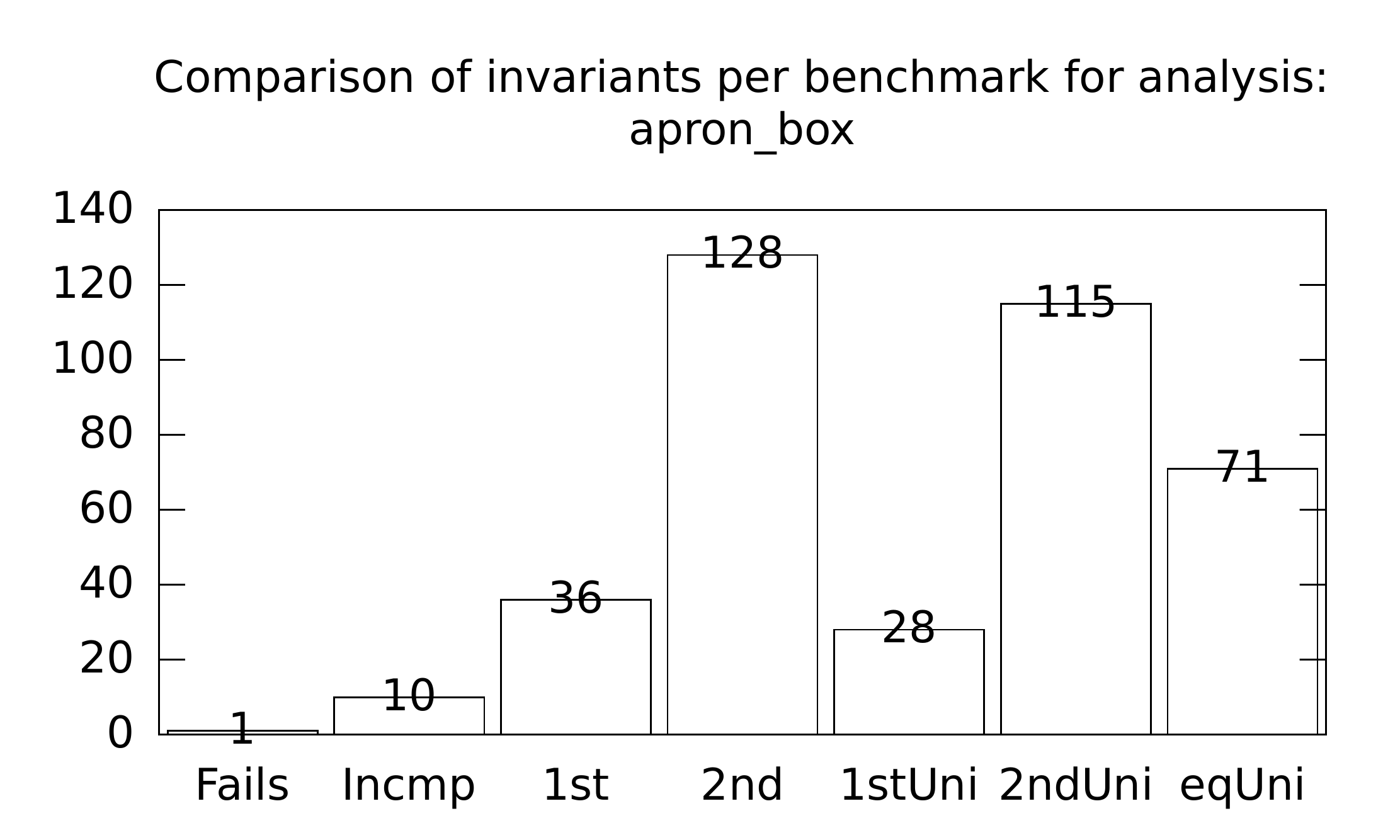}
\\
\includegraphics[scale=0.25]{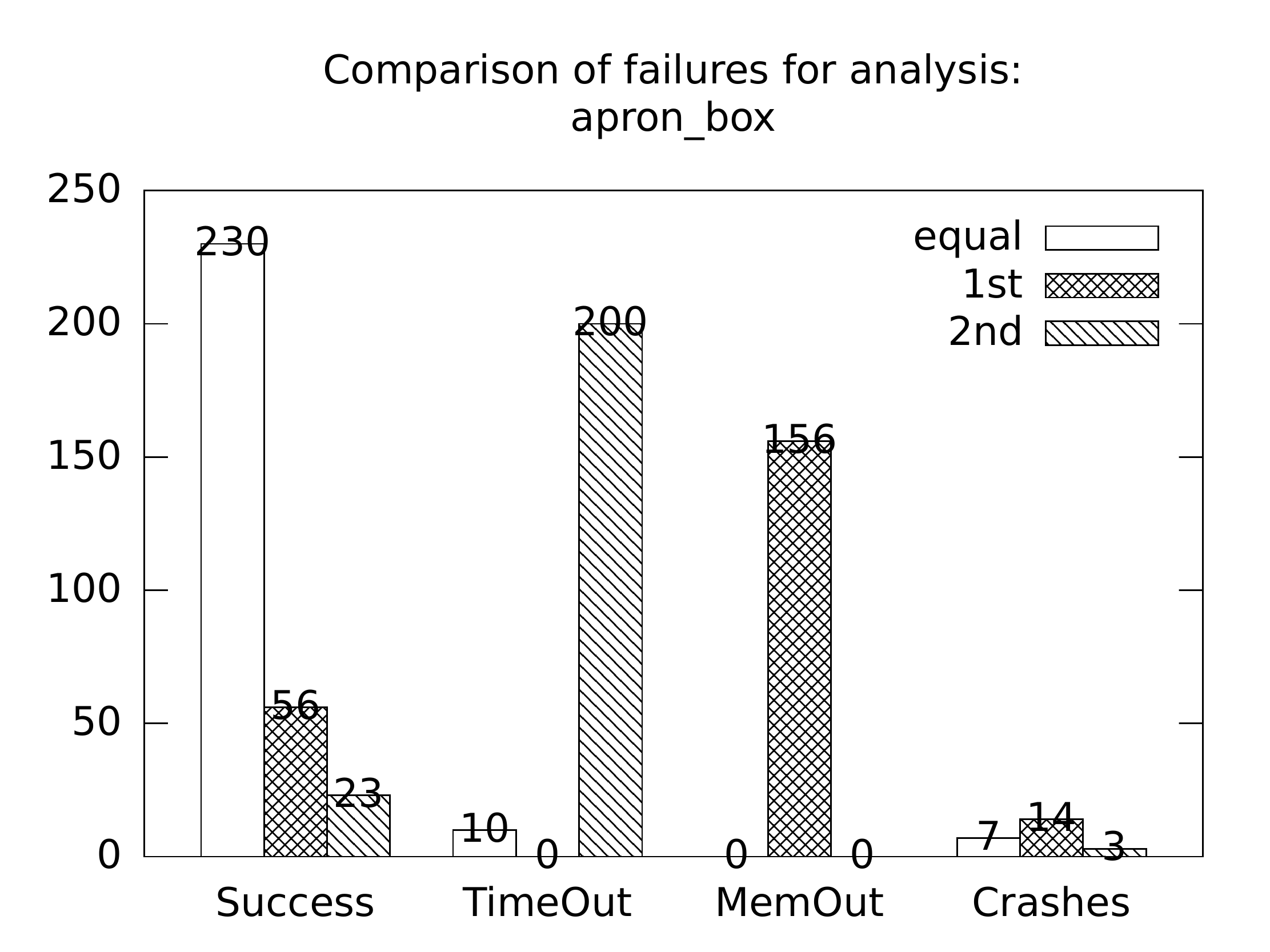}
&
\includegraphics[scale=0.25]{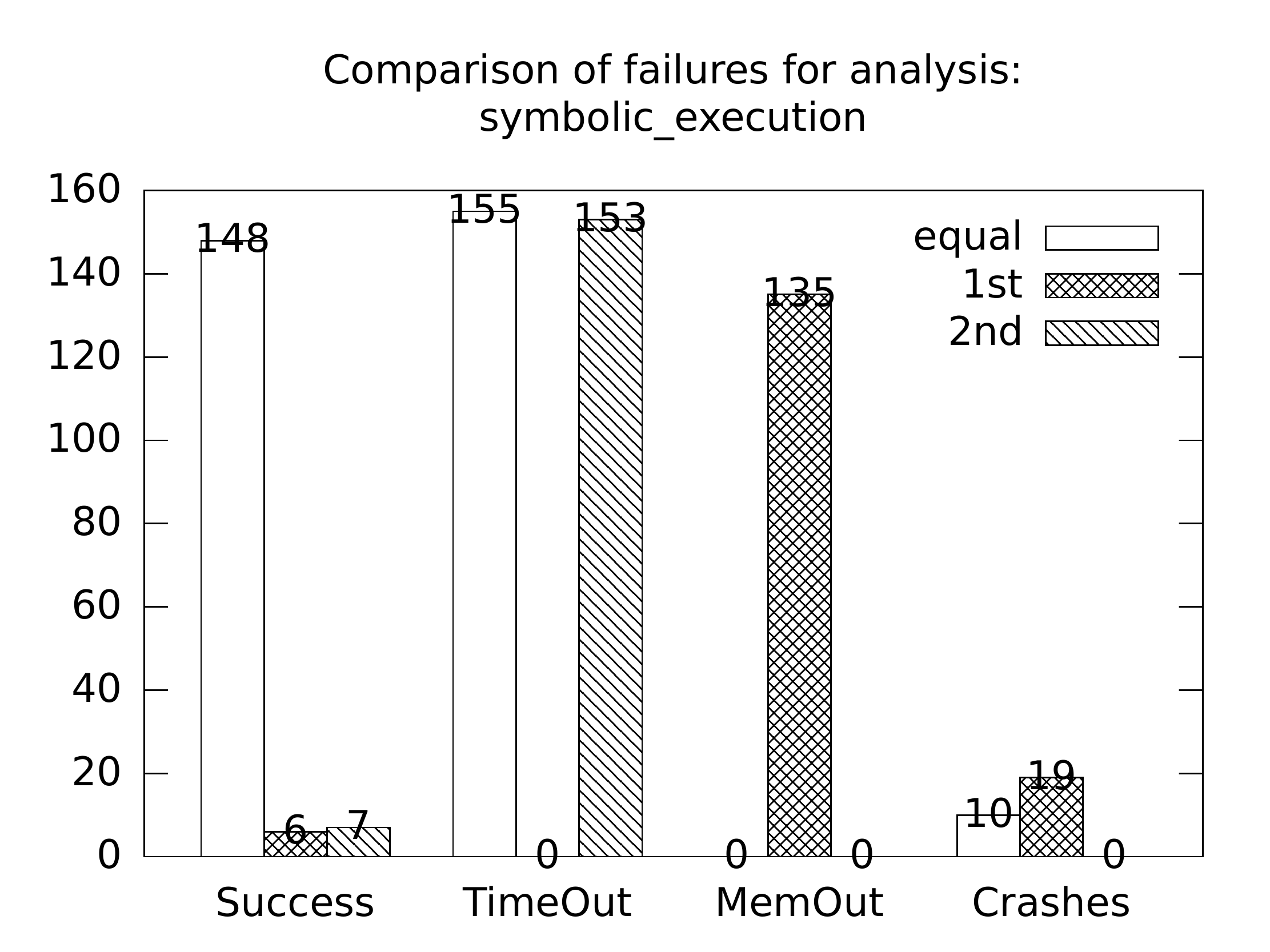}
\end{tabular}
\caption{Comparison:~~b*p*s~~vs.~~b+s.}
\end{figure}

\begin{figure}
\begin{tabular}{cc}
\includegraphics[scale=0.25]{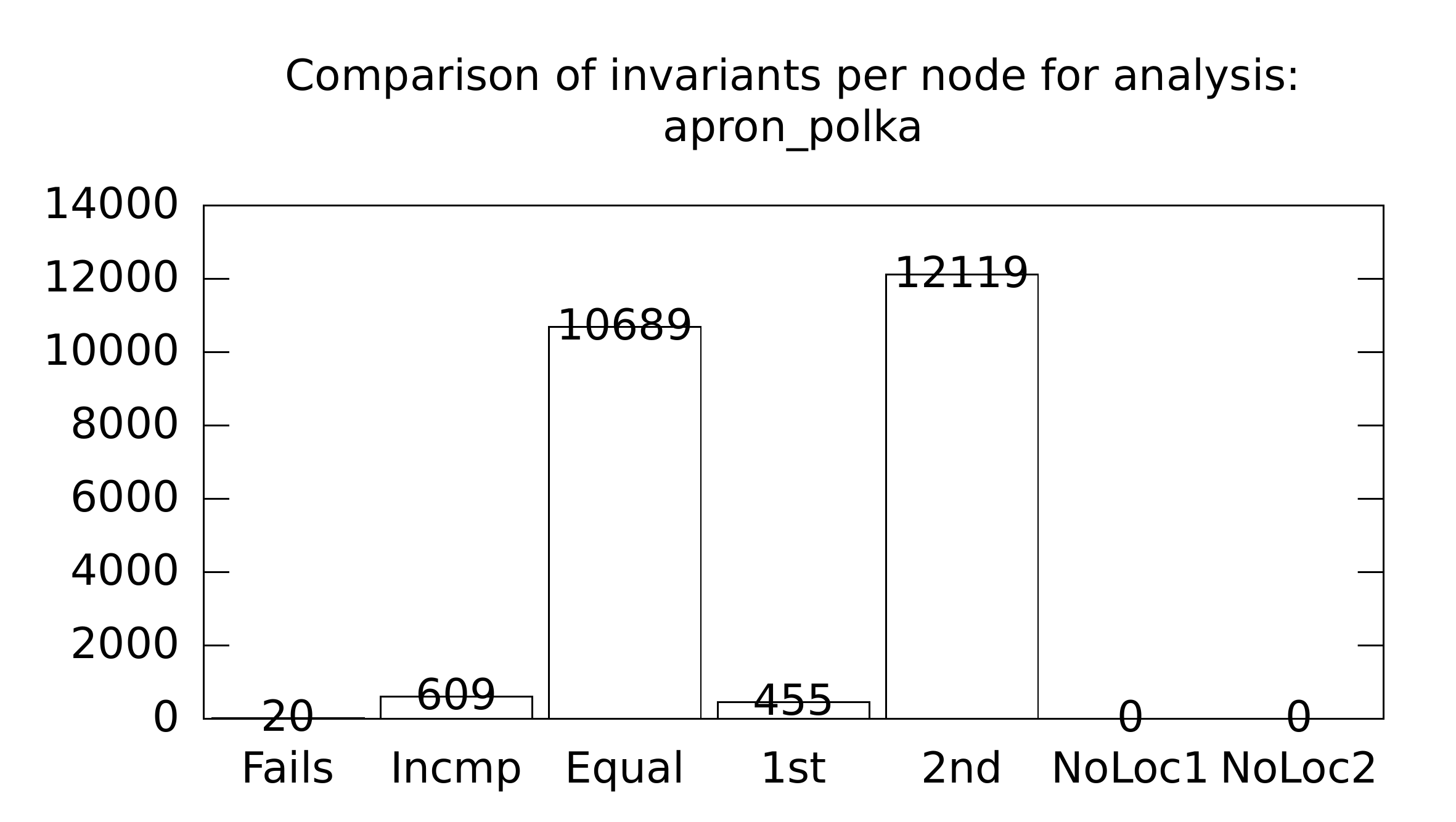}
&
\includegraphics[scale=0.25]{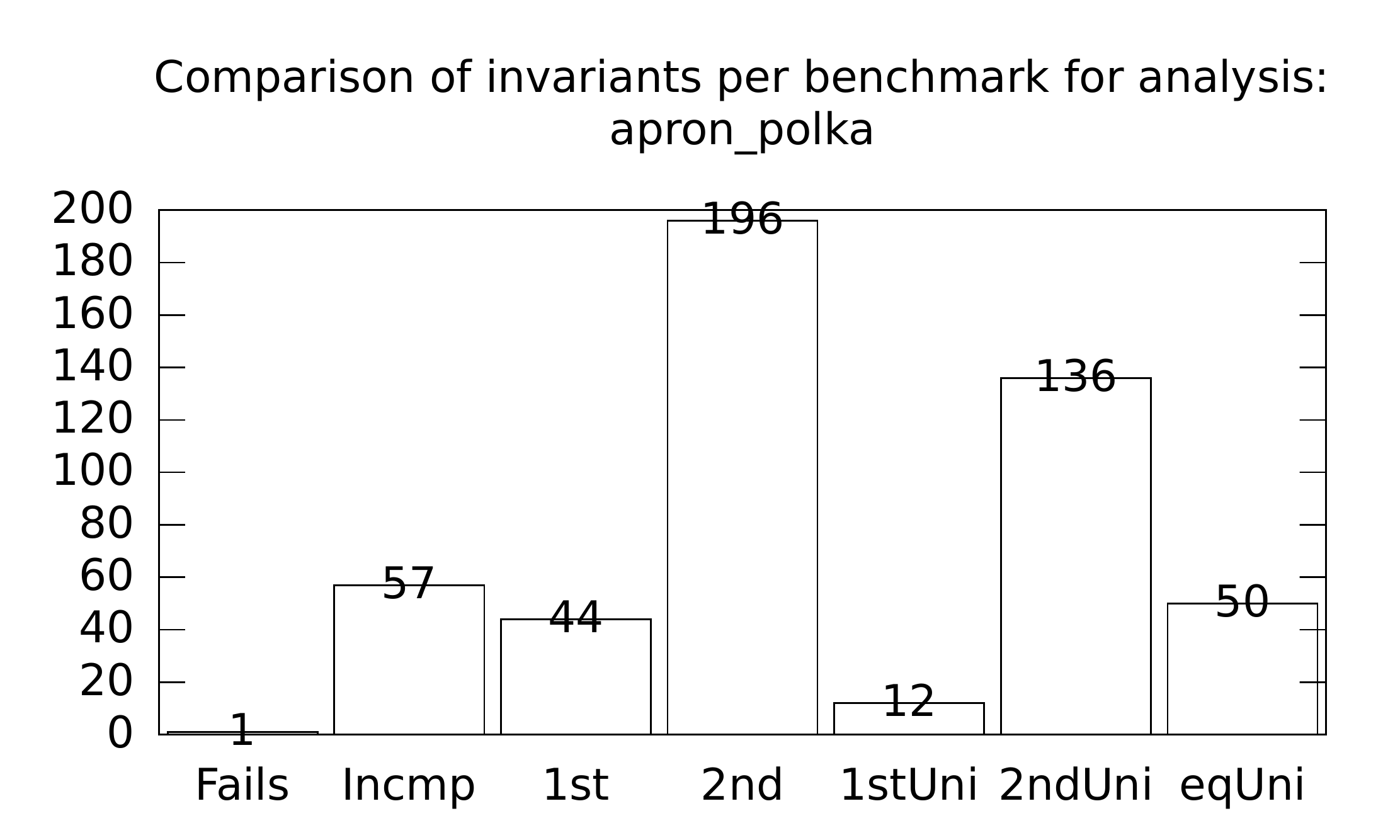}
\\
\includegraphics[scale=0.25]{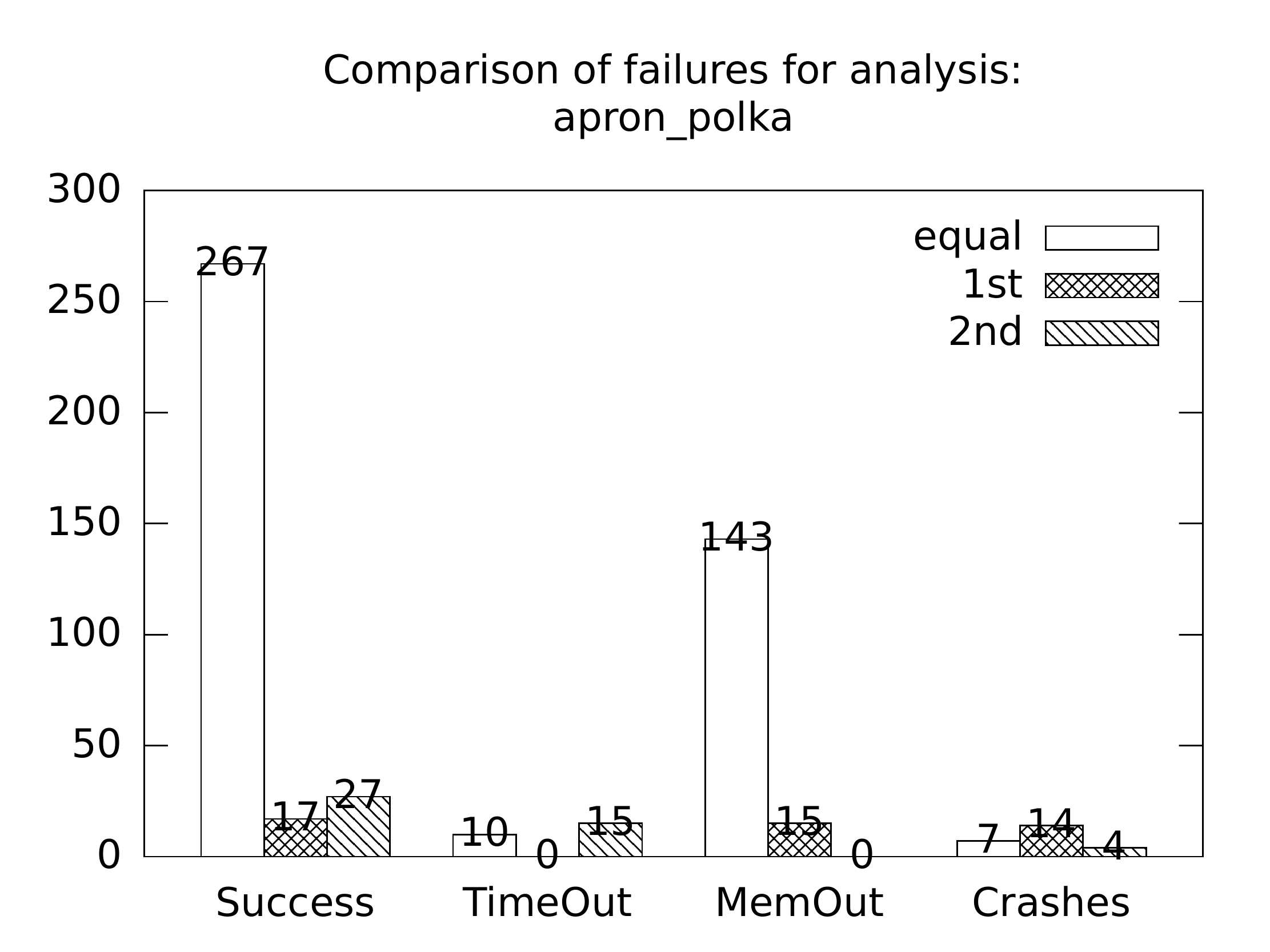}
&
\includegraphics[scale=0.25]{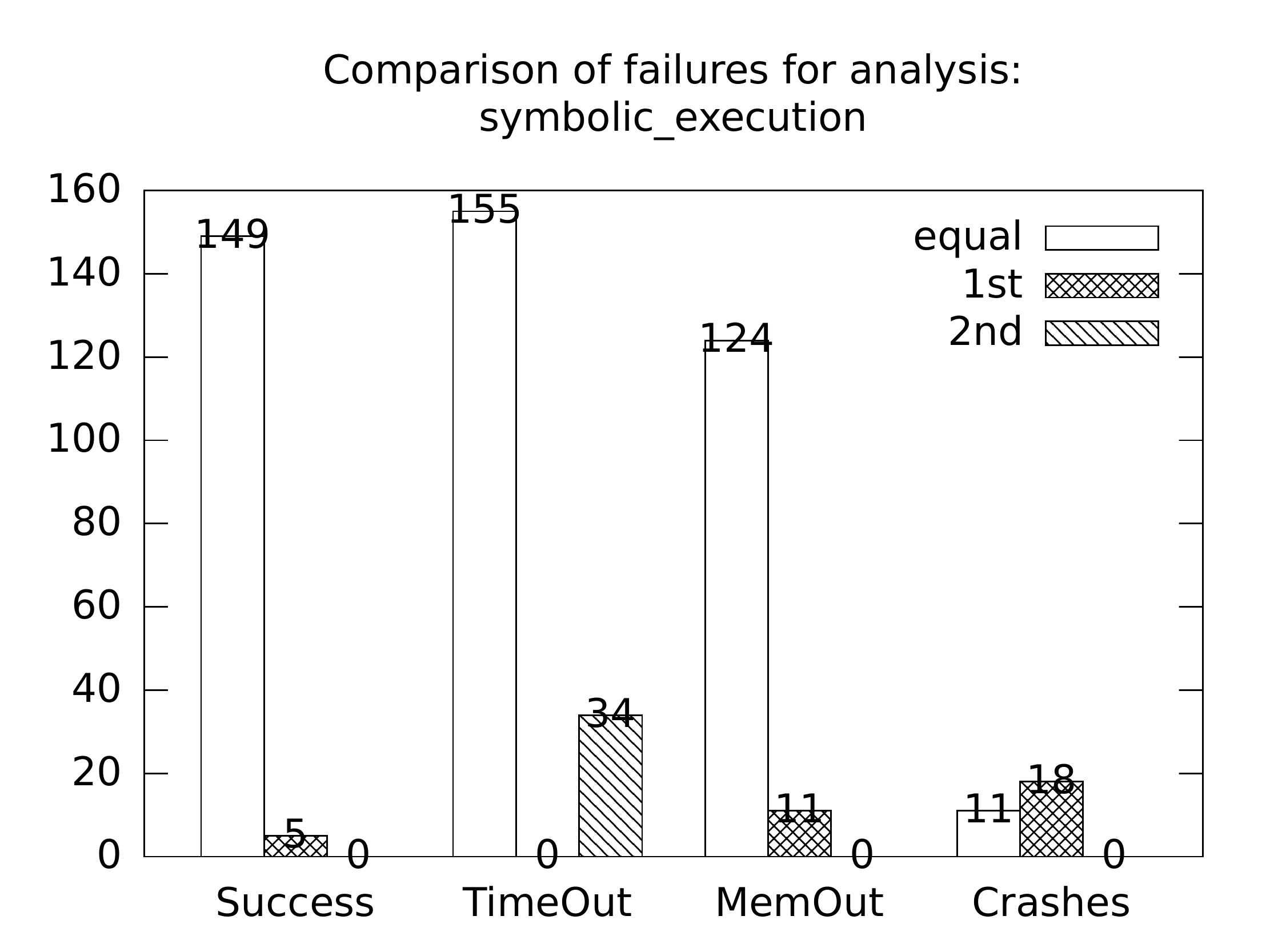}
\end{tabular}
\caption{Comparison:~~b*p*s~~vs.~~p+s.}
\end{figure}

\clearpage

\section{Access to the tool and Evaluation}
\label{sec:AccessToStatorEvaluation}

The binary of the tool and all results of the evaluation are available here
\cite{StatorURL}. The binary can be run Linux 64 bit operating system. The tool
was tested on Linux Mint 17 and Debian 4.6.3. The binary also runs on Ubuntu
14.04.

Installation is very simple. Download the ZIP package, unzip it into some
directory, and the tool is ready for an execution. The tool is located in
\texttt{STATOR-tool} sub-directory, the result from our evaluation in
\texttt{SVCOMP-repo.ORIG.EVAL} sub-directory. Note that SV-COMP benchmarks are
not included in the package. They have to be downloaded separately. All
necessary info about this can be found in Appendix~\ref{sec:BenchList}.

Use Python script \texttt{STATOR-tool/start.py} to run the tool on a single C
file you have to execute . Use the option \texttt{--help} to list detailed info
of usage. In order to run the evaluation for several benchmarks you have to
execute Python script \texttt{STATOR-tool/start\_evaluation.py}. Use the option
\texttt{--help} to list detailed info of usage. In all cases input C file(s)
have to always be preprocessed.

In case you want to recompute results of our evaluation we prepared a dedicated
shell script \texttt{STATOR\_SVCOMP\_evaluate.sh} located in the root directory
which calls \texttt{STATOR-tool/start\_evaluation.py} with our settings. Note
the shell script assumes that SV-COMP repository was cloned into sub-directory
\texttt{SVCOMP-repo} (which is empty in our package).

\section{List of Benchmarks}
\label{sec:BenchList}

Here we list all benchmarks we used in evaluations. The list references to
SV-COMP~2015~\cite{SVCOMPURL} benchmark suite, revision 571.

\noindent
Summary info:
\begin{itemize}
\item URL: https://svn.sosy-lab.org/software/sv-benchmarks/tags/svcomp15
\item revision: 571
\item Number of current benchmarks: 473
\item Number of current directories: 48
\begin{itemize}
\item     array-examples [10 benchmarks]
\item     bitvector [10 benchmarks]
\item   * bitvector-regression [9 benchmarks]
\item     ddv-machzwd [10 benchmarks]
\item     eca-rers2012 [10 benchmarks]
\item     float-benchs [10 benchmarks]
\item     floats-cbmc-regression [10 benchmarks]
\item     floats-cdfpl [10 benchmarks]
\item   * heap-manipulation [8 benchmarks]
\item     ldv-commit-tester [10 benchmarks]
\item     ldv-consumption [10 benchmarks]
\item     ldv-linux-3.0 [10 benchmarks]
\item     ldv-linux-3.12-rc1 [10 benchmarks]
\item     ldv-linux-3.16-rc1 [10 benchmarks]
\item     ldv-linux-3.4-simple [10 benchmarks]
\item     ldv-linux-3.7.3 [10 benchmarks]
\item     ldv-regression [10 benchmarks]
\item     ldv-validator-v0.6 [10 benchmarks]
\item     list-ext-properties [10 benchmarks]
\item     list-properties [10 benchmarks]
\item     locks [10 benchmarks]
\item     loop-acceleration [10 benchmarks]
\item     loop-invgen [10 benchmarks]
\item     loop-lit [10 benchmarks]
\item   * loop-new [8 benchmarks]
\item     loops [10 benchmarks]
\item     memory-alloca [10 benchmarks]
\item     memsafety [10 benchmarks]
\item   * memsafety-ext [8 benchmarks]
\item     ntdrivers [10 benchmarks]
\item     ntdrivers-simplified [10 benchmarks]
\item     product-lines [10 benchmarks]
\item     pthread [10 benchmarks]
\item     pthread-atomic [10 benchmarks]
\item     pthread-ext [10 benchmarks]
\item     pthread-lit [10 benchmarks]
\item     pthread-wmm [10 benchmarks]
\item     recursive [10 benchmarks]
\item     recursive-simple [10 benchmarks]
\item     seq-mthreaded [10 benchmarks]
\item     seq-pthread [10 benchmarks]
\item     ssh [10 benchmarks]
\item     ssh-simplified [10 benchmarks]
\item     systemc [10 benchmarks]
\item     termination-crafted [10 benchmarks]
\item     termination-crafted-lit [10 benchmarks]
\item     termination-memory-alloca [10 benchmarks]
\item     termination-numeric [10 benchmarks]
\end{itemize}
\end{itemize}

\noindent
List of benchmarks:
\begin{itemize}
\item ./array-examples/data\_structures\_set\_multi\_proc\_false-unreach-call\_ground.i
\item ./array-examples/data\_structures\_set\_multi\_proc\_true-unreach-call\_ground.i
\item ./array-examples/sanfoundry\_24\_true-unreach-call.i
\item ./array-examples/sorting\_bubblesort\_false-unreach-call\_ground.i
\item ./array-examples/standard\_compare\_true-unreach-call\_ground.i
\item ./array-examples/standard\_copy2\_true-unreach-call\_ground.i
\item ./array-examples/standard\_init3\_true-unreach-call\_ground.i
\item ./array-examples/standard\_partition\_false-unreach-call\_ground.i
\item ./array-examples/standard\_two\_index\_08\_true-unreach-call.i
\item ./array-examples/standard\_vector\_difference\_true-unreach-call\_ground.i

\item ./bitvector-regression/implicitfloatconversion\_false-unreach-call.i
\item ./bitvector-regression/implicitunsignedconversion\_false-unreach-call.i
\item ./bitvector-regression/implicitunsignedconversion\_true-unreach-call.i
\item ./bitvector-regression/integerpromotion\_false-unreach-call.i
\item ./bitvector-regression/integerpromotion\_true-unreach-call.i
\item ./bitvector-regression/signextension2\_false-unreach-call.i
\item ./bitvector-regression/signextension2\_true-unreach-call.i
\item ./bitvector-regression/signextension\_false-unreach-call.i
\item ./bitvector-regression/signextension\_true-unreach-call.i

\item ./bitvector/byte\_add\_false-unreach-call.i
\item ./bitvector/gcd\_4\_true-unreach-call.i
\item ./bitvector/interleave\_bits\_true-unreach-call.i
\item ./bitvector/jain\_7\_true-unreach-call.i
\item ./bitvector/modulus\_true-unreach-call.i
\item ./bitvector/num\_conversion\_1\_true-unreach-call.i
\item ./bitvector/parity\_true-unreach-call.i
\item ./bitvector/s3\_clnt\_1\_false-unreach-call.BV.c.cil.c
\item ./bitvector/soft\_float\_5\_true-unreach-call.c.cil.c
\item ./bitvector/sum02\_true-unreach-call.i

\item ./ddv-machzwd/ddv\_machzwd\_all\_false-unreach-call.i
\item ./ddv-machzwd/ddv\_machzwd\_inb\_p\_true-unreach-call.i
\item ./ddv-machzwd/ddv\_machzwd\_inb\_true-unreach-call.i
\item ./ddv-machzwd/ddv\_machzwd\_inl\_p\_true-unreach-call.i
\item ./ddv-machzwd/ddv\_machzwd\_inl\_true-unreach-call.i
\item ./ddv-machzwd/ddv\_machzwd\_inw\_p\_true-unreach-call.i
\item ./ddv-machzwd/ddv\_machzwd\_outb\_false-unreach-call.i
\item ./ddv-machzwd/ddv\_machzwd\_outb\_p\_true-unreach-call.i
\item ./ddv-machzwd/ddv\_machzwd\_outl\_true-unreach-call.i
\item ./ddv-machzwd/ddv\_machzwd\_outw\_p\_true-unreach-call.i

\item ./eca-rers2012/Problem01\_label01\_true-unreach-call.c
\item ./eca-rers2012/Problem01\_label03\_true-unreach-call.c
\item ./eca-rers2012/Problem01\_label10\_true-unreach-call.c
\item ./eca-rers2012/Problem05\_label48\_false-unreach-call.c
\item ./eca-rers2012/Problem10\_label57\_false-unreach-call.c
\item ./eca-rers2012/Problem10\_label58\_false-unreach-call.c
\item ./eca-rers2012/Problem11\_label15\_false-unreach-call.c
\item ./eca-rers2012/Problem11\_label18\_true-unreach-call.c
\item ./eca-rers2012/Problem19\_label52\_true-unreach-call.c
\item ./eca-rers2012/Problem19\_label59\_false-unreach-call.c

\item ./float-benchs/float\_int\_inv\_square\_false-unreach-call.c
\item ./float-benchs/inv\_square\_int\_true-unreach-call.c
\item ./float-benchs/inv\_square\_true-unreach-call.c
\item ./float-benchs/nan\_double\_false-unreach-call.c
\item ./float-benchs/nan\_double\_union\_true-unreach-call.c
\item ./float-benchs/nan\_float\_mask\_true-unreach-call.c
\item ./float-benchs/nan\_float\_range\_true-unreach-call.c
\item ./float-benchs/nan\_float\_union\_true-unreach-call.c
\item ./float-benchs/sin\_interpolated\_index\_false-unreach-call.c
\item ./float-benchs/sin\_interpolated\_smallrange\_true-unreach-call.c

\item ./floats-cbmc-regression/float-flags-simp1\_true-unreach-call.i
\item ./floats-cbmc-regression/float-no-simp1\_true-unreach-call.i
\item ./floats-cbmc-regression/float-no-simp2\_true-unreach-call.i
\item ./floats-cbmc-regression/float-zero-sum1\_true-unreach-call.i
\item ./floats-cbmc-regression/float11\_true-unreach-call.i
\item ./floats-cbmc-regression/float14\_true-unreach-call.i
\item ./floats-cbmc-regression/float22\_true-unreach-call.i
\item ./floats-cbmc-regression/float3\_true-unreach-call.i
\item ./floats-cbmc-regression/float6\_true-unreach-call.i
\item ./floats-cbmc-regression/float8\_true-unreach-call.i

\item ./floats-cdfpl/newton\_1\_4\_false-unreach-call.i
\item ./floats-cdfpl/newton\_2\_1\_true-unreach-call.i
\item ./floats-cdfpl/newton\_2\_7\_false-unreach-call.i
\item ./floats-cdfpl/newton\_3\_8\_false-unreach-call.i
\item ./floats-cdfpl/sine\_3\_false-unreach-call.i
\item ./floats-cdfpl/sine\_4\_true-unreach-call.i
\item ./floats-cdfpl/sine\_8\_true-unreach-call.i
\item ./floats-cdfpl/square\_1\_false-unreach-call.i
\item ./floats-cdfpl/square\_7\_true-unreach-call.i
\item ./floats-cdfpl/square\_8\_true-unreach-call.i

\item ./heap-manipulation/bubble\_sort\_linux\_false-unreach-call.i
\item ./heap-manipulation/bubble\_sort\_linux\_true-unreach-call.i
\item ./heap-manipulation/dll\_of\_dll\_false-unreach-call.i
\item ./heap-manipulation/dll\_of\_dll\_true-unreach-call.i
\item ./heap-manipulation/merge\_sort\_false-unreach-call.i
\item ./heap-manipulation/merge\_sort\_true-unreach-call.i
\item ./heap-manipulation/sll\_to\_dll\_rev\_false-unreach-call.i
\item ./heap-manipulation/sll\_to\_dll\_rev\_true-unreach-call.i

\item ./ldv-commit-tester/m0\_false-unreach-call\_drivers-media-radio-si4713-i2c-ko-{}-111\_1a-{}-064368f-1.c
\item ./ldv-commit-tester/m0\_true-unreach-call\_drivers-hwmon-s3c-hwmon-ko-{}-130\_7a-{}-af3071a.c
\item ./ldv-commit-tester/m0\_true-unreach-call\_drivers-media-video-cx88-cx88-blackbird-ko-{}-32\_7a-{}-d47b389.c
\item ./ldv-commit-tester/m0\_true-unreach-call\_drivers-media-video-cx88-cx88-dvb-ko-{}-32\_7a-{}-d47b389-1.c
\item ./ldv-commit-tester/m0\_true-unreach-call\_drivers-media-video-cx88-cx8802-ko-{}-32\_7a-{}-d47b389.c
\item ./ldv-commit-tester/m0\_true-unreach-call\_drivers-net-forcedeth-ko-{}-114\_1a-{}-fea891e-1.c
\item ./ldv-commit-tester/main2\_true-unreach-call\_drivers-media-video-tlg2300-poseidon-ko-{}-32\_7a-{}-4a349aa.c
\item ./ldv-commit-tester/main3\_true-unreach-call\_arch-x86-oprofile-oprofile-ko-{}-131\_1a-{}-79db8ef.c
\item ./ldv-commit-tester/main4\_true-unreach-call\_arch-x86-oprofile-oprofile-ko-{}-131\_1a-{}-79db8ef-1.c
\item ./ldv-commit-tester/main7\_true-unreach-call\_sound-oss-sound-ko-{}-32\_7a-{}-c4cb1dd-1.c

\item ./ldv-consumption/32\_7a\_cilled\_false-unreach-call\_linux-3.8-rc1-32\_7a-drivers-{}-ata-{}-libata.ko-ldv\_main4\_sequence\_infinite\_withcheck\_stateful.cil.out.c
\item ./ldv-consumption/32\_7a\_cilled\_false-unreach-call\_linux-3.8-rc1-32\_7a-fs-{}-ceph-{}-ceph.ko-ldv\_main11\_sequence\_infinite\_withcheck\_stateful.cil.out.c
\item ./ldv-consumption/32\_7a\_cilled\_true-unreach-call\_linux-3.8-rc1-32\_7a-drivers-{}-block-{}-paride-{}-pf.ko-ldv\_main0\_sequence\_infinite\_withcheck\_stateful.cil.out.c
\item ./ldv-consumption/32\_7a\_cilled\_true-unreach-call\_linux-3.8-rc1-32\_7a-drivers-{}-block-{}-paride-{}-pt.ko-ldv\_main0\_sequence\_infinite\_withcheck\_stateful.cil.out.c
\item ./ldv-consumption/32\_7a\_cilled\_true-unreach-call\_linux-3.8-rc1-32\_7a-drivers-{}-usb-{}-host-{}-xhci-hcd.ko-ldv\_main5\_sequence\_infinite\_withcheck\_stateful.cil.out.c
\item ./ldv-consumption/32\_7a\_cilled\_true-unreach-call\_linux-3.8-rc1-32\_7a-fs-{}-nfs-{}-nfsv4.ko-ldv\_main4\_sequence\_infinite\_withcheck\_stateful.cil.out.c
\item ./ldv-consumption/32\_7a\_cilled\_true-unreach-call\_linux-3.8-rc1-drivers-{}-block-{}-paride-{}-pt.ko-main.cil.out.c
\item ./ldv-consumption/32\_7a\_cilled\_true-unreach-call\_linux-3.8-rc1-drivers-{}-vfio-{}-pci-{}-vfio-pci.ko-main.cil.out.c
\item ./ldv-consumption/linux-3.8-rc1-32\_7a-drivers-{}-media-{}-usb-{}-em28xx-{}-em28xx.ko-ldv\_main0\_true-unreach-call.cil.out.c
\item ./ldv-consumption/linux-3.8-rc1-32\_7a-drivers-{}-usb-{}-core-{}-usbcore.ko-ldv\_main13\_false-unreach-call.cil.out.c

\item ./ldv-linux-3.0/module\_get\_put-drivers-atm-eni.ko\_true-unreach-call.cil.out.i.pp.i
\item ./ldv-linux-3.0/module\_get\_put-drivers-block-drbd-drbd.ko\_true-unreach-call.cil.out.i.pp.i
\item ./ldv-linux-3.0/module\_get\_put-drivers-net-ppp\_generic.ko\_false-unreach-call.cil.out.i.pp.i
\item ./ldv-linux-3.0/module\_get\_put-drivers-net-wan-farsync.ko\_false-unreach-call.cil.out.i.pp.i
\item ./ldv-linux-3.0/usb\_urb-drivers-input-misc-keyspan\_remote.ko\_false-unreach-call.cil.out.i.pp.i
\item ./ldv-linux-3.0/usb\_urb-drivers-input-tablet-kbtab.ko\_true-unreach-call.cil.out.i.pp.i
\item ./ldv-linux-3.0/usb\_urb-drivers-media-dvb-ttusb-dec-ttusb\_dec.ko\_false-unreach-call.cil.out.i.pp.i
\item ./ldv-linux-3.0/usb\_urb-drivers-media-video-c-qcam.ko\_true-unreach-call.cil.out.i.pp.i
\item ./ldv-linux-3.0/usb\_urb-drivers-net-usb-catc.ko\_false-unreach-call.cil.out.i.pp.i
\item ./ldv-linux-3.0/usb\_urb-drivers-vhost-vhost\_net.ko\_true-unreach-call.cil.out.i.pp.i

\item ./ldv-linux-3.12-rc1/linux-3.12-rc1.tar.xz-144\_2a-drivers-{}-input-{}-misc-{}-ims-pcu.ko-entry\_point\_false-unreach-call.cil.out.c
\item ./ldv-linux-3.12-rc1/linux-3.12-rc1.tar.xz-144\_2a-drivers-{}-isdn-{}-hisax-{}-hisax\_st5481.ko-entry\_point\_false-unreach-call.cil.out.c
\item ./ldv-linux-3.12-rc1/linux-3.12-rc1.tar.xz-144\_2a-drivers-{}-media-{}-rc-{}-imon.ko-entry\_point\_false-unreach-call.cil.out.c
\item ./ldv-linux-3.12-rc1/linux-3.12-rc1.tar.xz-144\_2a-drivers-{}-media-{}-usb-{}-stk1160-{}-stk1160.ko-entry\_point\_false-unreach-call.cil.out.c
\item ./ldv-linux-3.12-rc1/linux-3.12-rc1.tar.xz-144\_2a-drivers-{}-media-{}-usb-{}-usbvision-{}-usbvision.ko-entry\_point\_false-unreach-call.cil.out.c
\item ./ldv-linux-3.12-rc1/linux-3.12-rc1.tar.xz-144\_2a-drivers-{}-net-{}-can-{}-usb-{}-esd\_usb2.ko-entry\_point\_true-unreach-call.cil.out.c
\item ./ldv-linux-3.12-rc1/linux-3.12-rc1.tar.xz-144\_2a-drivers-{}-net-{}-usb-{}-cdc\_mbim.ko-entry\_point\_true-unreach-call.cil.out.c
\item ./ldv-linux-3.12-rc1/linux-3.12-rc1.tar.xz-144\_2a-drivers-{}-net-{}-usb-{}-smsc95xx.ko-entry\_point\_false-unreach-call.cil.out.c
\item ./ldv-linux-3.12-rc1/linux-3.12-rc1.tar.xz-144\_2a-drivers-{}-staging-{}-gdm72xx-{}-gdmwm.ko-entry\_point\_false-unreach-call.cil.out.c
\item ./ldv-linux-3.12-rc1/linux-3.12-rc1.tar.xz-144\_2a-drivers-{}-usb-{}-misc-{}-idmouse.ko-entry\_point\_false-unreach-call.cil.out.c

\item ./ldv-linux-3.16-rc1/205\_9a\_array\_unsafes\_linux-3.16-rc1.tar.xz-205\_9a-drivers-{}-net-{}-wan-{}-lapbether.ko-entry\_point\_false-unreach-call.cil.out.c
\item ./ldv-linux-3.16-rc1/205\_9a\_array\_unsafes\_linux-3.16-rc1.tar.xz-205\_9a-drivers-{}-net-{}-wireless-{}-ath-{}-ath6kl-{}-ath6kl\_usb.ko-entry\_point\_false-unreach-call.cil.out.c
\item ./ldv-linux-3.16-rc1/205\_9a\_array\_unsafes\_linux-3.16-rc1.tar.xz-205\_9a-drivers-{}-net-{}-wireless-{}-ath-{}-wcn36xx-{}-wcn36xx.ko-entry\_point\_false-unreach-call.cil.out.c
\item ./ldv-linux-3.16-rc1/205\_9a\_array\_unsafes\_linux-3.16-rc1.tar.xz-205\_9a-drivers-{}-net-{}-wireless-{}-hostap-{}-hostap\_plx.ko-entry\_point\_false-unreach-call.cil.out.c
\item ./ldv-linux-3.16-rc1/43\_2a\_consumption\_linux-3.16-rc1.tar.xz-43\_2a-drivers-{}-input-{}-gameport-{}-ns558.ko-entry\_point\_true-unreach-call.cil.out.c
\item ./ldv-linux-3.16-rc1/43\_2a\_consumption\_linux-3.16-rc1.tar.xz-43\_2a-drivers-{}-target-{}-sbp-{}-sbp\_target.ko-entry\_point\_true-unreach-call.cil.out.c
\item ./ldv-linux-3.16-rc1/43\_2a\_consumption\_linux-3.16-rc1.tar.xz-43\_2a-drivers-{}-tty-{}-isicom.ko-entry\_point\_true-unreach-call.cil.out.c
\item ./ldv-linux-3.16-rc1/43\_2a\_consumption\_linux-3.16-rc1.tar.xz-43\_2a-drivers-{}-tty-{}-serial-{}-8250-{}-8250.ko-entry\_point\_true-unreach-call.cil.out.c
\item ./ldv-linux-3.16-rc1/43\_2a\_consumption\_linux-3.16-rc1.tar.xz-43\_2a-drivers-{}-tty-{}-serial-{}-8250-{}-8250\_pci.ko-entry\_point\_true-unreach-call.cil.out.c
\item ./ldv-linux-3.16-rc1/43\_2a\_consumption\_linux-3.16-rc1.tar.xz-43\_2a-drivers-{}-video-{}-fbdev-{}-via-{}-viafb.ko-entry\_point\_true-unreach-call.cil.out.c

\item ./ldv-linux-3.4-simple/32\_1\_cilled\_true-unreach-call\_ok\_nondet\_linux-3.4-32\_1-drivers-{}-acpi-{}-bgrt.ko-ldv\_main0\_sequence\_infinite\_withcheck\_stateful.cil.out.c
\item ./ldv-linux-3.4-simple/32\_1\_cilled\_true-unreach-call\_ok\_nondet\_linux-3.4-32\_1-drivers-{}-media-{}-dvb-{}-dvb-usb-{}-dvb-usb-dibusb-mc.ko-ldv\_main0\_sequence\_infinite\_withcheck\_stateful.cil.out.c
\item ./ldv-linux-3.4-simple/32\_1\_cilled\_true-unreach-call\_ok\_nondet\_linux-3.4-32\_1-drivers-{}-media-{}-dvb-{}-dvb-usb-{}-dvb-usb-digitv.ko-ldv\_main0\_sequence\_infinite\_withcheck\_stateful.cil.out.c
\item ./ldv-linux-3.4-simple/43\_1a\_cilled\_true-unreach-call\_ok\_nondet\_linux-43\_1a-drivers-{}-media-{}-dvb-{}-dvb-usb-{}-dvb-usb-vp702x.ko-ldv\_main1\_sequence\_infinite\_withcheck\_stateful.cil.out.c
\item ./ldv-linux-3.4-simple/43\_1a\_cilled\_true-unreach-call\_ok\_nondet\_linux-43\_1a-drivers-{}-media-{}-dvb-{}-frontends-{}-dib3000mc.ko-ldv\_main0\_sequence\_infinite\_withcheck\_stateful.cil.out.c
\item ./ldv-linux-3.4-simple/43\_1a\_cilled\_true-unreach-call\_ok\_nondet\_linux-43\_1a-drivers-{}-media-{}-video-{}-gspca-{}-gspca\_jl2005bcd.ko-ldv\_main0\_sequence\_infinite\_withcheck\_stateful.cil.out.c
\item ./ldv-linux-3.4-simple/43\_1a\_cilled\_true-unreach-call\_ok\_nondet\_linux-43\_1a-drivers-{}-media-{}-video-{}-gspca-{}-gspca\_pac207.ko-ldv\_main0\_sequence\_infinite\_withcheck\_stateful.cil.out.c
\item ./ldv-linux-3.4-simple/43\_1a\_cilled\_true-unreach-call\_ok\_nondet\_linux-43\_1a-drivers-{}-media-{}-video-{}-gspca-{}-gspca\_stv0680.ko-ldv\_main0\_sequence\_infinite\_withcheck\_stateful.cil.out.c
\item ./ldv-linux-3.4-simple/43\_1a\_cilled\_true-unreach-call\_ok\_nondet\_linux-43\_1a-drivers-{}-watchdog-{}-wdt\_pci.ko-ldv\_main0\_sequence\_infinite\_withcheck\_stateful.cil.out.c
\item ./ldv-linux-3.4-simple/43\_1a\_cilled\_true-unreach-call\_ok\_nondet\_linux-43\_1a-drivers-{}-xen-{}-xenfs-{}-xenfs.ko-ldv\_main0\_sequence\_infinite\_withcheck\_stateful.cil.out.c

\item ./ldv-linux-3.7.3/linux-3.10-rc1-43\_1a-bitvector-drivers-{}-atm-{}-he.ko-ldv\_main0\_true-unreach-call.cil.out.c
\item ./ldv-linux-3.7.3/main0\_false-unreach-call\_drivers-{}-media-{}-dvb-frontends-{}-stv090x-ko-{}--32\_7a-{}-linux-3.7.3.c
\item ./ldv-linux-3.7.3/main0\_false-unreach-call\_drivers-net-wireless-mwl8k-ko-{}--32\_7a-{}-linux-3.7.3.c
\item ./ldv-linux-3.7.3/main11\_false-unreach-call\_drivers-usb-core-usbcore-ko-{}-32\_7a-{}-linux-3.7.3.c
\item ./ldv-linux-3.7.3/main15\_false-unreach-call\_drivers-usb-core-usbcore-ko-{}-32\_7a-{}-linux-3.7.3.c
\item ./ldv-linux-3.7.3/main17\_false-unreach-call\_drivers-gpu-drm-vmwgfx-vmwgfx-ko-{}-32\_7a-{}-linux-3.5.c
\item ./ldv-linux-3.7.3/main1\_false-unreach-call\_drivers-usb-core-usbcore-ko-{}-32\_7a-{}-linux-3.7.3.c
\item ./ldv-linux-3.7.3/main1\_false-unreach-call\_drivers-vhost-vhost\_net-ko-{}-32\_7a-{}-linux-3.7.3.c
\item ./ldv-linux-3.7.3/main3\_false-unreach-call\_drivers-gpu-drm-vmwgfx-vmwgfx-ko-{}-32\_7a-{}-linux-3.5.c
\item ./ldv-linux-3.7.3/main4\_false-unreach-call\_drivers-scsi-mpt2sas-mpt2sas-ko-{}-32\_7a-{}-linux-3.7.3.c

\item ./ldv-regression/alias\_of\_return\_2.c\_true-unreach-call\_1.i
\item ./ldv-regression/nested\_structure\_ptr\_true-unreach-call.i
\item ./ldv-regression/nested\_structure\_true-unreach-call.i
\item ./ldv-regression/sizeofparameters\_test.c\_true-unreach-call.i
\item ./ldv-regression/stateful\_check\_false-unreach-call.i
\item ./ldv-regression/test\_address.c\_true-unreach-call.i
\item ./ldv-regression/test\_cut\_trace.c\_true-unreach-call.i
\item ./ldv-regression/test\_union\_cast.c\_true-unreach-call.i
\item ./ldv-regression/test\_union\_cast.c\_true-unreach-call\_1.i
\item ./ldv-regression/volatile\_alias.c\_true-unreach-call.i

\item ./ldv-validator-v0.6/linux-stable-1575714-1-150\_1a-drivers-{}-net-{}-wireless-{}-b43-{}-b43.ko-entry\_point\_false-unreach-call.cil.out.c
\item ./ldv-validator-v0.6/linux-stable-1b0b0ac-1-108\_1a-drivers-{}-net-{}-slip.ko-entry\_point\_false-unreach-call.cil.out.c
\item ./ldv-validator-v0.6/linux-stable-42f9f8d-1-111\_1a-sound-{}-oss-{}-opl3.ko-entry\_point\_false-unreach-call.cil.out.c
\item ./ldv-validator-v0.6/linux-stable-431e8d4-1-102\_1a-drivers-{}-net-{}-r8169.ko-entry\_point\_false-unreach-call.cil.out.c
\item ./ldv-validator-v0.6/linux-stable-4a349aa-1-32\_7a-drivers-{}-media-{}-video-{}-tlg2300-{}-poseidon.ko-entry\_point\_false-unreach-call.cil.out.c
\item ./ldv-validator-v0.6/linux-stable-4ed3cba-1-100\_1a-drivers-{}-usb-{}-serial-{}-qcserial.ko-entry\_point\_false-unreach-call.cil.out.c
\item ./ldv-validator-v0.6/linux-stable-5934df9-1-111\_1a-drivers-{}-scsi-{}-gdth.ko-entry\_point\_false-unreach-call.cil.out.c
\item ./ldv-validator-v0.6/linux-stable-90a4845-1-110\_1a-drivers-{}-char-{}-ipmi-{}-ipmi\_si.ko-entry\_point\_false-unreach-call.cil.out.c
\item ./ldv-validator-v0.6/linux-stable-c0cc359-104\_1a-drivers-{}-usb-{}-serial-{}-qcserial.ko-entry\_point\_false-unreach-call.cil.out.c
\item ./ldv-validator-v0.6/linux-torvalds-645ef9e-32\_7a-sound-{}-oss-{}-sound.ko-entry\_point\_false-unreach-call.cil.out.c

\item ./list-ext-properties/960521-1\_1\_false-valid-deref.i
\item ./list-ext-properties/960521-1\_1\_true-valid-memsafety.i
\item ./list-ext-properties/list-ext\_1\_true-valid-memsafety.i
\item ./list-ext-properties/simple-ext\_1\_true-valid-memsafety.i
\item ./list-ext-properties/test-0019\_1\_true-valid-memsafety.i
\item ./list-ext-properties/test-0158\_1\_false-valid-free.i
\item ./list-ext-properties/test-0214\_1\_true-valid-memsafety.i
\item ./list-ext-properties/test-0232\_1\_false-valid-memtrack.i
\item ./list-ext-properties/test-0232\_1\_true-valid-memsafety.i
\item ./list-ext-properties/test-0504\_1\_true-valid-memsafety.i

\item ./list-properties/alternating\_list\_true-unreach-call.i
\item ./list-properties/list\_false-unreach-call.i
\item ./list-properties/list\_flag\_false-unreach-call.i
\item ./list-properties/list\_flag\_true-unreach-call.i
\item ./list-properties/list\_search\_false-unreach-call.i
\item ./list-properties/list\_search\_true-unreach-call.i
\item ./list-properties/simple\_built\_from\_end\_true-unreach-call.i
\item ./list-properties/simple\_false-unreach-call.i
\item ./list-properties/simple\_true-unreach-call.i
\item ./list-properties/splice\_false-unreach-call.i

\item ./locks/test\_locks\_10\_true-unreach-call.c
\item ./locks/test\_locks\_11\_true-unreach-call\_false-termination.c
\item ./locks/test\_locks\_13\_true-unreach-call.c
\item ./locks/test\_locks\_14\_false-unreach-call.c
\item ./locks/test\_locks\_14\_true-unreach-call.c
\item ./locks/test\_locks\_15\_true-unreach-call\_false-termination.c
\item ./locks/test\_locks\_5\_true-unreach-call\_false-termination.c
\item ./locks/test\_locks\_6\_true-unreach-call\_false-termination.c
\item ./locks/test\_locks\_8\_true-unreach-call\_false-termination.c
\item ./locks/test\_locks\_9\_true-unreach-call.c

\item ./loop-acceleration/array\_false-unreach-call2.i
\item ./loop-acceleration/array\_false-unreach-call4.i
\item ./loop-acceleration/diamond\_false-unreach-call2.i
\item ./loop-acceleration/diamond\_true-unreach-call1.i
\item ./loop-acceleration/diamond\_true-unreach-call2.i
\item ./loop-acceleration/multivar\_true-unreach-call1.i
\item ./loop-acceleration/nested\_false-unreach-call1.i
\item ./loop-acceleration/phases\_false-unreach-call2.i
\item ./loop-acceleration/simple\_true-unreach-call4.i
\item ./loop-acceleration/underapprox\_true-unreach-call2.i

\item ./loop-invgen/NetBSD\_loop\_false-unreach-call.i
\item ./loop-invgen/SpamAssassin-loop\_false-unreach-call.i
\item ./loop-invgen/down\_true-unreach-call.i
\item ./loop-invgen/fragtest\_simple\_true-unreach-call.i
\item ./loop-invgen/half\_2\_true-unreach-call.i
\item ./loop-invgen/heapsort\_true-unreach-call.i
\item ./loop-invgen/id\_build\_true-unreach-call.i
\item ./loop-invgen/nested9\_true-unreach-call.i
\item ./loop-invgen/seq\_true-unreach-call.i
\item ./loop-invgen/string\_concat-noarr\_true-unreach-call.i

\item ./loop-lit/afnp2014\_true-unreach-call.c.i
\item ./loop-lit/cggmp2005\_true-unreach-call.c.i
\item ./loop-lit/cggmp2005b\_true-unreach-call.c.i
\item ./loop-lit/css2003\_true-unreach-call.c.i
\item ./loop-lit/ddlm2013\_true-unreach-call.c.i
\item ./loop-lit/gj2007b\_true-unreach-call.c.i
\item ./loop-lit/gsv2008\_true-unreach-call.c.i
\item ./loop-lit/jm2006\_true-unreach-call.c.i
\item ./loop-lit/jm2006\_variant\_true-unreach-call.c.i
\item ./loop-lit/mcmillan2006\_true-unreach-call.c.i

\item ./loop-new/count\_by\_1\_true-unreach-call.i
\item ./loop-new/count\_by\_1\_variant\_true-unreach-call.i
\item ./loop-new/count\_by\_2\_true-unreach-call.i
\item ./loop-new/count\_by\_k\_true-unreach-call.i
\item ./loop-new/count\_by\_nondet\_true-unreach-call.i
\item ./loop-new/gauss\_sum\_true-unreach-call.i
\item ./loop-new/half\_true-unreach-call.i
\item ./loop-new/nested\_true-unreach-call.i

\item ./loops/array\_false-unreach-call.i
\item ./loops/insertion\_sort\_true-unreach-call.i
\item ./loops/nec20\_false-unreach-call.i
\item ./loops/sum01\_bug02\_sum01\_bug02\_base.case\_false-unreach-call\_true-termination.i
\item ./loops/terminator\_02\_false-unreach-call\_true-termination.i
\item ./loops/terminator\_03\_true-unreach-call\_true-termination.i
\item ./loops/trex03\_false-unreach-call\_true-termination.i
\item ./loops/veris.c\_OpenSER\_\_cases1\_stripFullBoth\_arr\_true-unreach-call.i
\item ./loops/vogal\_false-unreach-call.i
\item ./loops/while\_infinite\_loop\_2\_true-unreach-call\_false-termination.i

\item ./memory-alloca/HarrisLalNoriRajamani-2010SAS-Fig3-alloca\_true-valid-memsafety.i
\item ./memory-alloca/array02-alloca\_true-valid-memsafety.i
\item ./memory-alloca/cstrcmp-alloca\_true-valid-memsafety.i
\item ./memory-alloca/cstrcpy-alloca\_true-valid-memsafety.i
\item ./memory-alloca/cstrcpy\_unsafe\_false-valid-deref.i
\item ./memory-alloca/cstrlen\_unsafe\_false-valid-deref.i
\item ./memory-alloca/openbsd\_cstrcpy-alloca\_true-valid-memsafety.i
\item ./memory-alloca/reverse\_array\_alloca\_unsafe\_false-valid-deref.i
\item ./memory-alloca/reverse\_array\_unsafe\_false-valid-deref.i
\item ./memory-alloca/selection\_sort\_unsafe\_false-valid-deref.i

\item ./memsafety-ext/dll\_extends\_pointer\_true-valid-memsafety.i
\item ./memsafety-ext/skiplist\_2lvl\_true-valid-memsafety.i
\item ./memsafety-ext/skiplist\_3lvl\_true-valid-memsafety.i
\item ./memsafety-ext/tree\_cnstr\_true-valid-memsafety.i
\item ./memsafety-ext/tree\_dsw\_true-valid-memsafety.i
\item ./memsafety-ext/tree\_of\_cslls\_true-valid-memsafety.i
\item ./memsafety-ext/tree\_parent\_ptr\_true-valid-memsafety.i
\item ./memsafety-ext/tree\_stack\_true-valid-memsafety.i

\item ./memsafety/960521-1\_false-valid-free.i
\item ./memsafety/lockfree-3.0\_true-valid-memsafety.i
\item ./memsafety/lockfree-3.3\_false-valid-memtrack.i
\item ./memsafety/test-0019\_false-valid-memtrack.i
\item ./memsafety/test-0137\_false-valid-deref.i
\item ./memsafety/test-0158\_false-valid-memtrack.i
\item ./memsafety/test-0219\_true-valid-memsafety.i
\item ./memsafety/test-0220\_false-valid-memtrack.i
\item ./memsafety/test-0236\_true-valid-memsafety.i
\item ./memsafety/test-0504\_true-valid-memsafety.i

\item ./ntdrivers-simplified/cdaudio\_simpl1\_false-unreach-call\_true-termination.cil.c
\item ./ntdrivers-simplified/cdaudio\_simpl1\_true-unreach-call\_true-termination.cil.c
\item ./ntdrivers-simplified/diskperf\_simpl1\_true-unreach-call\_true-termination.cil.c
\item ./ntdrivers-simplified/floppy\_simpl3\_false-unreach-call\_true-termination.cil.c
\item ./ntdrivers-simplified/floppy\_simpl3\_true-unreach-call\_true-termination.cil.c
\item ./ntdrivers-simplified/floppy\_simpl4\_false-unreach-call\_true-termination.cil.c
\item ./ntdrivers-simplified/floppy\_simpl4\_true-unreach-call\_true-termination.cil.c
\item ./ntdrivers-simplified/kbfiltr\_simpl1\_true-unreach-call\_true-termination.cil.c
\item ./ntdrivers-simplified/kbfiltr\_simpl2\_false-unreach-call\_true-termination.cil.c
\item ./ntdrivers-simplified/kbfiltr\_simpl2\_true-unreach-call\_true-termination.cil.c

\item ./ntdrivers/cdaudio\_false-unreach-call.i.cil.c
\item ./ntdrivers/cdaudio\_true-unreach-call.i.cil.c
\item ./ntdrivers/diskperf\_false-unreach-call.i.cil.c
\item ./ntdrivers/diskperf\_true-unreach-call.i.cil.c
\item ./ntdrivers/floppy2\_true-unreach-call.i.cil.c
\item ./ntdrivers/floppy\_false-unreach-call.i.cil.c
\item ./ntdrivers/floppy\_true-unreach-call.i.cil.c
\item ./ntdrivers/kbfiltr\_false-unreach-call.i.cil.c
\item ./ntdrivers/parport\_false-unreach-call.i.cil.c
\item ./ntdrivers/parport\_true-unreach-call.i.cil.c

\item ./product-lines/elevator\_spec13\_product21\_true-unreach-call.cil.c
\item ./product-lines/email\_spec3\_product19\_false-unreach-call.cil.c
\item ./product-lines/email\_spec4\_product17\_true-unreach-call.cil.c
\item ./product-lines/email\_spec6\_product15\_false-unreach-call.cil.c
\item ./product-lines/email\_spec9\_product21\_false-unreach-call.cil.c
\item ./product-lines/minepump\_spec2\_product34\_false-unreach-call.cil.c
\item ./product-lines/minepump\_spec2\_product38\_true-unreach-call.cil.c
\item ./product-lines/minepump\_spec2\_product60\_true-unreach-call.cil.c
\item ./product-lines/minepump\_spec5\_product20\_true-unreach-call.cil.c
\item ./product-lines/minepump\_spec5\_product61\_true-unreach-call.cil.c

\item ./pthread-atomic/dekker\_true-unreach-call.i
\item ./pthread-atomic/lamport\_true-unreach-call.i
\item ./pthread-atomic/peterson\_true-unreach-call.i
\item ./pthread-atomic/qrcu\_false-unreach-call.i
\item ./pthread-atomic/qrcu\_true-unreach-call.i
\item ./pthread-atomic/read\_write\_lock\_false-unreach-call.i
\item ./pthread-atomic/read\_write\_lock\_true-unreach-call.i
\item ./pthread-atomic/scull\_true-unreach-call.i
\item ./pthread-atomic/szymanski\_true-unreach-call.i
\item ./pthread-atomic/time\_var\_mutex\_true-unreach-call.i

\item ./pthread-ext/01\_inc\_true-unreach-call.i
\item ./pthread-ext/04\_incdec\_cas\_true-unreach-call.i
\item ./pthread-ext/05\_tas\_true-unreach-call.i
\item ./pthread-ext/08\_rand\_cas\_true-unreach-call.i
\item ./pthread-ext/18\_read\_write\_lock\_true-unreach-call.i
\item ./pthread-ext/19\_time\_var\_mutex\_true-unreach-call.i
\item ./pthread-ext/23\_lu-fig2.fixed\_true-unreach-call.i
\item ./pthread-ext/25\_stack\_longest\_true-unreach-call.i
\item ./pthread-ext/27\_Boop\_simple\_vf\_false-unreach-call.i
\item ./pthread-ext/30\_Function\_Pointer3\_vs\_true-unreach-call.i

\item ./pthread-lit/fk2012\_true-unreach-call.i
\item ./pthread-lit/fkp2013\_false-unreach-call.i
\item ./pthread-lit/fkp2013\_true-unreach-call.i
\item ./pthread-lit/fkp2013\_variant\_true-unreach-call.i
\item ./pthread-lit/fkp2014\_true-unreach-call.i
\item ./pthread-lit/qw2004\_false-unreach-call.i
\item ./pthread-lit/qw2004\_true-unreach-call.i
\item ./pthread-lit/qw2004\_variant\_true-unreach-call.i
\item ./pthread-lit/sssc12\_true-unreach-call.i
\item ./pthread-lit/sssc12\_variant\_true-unreach-call.i

\item ./pthread-wmm/mix000\_power.oepc\_false-unreach-call.i
\item ./pthread-wmm/mix008\_pso.oepc\_false-unreach-call.i
\item ./pthread-wmm/mix041\_power.oepc\_false-unreach-call.i
\item ./pthread-wmm/mix054\_tso.opt\_false-unreach-call.i
\item ./pthread-wmm/mix055\_power.opt\_false-unreach-call.i
\item ./pthread-wmm/rfi009\_rmo.oepc\_false-unreach-call.i
\item ./pthread-wmm/rfi009\_rmo.opt\_false-unreach-call.i
\item ./pthread-wmm/rfi009\_tso.oepc\_false-unreach-call.i
\item ./pthread-wmm/safe037\_tso.oepc\_true-unreach-call.i
\item ./pthread-wmm/thin002\_pso.oepc\_true-unreach-call.i

\item ./pthread/fib\_bench\_longer\_false-unreach-call.i
\item ./pthread/fib\_bench\_longer\_true-unreach-call.i
\item ./pthread/fib\_bench\_longest\_true-unreach-call.i
\item ./pthread/queue\_longest\_false-unreach-call.i
\item ./pthread/queue\_ok\_longer\_true-unreach-call.i
\item ./pthread/queue\_ok\_longest\_true-unreach-call.i
\item ./pthread/queue\_ok\_true-unreach-call.i
\item ./pthread/sigma\_false-unreach-call.i
\item ./pthread/singleton\_false-unreach-call.i
\item ./pthread/stateful01\_false-unreach-call.i

\item ./recursive-simple/afterrec\_2calls\_false-unreach-call.c
\item ./recursive-simple/afterrec\_true-unreach-call.c
\item ./recursive-simple/fibo\_2calls\_2\_false-unreach-call.c
\item ./recursive-simple/fibo\_2calls\_5\_false-unreach-call.c
\item ./recursive-simple/id\_b5\_o10\_true-unreach-call.c
\item ./recursive-simple/id\_i10\_o10\_true-unreach-call.c
\item ./recursive-simple/id\_o10\_false-unreach-call.c
\item ./recursive-simple/id\_o200\_false-unreach-call.c
\item ./recursive-simple/sum\_2x3\_false-unreach-call.c
\item ./recursive-simple/sum\_non\_true-unreach-call.c

\item ./recursive/Ackermann03\_true-unreach-call.c
\item ./recursive/Addition01\_true-unreach-call\_true-termination.c
\item ./recursive/Addition03\_false-unreach-call.c
\item ./recursive/EvenOdd03\_false-unreach-call\_false-termination.c
\item ./recursive/Fibonacci01\_true-unreach-call.c
\item ./recursive/Fibonacci02\_true-unreach-call\_true-termination.c
\item ./recursive/McCarthy91\_true-unreach-call.c
\item ./recursive/Primes\_true-unreach-call.c
\item ./recursive/gcd02\_true-unreach-call.c
\item ./recursive/recHanoi03\_true-unreach-call\_true-termination.c

\item ./seq-mthreaded/pals\_STARTPALS\_Triplicated\_true-unreach-call.ufo.BOUNDED-10.pals.c
\item ./seq-mthreaded/pals\_floodmax.5\_true-unreach-call.ufo.BOUNDED-10.pals.c
\item ./seq-mthreaded/pals\_lcr.7\_true-unreach-call.ufo.UNBOUNDED.pals.c
\item ./seq-mthreaded/pals\_opt-floodmax.5\_false-unreach-call.1.ufo.BOUNDED-10.pals.c
\item ./seq-mthreaded/pals\_opt-floodmax.5\_false-unreach-call.1.ufo.UNBOUNDED.pals.c
\item ./seq-mthreaded/pals\_opt-floodmax.5\_false-unreach-call.2.ufo.BOUNDED-10.pals.c
\item ./seq-mthreaded/rekh\_ctm\_true-unreach-call.2.c
\item ./seq-mthreaded/rekh\_ctm\_true-unreach-call.4.c
\item ./seq-mthreaded/rekh\_nxt\_false-unreach-call.2.M4.c
\item ./seq-mthreaded/rekh\_nxt\_true-unreach-call.3.M1.c

\item ./seq-pthread/cs\_fib\_longer\_false-unreach-call.i
\item ./seq-pthread/cs\_fib\_longer\_true-unreach-call.i
\item ./seq-pthread/cs\_fib\_true-unreach-call.i
\item ./seq-pthread/cs\_lamport\_true-unreach-call.i
\item ./seq-pthread/cs\_peterson\_true-unreach-call.i
\item ./seq-pthread/cs\_queue\_false-unreach-call.i
\item ./seq-pthread/cs\_stack\_false-unreach-call.i
\item ./seq-pthread/cs\_stateful\_false-unreach-call.i
\item ./seq-pthread/cs\_sync\_true-unreach-call.i
\item ./seq-pthread/cs\_szymanski\_true-unreach-call.i

\item ./ssh-simplified/s3\_clnt\_1\_true-unreach-call.cil.c
\item ./ssh-simplified/s3\_clnt\_2\_true-unreach-call\_true-termination.cil.c
\item ./ssh-simplified/s3\_clnt\_4\_false-unreach-call.cil.c
\item ./ssh-simplified/s3\_clnt\_4\_true-unreach-call.cil.c
\item ./ssh-simplified/s3\_srvr\_11\_false-unreach-call.cil.c
\item ./ssh-simplified/s3\_srvr\_14\_false-unreach-call.cil.c
\item ./ssh-simplified/s3\_srvr\_1\_false-unreach-call.cil.c
\item ./ssh-simplified/s3\_srvr\_1\_true-unreach-call.cil.c
\item ./ssh-simplified/s3\_srvr\_4\_true-unreach-call.cil.c
\item ./ssh-simplified/s3\_srvr\_6\_true-unreach-call.cil.c

\item ./ssh/s3\_srvr.blast.01\_true-unreach-call.i.cil.c
\item ./ssh/s3\_srvr.blast.02\_false-unreach-call.i.cil.c
\item ./ssh/s3\_srvr.blast.03\_false-unreach-call.i.cil.c
\item ./ssh/s3\_srvr.blast.04\_false-unreach-call.i.cil.c
\item ./ssh/s3\_srvr.blast.06\_true-unreach-call.i.cil.c
\item ./ssh/s3\_srvr.blast.07\_true-unreach-call.i.cil.c
\item ./ssh/s3\_srvr.blast.08\_false-unreach-call.i.cil.c
\item ./ssh/s3\_srvr.blast.10\_true-unreach-call.i.cil.c
\item ./ssh/s3\_srvr.blast.12\_true-unreach-call.i.cil.c
\item ./ssh/s3\_srvr.blast.13\_false-unreach-call.i.cil.c

\item ./systemc/kundu2\_false-unreach-call\_false-termination.cil.c
\item ./systemc/kundu\_true-unreach-call\_false-termination.cil.c
\item ./systemc/pc\_sfifo\_2\_false-unreach-call\_false-termination.cil.c
\item ./systemc/token\_ring.03\_false-unreach-call\_false-termination.cil.c
\item ./systemc/token\_ring.12\_true-unreach-call\_false-termination.cil.c
\item ./systemc/token\_ring.13\_false-unreach-call\_false-termination.cil.c
\item ./systemc/token\_ring.14\_false-unreach-call\_false-termination.cil.c
\item ./systemc/toy2\_false-unreach-call\_false-termination.cil.c
\item ./systemc/toy\_true-unreach-call\_false-termination.cil.c
\item ./systemc/transmitter.12\_false-unreach-call\_false-termination.cil.c

\item ./termination-crafted-lit/AliasDarteFeautrierGonnord-SAS2010-Fig2a\_true-termination.c
\item ./termination-crafted-lit/HeizmannHoenickeLeikePodelski-ATVA2013-Fig1\_true-termination.c
\item ./termination-crafted-lit/LarrazOliverasRodriguez-CarbonellRubio-FMCAD2013-Fig1\_true-termination.c
\item ./termination-crafted-lit/PodelskiRybalchenko-TACAS2011-Fig4\_true-termination.c
\item ./termination-crafted-lit/PodelskiRybalchenko-VMCAI2004-Ex1\_true-termination.c
\item ./termination-crafted-lit/Velroyen\_false-termination.c
\item ./termination-crafted-lit/cstrcspn\_true-termination.c
\item ./termination-crafted-lit/cstrpbrk\_true-termination.c
\item ./termination-crafted-lit/cstrspn\_true-termination.c
\item ./termination-crafted-lit/gcd1\_true-termination.c

\item ./termination-crafted/4BitCounterPointer\_true-termination.c
\item ./termination-crafted/Cairo\_true-termination.c
\item ./termination-crafted/Copenhagen\_true-termination.c
\item ./termination-crafted/NonTermination3\_false-termination.c
\item ./termination-crafted/NonTermination4\_false-termination.c
\item ./termination-crafted/NonTerminationSimple9\_false-termination.c
\item ./termination-crafted/Parallel\_true-termination.c
\item ./termination-crafted/Pure2Phase\_true-termination.c
\item ./termination-crafted/RecursiveNonterminating\_false-termination.c
\item ./termination-crafted/Stockholm\_true-termination.c

\item ./termination-memory-alloca/b.09\_assume-alloca\_true-termination.c.i
\item ./termination-memory-alloca/b.10-alloca\_true-termination.c.i
\item ./termination-memory-alloca/b.13-alloca\_true-termination.c.i
\item ./termination-memory-alloca/cstrncat-alloca\_true-termination.c.i
\item ./termination-memory-alloca/java\_Break-alloca\_true-termination.c.i
\item ./termination-memory-alloca/java\_LogBuiltIn-alloca\_true-termination.c.i
\item ./termination-memory-alloca/java\_Nested-alloca\_true-termination.c.i
\item ./termination-memory-alloca/java\_Sequence-alloca\_true-termination.c.i
\item ./termination-memory-alloca/openbsd\_cbzero-alloca\_true-termination.c.i
\item ./termination-memory-alloca/openbsd\_cstrcmp-alloca\_true-termination.c.i

\item ./termination-numeric/Avg\_true\_true-termination.c
\item ./termination-numeric/Binomial\_true-termination.c
\item ./termination-numeric/Et1\_true\_true-termination.c
\item ./termination-numeric/Parts\_true-termination.c
\item ./termination-numeric/TwoWay\_true-termination.c
\item ./termination-numeric/b.03-no-inv\_assume\_true-termination.c
\item ./termination-numeric/easySum\_true-termination.c
\item ./termination-numeric/java\_LogBuiltIn\_true-termination.c
\item ./termination-numeric/java\_Nested\_true-termination.c
\item ./termination-numeric/rec\_counter1\_true-termination.c
\end{itemize}

\end{document}